\documentclass{aa}

\usepackage{graphicx}
\usepackage{txfonts}
\usepackage{hyperref}
\usepackage{multirow}

\hypersetup{colorlinks = true, citecolor = {blue}, urlcolor= {blue}}

\def\PGPU{$\varphi-$GPU}

\def\gapprox{\;\rlap{\lower 3.0pt                       % approximately smaller
        \hbox{$\sim$}}\raise 2.5pt\hbox{$>$}\;}
\def\lapprox{\;\rlap{\lower 3.1pt                       % approximately smaller
        \hbox{$\sim$}}\raise 2.7pt\hbox{$<$}\;}

\newcommand{\be}{ \begin{equation} }
\newcommand{\ee}{\end{equation}}

\newcommand{\ben}{\begin{enumerate}}
\newcommand{\een}{\end{enumerate}}

\renewenvironment{thebibliography}[1]{\begin{oldthebibliography}{#1}\setlength{\parskip}{0ex}\setlength{\itemsep}{0ex}}{\end{oldthebibliography}}

\usepackage{diagbox}
\usepackage{siunitx}
\newcommand{\orcid}[1]{\href{https://orcid.org/#1}{\protect\includegraphics[width=8pt]{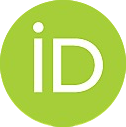}}}
\makeatletter
\renewcommand*\aa@pageof{, page \thepage{} of \pageref*{LastPage}}
\makeatother

\usepackage[dvipsnames]{xcolor}

\definecolor{darkgreen}{RGB}{31, 207, 31}

\begin{document}

\title{How our proto-nuclear star cluster formed and grew due to early globular cluster disruption} 

\subtitle{I. Case of low masses}

\author{
D.~Kuvatova
\inst{1,2,3}\orcid{0000-0002-5937-4985}
\and
M.~Ishchenko
\inst{4,1,2}\orcid{0000-0002-6961-8170}
\and
P.~Berczik
\inst{1,2,4,6}\orcid{0000-0003-4176-152X}
\and
O.~Veles
\inst{4,1}\orcid{0000-0001-5221-2513}
\and
O. Sobodar
\inst{4,1}\orcid{0000-0001-5788-9996}
\and
T.~Panamarev
\inst{5,2}\orcid{0000-0002-1090-4463}
}

\institute{Nicolaus Copernicus Astronomical Centre Polish Academy of Sciences, ul. Bartycka 18, 00-716 Warsaw, Poland
           \and
           Fesenkov Astrophysical Institute, Observatory 23, 050020 Almaty, Kazakhstan
           \and
           Faculty of Physics and Technology, Al-Farabi Kazakh National University,
           al-Farabi ave. 71, 050040 Almaty, Kazakhstan
           \and
           Main Astronomical Observatory, National Academy of Sciences of Ukraine,
           27 Akademika Zabolotnoho St, 03143 Kyiv, Ukraine  
           \and 
           Rudolf Peierls Centre for Theoretical Physics, Parks Road, OX1 3PU, Oxford, UK
           \and
           Szechenyi Istvan University, Space Technology and Space Law Research Center, H-9026 Gyor, Egyetem ter 1. Hungary \\
           }
   
\date{Received xxx / Accepted xxx}

\abstract
{To date, two main mechanisms have been proposed for the formation and growth of nuclear star clusters (NSCs) in galaxies. The first suggests in situ star formation from gas that has migrated to the central regions from the galaxy’s outskirts, while the second involves the accretion of stars from disrupted globular clusters (GCs) onto the galactic centre. However, the relative importance of these mechanisms in the evolution of NSCs across different galaxy morphologies remains an open question.}
{We investigate the accretion of GC stars on early cosmological timescales through detailed $N$-body simulations of theoretical GC models to assess the role of this mechanism in Milky Way-like (MW-like) galaxies.}
{For the dynamical modelling, we used the updated parallel $N$-body code $\varphi$-GPU, including stellar evolution. We prepared three sets of GC models with different half-mass radii (r$_{\rm hm}$), each consisting of 50 full N-body GC models, and integrated these models in an external, time-variable MW-like potential taken from the cosmological database IllustrisTNG-100. The simulations cover the time interval from -10 Gyr to -5 Gyr, enabling us to assess the rate of early stellar accretion onto the proto-NSC.}
{We find that GC models with average orbital eccentricities of 0.4–0.5 and orbits oriented perpendicular to the galactic disc contribute most significantly to the mass of the proto-NSC formation. Accretion is especially efficient in the first billion years (Gyr) and in compact GC models with r$_{\rm hm}$ = 1 pc. In all sets, the dominant accreted stellar population consists of low-mass stars ($\approx$0.33 M$_{\odot}$). However, the accreted mass alone is insufficient to fully account for the current NSC mass.}
{Based on our extended set of numerical simulations, we obtained an average lower limit of mass contribution ($\approx$ 6\%) to the NSC from investigated  GCs. The fraction of mass contribution from individual disrupted GCs can significantly vary from 0.1\% up to 90\%. Generally, we conclude that the GC stellar accretion channel alone might not be sufficient to ensure the present-day MW galaxy's NSC mass budget.}

\keywords{Galaxy: globular clusters: general - Galaxy: centre - Galaxy: nucleus - Galaxy: evolution - Methods: numerical}

\titlerunning{NSC formation and growth due to early GCs disruption}
\authorrunning{D.~Kuvatova et al.}
\maketitle

%%%%%%%%%%%%%%%%%%%%%%%%%%%%%%%%%%%%%%%%%%%%%%%%%%%%%%%%%%%%%%%%%%%%
\section{Introduction}\label{sec:Intr}
%%%%%%%%%%%%%%%%%%%%%%%%%%%%%%%%%%%%%%%%%%%%%%%%%%%%%%%%%%%%%%%%%%%%
At the centres of most galaxies, there lie a supermassive black hole (SMBH) and a massive, dense stellar cluster \citep{Kormendy2013, Neumayer2020}. Unlike the SMBHs, the clusters are directly observable, particularly in late-type galaxies \citep{Neumayer2020}. The sizes of these clusters, known as nuclear star clusters (NSCs), are comparable to those of typical globular clusters (GCs), with a half-light radius of approximately 3.5--5.5 pc \citep{Boker2004, Baumgardt2021}. However, NSCs significantly surpass GCs in brightness \citep[$\sim \! 10^7 \text{ L}_\odot$;][]{Matthews1997, Boker2002} due to their greater mass ($\sim \! 10^6$--$10^7 \text{ M}_\odot$) and the presence of both young and old stellar populations \citep{Ho1995, Boker2002, Lotz2004} with unique evolutionary features owing to their dense environments, such as stellar collisions and mergers leading to rejuvenated stars, and the depletion of late-type giants due to envelope stripping \citep{Genzel2003}. Certain correlations have been observed between NSCs and their host galaxies \citep{Cote2006, Wehner2006, Ferrarese2006, Rossa2006}. In particular, a number of studies have noted a correlation between the velocity dispersion of NSCs and that of their host galaxies, as well as between NSC masses and the bulge masses. Some of these relations are also present for SMBHs, raising questions about the mutual role of NSCs and SMBHs in their evolutionary processes.

The morphology of NSCs varies significantly between galaxies and can include both disc-like and spherical components \citep{Seth2006}. Apparently, the morphology of an NSC partially or fully reflects that of its surrounding environment \citep{Capuzzo-Dolcetta2008ApJ}. Understanding the morphology and kinematics of NSCs plays a key role in deciphering the mechanisms of their formation and growth \citep{DeLorenzi2013}. Overall, NSC formation is thought to occur via two main scenarios: 
\begin{itemize}
    \item The accretion of stellar clusters onto galactic centre \citep[often referred to as the dissipationless, cluster-inspiral, or dry-merger scenarios;][]{Tremaine1975, Lotz2001, Capuzzo-Dolcetta2008ApJ, Capuzzo-Dolcetta2008MNRAS, Agarwal2011, Antonini2013, Gnedin2014, Tsatsi2017, Fahrion2022, Gao2024} 

    \item The formation of stars from gas that has previously migrated to the galactic centre \citep[also known as the dissipative, in situ formation, or wet formation scenarios;][]{Loose1982, Milosavljevic2004, Seth2006, Nayakshin2009, Aharon2015}. 
\end{itemize}

Both scenarios are likely valid \citep{Hartmann2011, Leigh2015}, with their relative contributions depending on the environment in which the NSC forms. As an example, \cite{Fahrion2022} showed that the dominant NSC formation channel depends on NSC (and host galaxy) mass. The primary question is whether these mechanisms are dominant and (if not) what their role is in the formation and growth of the NSC. Additionally, the mechanisms of NSC formation and subsequent growth may not necessarily be identical. In early-type galaxies, where gas concentration is low, the dominant mechanism for NSC growth is expected to be cluster-inspiral, although in situ formation cannot be excluded during the early stages of such galaxies’ history. The diversity of stellar populations,  manifested by the coexistence of different stellar generations and evidence for multiple star-formation episodes in NSCs, also supports the simultaneous operation of both mechanisms. Moreover, as the central region grows in mass, it can more efficiently attract gas, triggering new star formation or enhancing its rate  \citep{Schinnerer2006, Partmann2025}. NSC formation might also result from galaxy interactions and mergers during their dynamic evolution. It is logical to assume that during the accretion of randomly distributed stellar clusters onto the galactic centre would not impart their orbital angular momentum to the forming NSC. However, simulations of randomly distributed stellar cluster accretion have shown that special initial conditions are not required for NSC rotation to develop, although rotation becomes more pronounced if clusters infall with their orientations aligned with the galactic disc's rotation \citep{Seth2008}.

Unlike NSCs in other galaxies, the NSC of the Milky Way (MW) can be resolved into individual stars due to its relative proximity \citep[$\sim \! 8.1 \pm 0.1 \text{ kpc}$;][]{Gravity2019, Do2019}. However, interstellar matter along the line of sight to the Galactic centre (GalC) imposes limitations on observing certain NSC properties. The MW's NSC is thought to have a mass of $\sim \! 10^7 \text{ M}_\odot$ \citep{Schodel2008} and contains the SMBH (Sgr A*) with a mass of $\sim \! 4.3 \times 10^6 \text{ M}_\odot$ \citep{Gillessen2009}. Observations indicate an increase in star formation rates in the MW’s NSC over the past few hundred million years \citep{Blum2003, Pfuhl2011}. In the region within $\sim$ 0.5 pc of the GalC, young stars with ages around 6 Myr dominate \citep{Paumard2006, Feldmeier-Krause2015}, while the fraction of old stars comparable in age to MW GCs is extremely low \citep{Do2015}. This supports the dominance of the in situ scenario. However, in a $\sim$ 2.5 pc region around the GalC, 80 \% of stars are older than 5 Gyr \citep{Blum2003, Pfuhl2011}. The kinematic features of the MW’s NSC, such as the detection of a dynamically distinct, metal-poor component and the evidence for non-isotropic rotation, favour the cluster-inspiral scenario \citep{Do2020}. The deficit of GCs in the central region compared to the overall stellar distribution in the MW may also be a consequence of GC inspiralling and destruction during NSC formation \citep{Lotz2001, Capuzzo-Dolcetta2009}. Additionally, the MW’s NSC exhibits rotation aligned with the Galactic disc's rotation \citep{Trippe2008, Schodel2009}. The cluster-inspiral scenario could also partially or fully explain the $\gamma$-ray and X-ray excess observed in the GalC as numerous millisecond pulsars, cataclysmic variables, and black holes formed in dense star clusters may have been delivered to the GalC via inspiralling \citep{Arca-Sedda2018}.

To validate the cluster-inspiral mechanism for the formation and growth of NSCs, detailed numerical simulations of stellar cluster dynamics within the potential field of the host galaxy are typically performed, with special attention paid to cluster-GalC and cluster-cluster interactions \citep{Oh2000, Bekki2004, Hartmann2011}. A significant factor in these studies is the consideration of dynamics within a live, time-dependent potential, rather than a static one \citep{Bekki2010, Bekki2010MNRAS}. 

In this paper, we investigate the cluster-inspiral scenario for the formation and growth of the NSC via the accretion of the MW's GCs. Overall, GCs are tightly bound stellar conglomerations associated with all types of galaxies. The survivability of GCs during their passages near the GalC depends on their mass and concentration parameter, defined as the logarithm of the ratio of the tidal and core radii, $c=\log{(r_{\rm t}/r_{\rm c})}$. Larger GC mass and higher concentration allow a cluster to survive more passages near the GalC. For instance, \cite{Miocchi2006} simulated the dynamics of GCs in a triaxial system and found a correlation between the mass-loss timescale, $\tau$, and the concentration parameter, $\tau \simeq 14 c^{6.1} t_{\rm dyn}$. Orbital energy loss and structural changes in GCs (especially for the higher masses $\gtrsim$ 10$^6$ M$_\odot$) can occur due to the dynamical friction and tidal interactions with the galaxy's potential (tidal dissipation), resulting in GC deceleration and potential accretion onto the GalC, thereby contributing mass \citep{Pesce1992, Capuzzo-Dolcetta1993, Capuzzo-Dolcetta2005}.
Depending on the compactness of the GCs, the number required for NSC formation varies, ranging from several dozen for compact clusters to several hundred for looser ones. \cite{Antonini2013} used  simulations to find that the rapid formation of NSCs from GCs depends heavily on the presence or absence of a SMBH at the centre. \cite{Perets2014} conducted a series of simulations involving the sequential infall of 12 identical GCs (each -- $1.1 \times 10^6 \text{ M}_\odot$) onto the inner region of the Milky Way, which contains a SMBH ($\text{M}_\bullet = 4 \times 10^6 \text{ M}_\odot$), from circular orbits with a radius of 20 pc and random initial parameters. Their results yielded a total NSC mass comparable to observational data: $1.4 \times 10^7 \text{ M}_\odot$.

Our previous studies \citep{Ishchenko2023a, Ishchenko2023b} have demonstrated that during their dynamical evolution in a time-varying MW-like external potential from cosmological database IllustrisTNG-100 \citep{Nelson2019}, some MW GCs might closely approach the GalC on a cosmological timescale, potentially contributing to the growth of the NSC's mass. The frequency of close GC passages near the central region of the Galaxy was estimated and found to be too low to fully account for the NSC's mass if we are considering only the periodic capture of stars during such encounters. This led us to conclude that it is essential to consider that some GCs may not have survived to the present day, having been tidally disrupted through interactions with the NSC and thereby contributing significantly more to its mass.

The primary objective of this study is to perform detailed dynamic N-body modelling of theoretical GC models within a time-varying Milky Way-like potential, accounting for stellar evolution. This aims to evaluate the contribution of accreted mass from GCs with varying orbital parameters, including cases of complete cluster disruption.

The paper is organised as follows. Section \ref{sec:Mesh} describes the generation of initial conditions for point-mass models of GCs, integrated backward over 10 Gyrs to assess close encounters with the GalC. Section \ref{sec:integration} details the full N-body integration of theoretical GCs models, incorporating stellar evolution and  physical model parameters. Section \ref{sec:analysis} presents our analysis of stellar accretion onto the proto-NSC, while Section \ref{sec:dis-con} discusses and summarises the results.

%%%%%%%%%%%%%%%%%%%%%%%%%%%%%%%%%%%%%%%%%%%%%%%%%%%%%%%%%%%%%%%%%%%%%
\section{Mesh of initial condition distributions and GC orbital integration as a point mass}\label{sec:Mesh}
%%%%%%%%%%%%%%%%%%%%%%%%%%%%%%%%%%%%%%%%%%%%%%%%%%%%%%%%%%%%%%%%%%%%

%-------------------------------------------------------------------------%
\begin{figure}[htb!]
\centering
\includegraphics[width=0.99\linewidth]{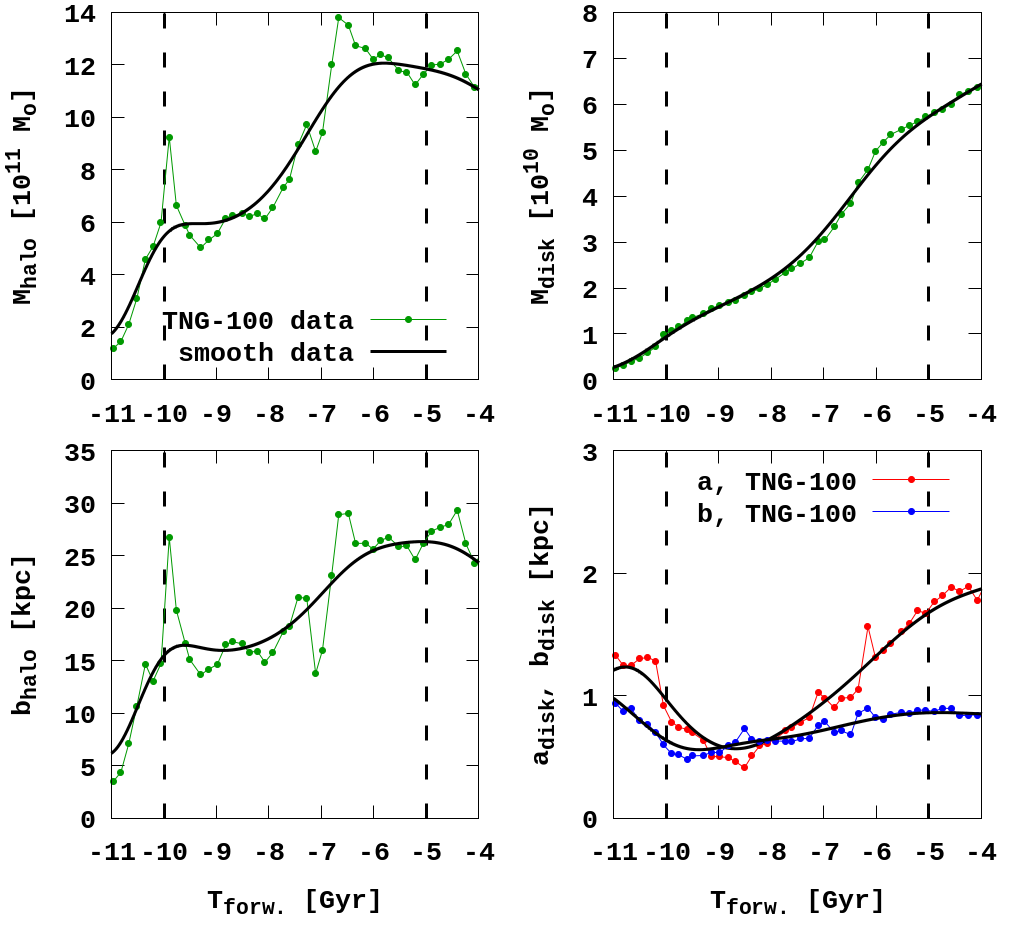}
\caption{Evolution of halo and disc masses, along with their characteristic scales for {\tt 411321} external potential. Green, red, and blue lines with dots show the parameters recovered from the IllustrisTNG-100 data. Black solid lines correspond to the values after the interpolation and smoothing with a 1~Myr time step that was used in the orbital integration. The dashed lines represent the 5 Gyr duration.}
\label{fig:subgalo}
\end{figure}
%-------------------------------------------------------------------------%

%-------------------------------------------------------------------------%
\begin{figure}[tb!]
\centering
\includegraphics[width=0.99\linewidth]{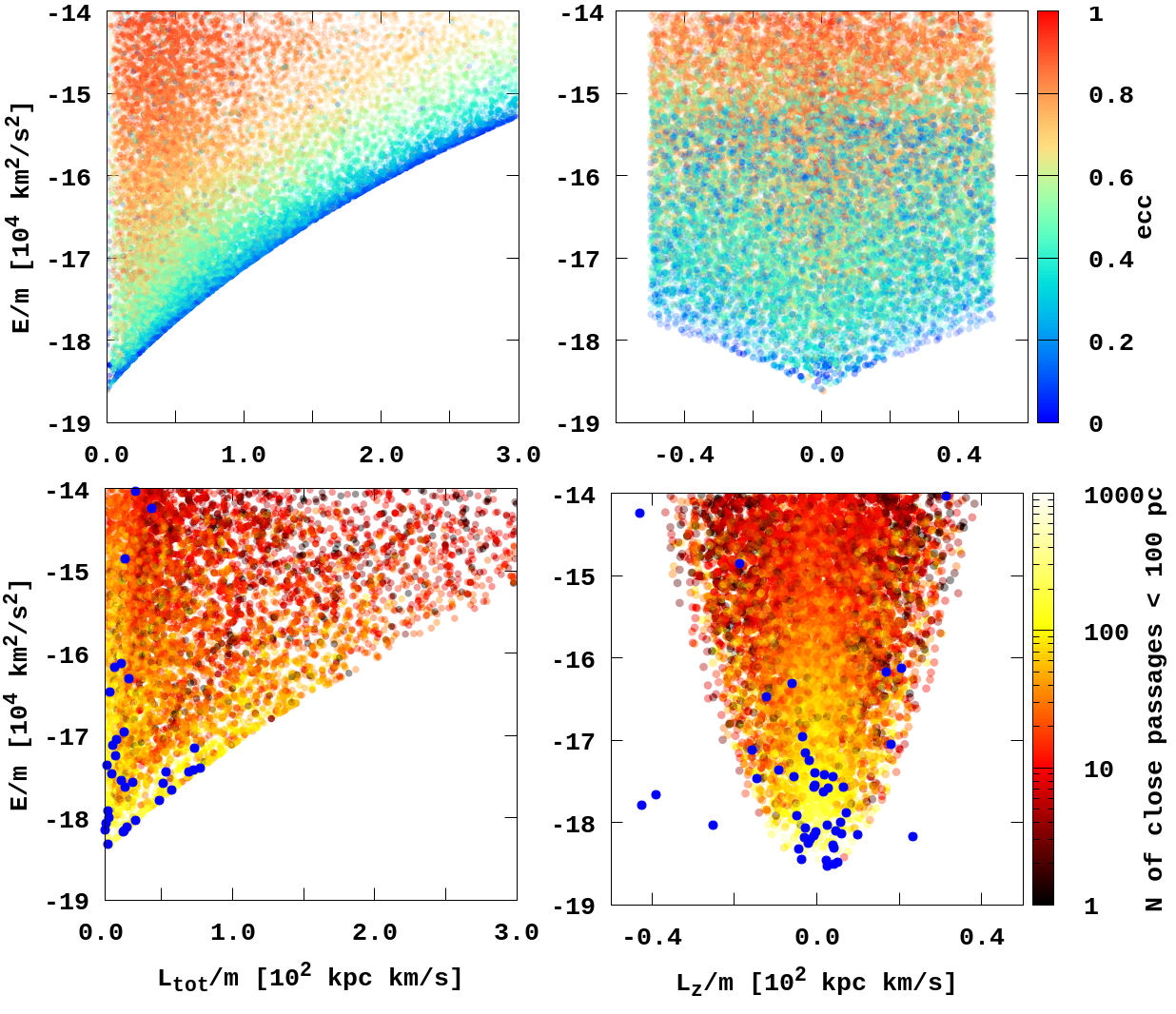}
\caption{Distribution of the GCs in phase space: E vs L$_{\rm tot}$, E vs L$_{\rm perp}$, and E vs L$_{\rm z}$ in {\tt 411321} external potential. Top: Initial phase space distribution, where the colour-coding shows orbital mean {\tt ecc}, based on 5 Gyr of integration. Bottom:\ Interactions (close passages) with the GalC, based on 5 Gyr of integration. Colour-coding represents the number of such events. Black dots represent selected GCs for future $N$-body simulations.}
\label{fig:init-grid-dr-6}
\end{figure}
%-------------------------------------------------------------------------%

According to our current understanding, it is clear that the initial condition phase space for Galactic GCs progenitors positions and velocities is quite broad, but obviously limited by the Galaxy assembling history in the early Universe  \citep{Kravtsov2005, Pagnini2023, Ishchenko2023a}. It is also evident that we cannot  numerically model the full set of GCs initial conditions phase space distribution. On this basis, we restricted our study of the potential NSC donor GCs with some limits in angular momentum and energy space. As a starting point, for our initial condition generation, we chose the well-studied {\tt 411321} TNG-TVP potential \citep{Ishchenko2023a}, as detailed in Fig. \ref{fig:subgalo}. 

To obtain the spatial scales of the discs and dark matter haloes, we decomposed the mass distribution using the Miyamoto--Nagai (MN)~$\Phi_{\rm d} (R,z)$ potential \citep{Miyamoto1975} and Navarro--Frenk--White (NFW) $\Phi_{\rm h} (R,z)$ potential \citep[][]{NFW1997}, expressed as 
\begin{equation}
\begin{split}
\Phi_{\rm tot} &= \Phi_{\rm d} (R,z) + \Phi_{\rm h} (R,z) = \\
&= - \frac{GM_{\rm d}}{\sqrt{R^{2}+\Bigl(a_{\rm d}+\sqrt{z^{2}+b^{2}_{\rm d}}\Bigr)^{2}}} - 
\frac{GM_{\rm h}\cdot{\rm ln}\Bigr(1+\frac{\sqrt{R^{2}+z^{2}}}{b_{\rm h}}\Bigl)}{\sqrt{R^{2}+z^{2}}},
\end{split}
\end{equation}
where 
$R=\sqrt{x^{2}+y^{2}}$ is the planar galactocentric radius, 
$z$ is the distance above the plane of the disc, 
$G$ is the gravitational constant, 
$a_{\rm d}$ is the disc scale length, 
$b_{\rm d,h}$ are the disc and halo scale heights, respectively, and 
$M_{\rm d}$ and 
$M_{\rm h}=4\pi\rho_{0}b^{3}_{\rm h}$ ($\rho_{0}$ is the central mass density of the halo) are the masses of the disc and halo, respectively. 

In this potential, we step back to the 10 Gyr look-back time and put our theoretical GC's centres in the Galactocentric Cartesian coordinate frame. In this potential for the selected time, we see that the minimum specific energy level is around -18.6 $\times$10$^{4}$ km$^{2}$ s$^{-2}$. Thus, in the generation routine, we started our energies from -19 to -14 $\times$10$^{4}$ km$^{2}$ s$^{-2}$. For the specific angular momentum space, we chose the limits: 0--3 $\times$10$^{2}$ kpc km s$^{-1}$ for L$_{\rm tot}$ to cover a full range of orbital {\tt ecc} almost from 0 to 1 and from -0.5 to +0.5 $\times$10$^{2}$ kpc km s$^{-1}$ for L$_{\rm z}$ to cover a significant range of all prograde and retrograde orbits of the currently observed and modelled Galactic GCs \citep{Ishchenko2023b, Ishchenko2024massloss}. 

Using the simple rejection method, inside this mesh of energy and angular momentum, we generated limits randomly with a uniform distribution of more than 16$k$ initial models. For each of these points in a phase space, E, L$_{\rm tot}$, L$_{\rm perp}$, L$_{\rm z}$, we find a corresponding random position and velocity for our theoretical GC centres in the Galactocentric Cartesian coordinate frame. In Fig. \ref{fig:init-grid-dr-6} (top), we present the initial phase space distribution of GCs in a specific energy and angular momentum space. We are aware of the limitations of our approach for the theoretical model GC centres' phase space distribution, which is connected with the underlying assumption of the similarities between currently observed and currently already tidally destroyed GCs. However, we use this assumption here as a first step to generate our set of theoretical GCs. In the future, we will also investigate other possibilities.

Using this set of initial positions and velocities, we ran our orbital integration for 5 Gyr, from -10 Gyr look-back time. For this task we used the high-order parallel dynamical $N$-body code \PGPU\footnote{$N$-body code \PGPU: \\~\url{ https://github.com/berczik/phi-GPU-mole}} \citep{Berczik2011,BSW2013}. The code is based on the fourth-order Hermite integration scheme with hierarchical individual block time steps. In Fig. \ref{fig:init-grid-dr-6}, we present mean values for orbital {\tt ecc} via colour palette, calculated during 5 Gyr of dynamical integration time. {The orbital eccentricity we calculated using the simple estimation based on the apo- and peri-centres of the particle orbit: {\tt ecc} = (r$_{\rm apo}$ - r$_{\rm per}$})/(r$_{\rm apo}$ + r$_{\rm per}$). In Fig. \ref{fig:orb-evol} as a illustration, we present orbits for each representatives of the {\tt ecc}, from 0.0 to 0.9.

In the next step, we analysed all GC models to see the number of close passages with relative distance less than 100 pc near the GalC during the full 5 Gyr of evolution. We chose this indicator as the first pre-selected option for future GCs list of potential NSC contributors, which will be simulated with full $N$-body particle systems. From the initial 16$k$ GCs models, we found $\sim$12$k$ models that have recorded at least one close pass. In Fig. \ref{fig:init-grid-dr-6} (bottom panels) via colour palette, we present the number of close passages. As we see from these plots, high numbers of close passages have models, which are distributed in low specific energy limits, from -17 to -18.5 (bright yellow dots). Models with a low probability of close passes are of little interest in our case of NSC formation via GC accretion (dark dots). In the next step, we prepare a set of models, which have more than 100 events of close passage near the GalC per model. We found 6621 such models.  

To reduce our number of models for future $N$-body simulations, we assumed the second pre-selection option, namely, {\tt ecc}. In this way, we selected five GC models, which are representatives for each {\tt ecc} bin in steps 0.1, from 0.0 to 0.9 (single blue dots in Fig. \ref{fig:init-grid-dr-6} in bottom panels). We note that for {\tt ecc} 0.9, there are no models that have at least one close passage. However, to maintain consistency, we have selected some of these models. In total, based on our two pre-selection options (described above), we selected 50 GC models for full $N$-body integration. In Fig. \ref{fig:init-grid-dr-6}, we present the initial distribution for these selected models in the phase space E versus L$_{\rm tot}$, E  L$_{\rm perp}$, and E versus L$_{\rm z}$ as blue dots. In physical space, these GC objects are in a range of distance from the GalC of 100 pc -- 5 kpc.

%%%%%%%%%%%%%%%%%%%%%%%%%%%%%%%%%%%%%%%%%%%%%%%%%%%%%%%%%%%%%%%%%%%%
\section{Full $N$-body integration of GC's}\label{sec:integration}
%%%%%%%%%%%%%%%%%%%%%%%%%%%%%%%%%%%%%%%%%%%%%%%%%%%%%%%%%%%%%%%%%%%%

To carry out the $N$-body dynamical modelling together with the stellar evolution for selected GCs, we assumed the initial conditions described below. The GCs centres initial coordinates and velocities at the 10 Gyr lookback time were taken from the orbital integration of our theoretical GC from the previous step (see Sect. \ref{sec:Mesh}). 

\begin{figure}[h]
\centering
\includegraphics[width=0.80\linewidth]{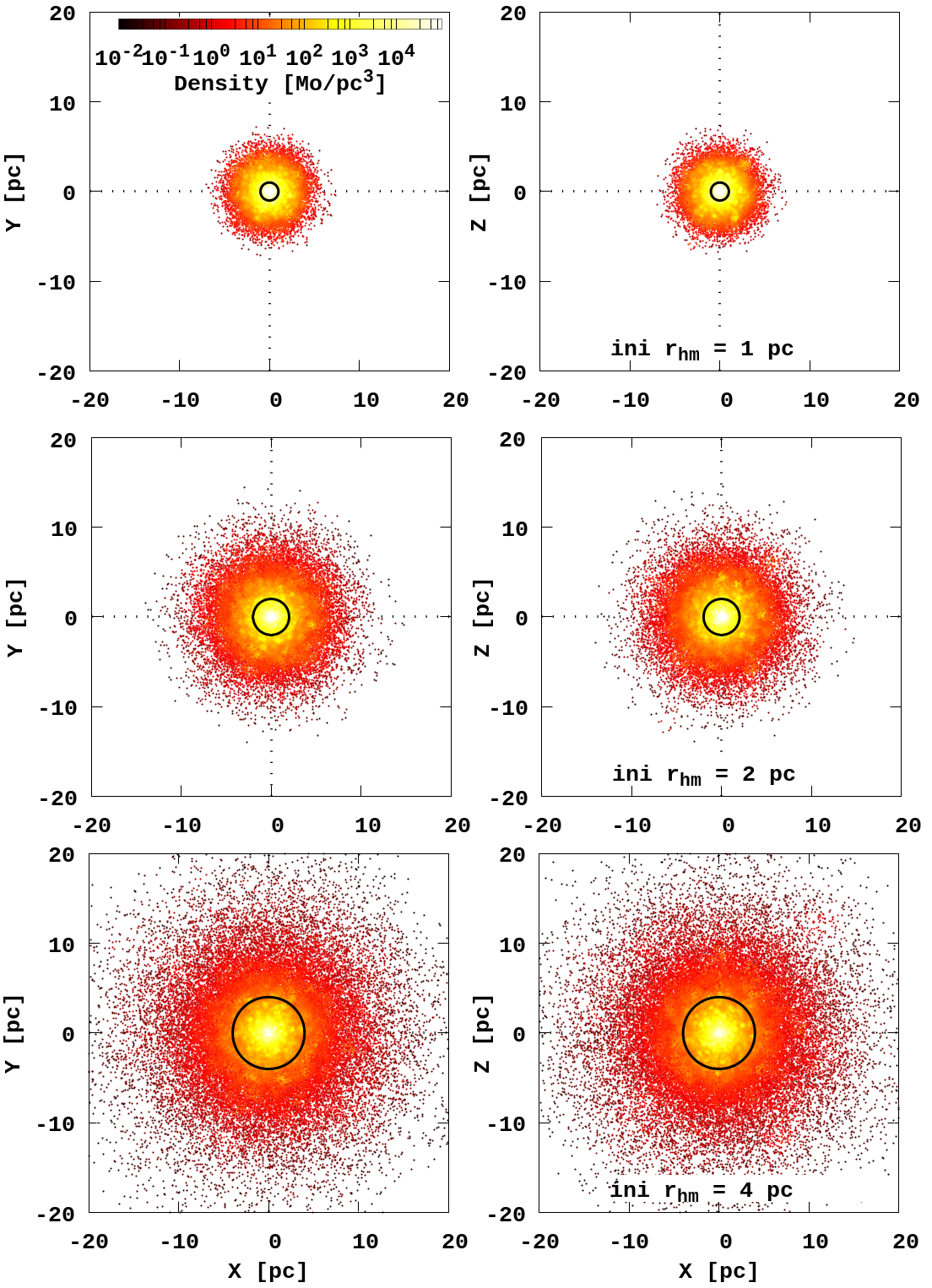}
\caption{Initial density distribution for our models with different initial half-mass radii.}
\label{fig:ini-dem}
\end{figure}

For our $N$-body modelling of the theoretical GCs, we created three sets of models with different half-mass radii: 1, 2, and 4 pc. Other physical parameters, such as initial masses and King concentration parameters, we assumed the same for all models: 60$\times$10$^3$ M$_{\odot}$ and W$_{\rm 0}$ = 8.0. We generated the initial positions and velocities of the stars inside the clusters using the nowadays very popular {\tt Agama} library \citep{VasAgama2019}. For the individual initial mass of the stars, we used the Kroupa mass function \citep{Kroupa2001} with lower--upper mass limits equal to 0.08--100~M$_{\odot}$. Each cluster was initially set in a dynamical equilibrium state with the distribution function of the theoretical King model \citep{King1966}. Because our GC models have relatively low mass, for survivability we try to compensate for this with the higher King concentration parameter for all the model systems. In Table \ref{tab:init-rhm-mass} we summarise these parameters.  In Fig. \ref{fig:ini-dem}, we show initial density distribution for our models with different initial half-mass radii.

%-------------------------------------------------------------------------%
\begin{table}[h]
\caption{Initial physical parameters for GC's $N$-body modelling.}
\centering
\begin{tabular}{ccccc}
\hline
\hline
No & M, M$_{\odot}$ & N & r$_{\rm hm}$, pc & W$_{\rm 0}$  \\
\hline
\hline
50  & 60 000 & 104 554  & 1 & 8.0 \\
50  & 60 000 & 104 554  & 2 & 8.0 \\
50  & 60 000 & 104 554  & 4 & 8.0 \\
\hline
4   & 120 000 & 209 108  & 2 & 8.0 \\
4   & 180 000 & 313 662  & 2 & 8.0 \\
\hline
\end{tabular}
\label{tab:init-rhm-mass}
\end{table} 
%-------------------------------------------------------------------------%

%-------------------------------------------------------------------------%
%--------------------------

\begin{figure}[h]
\centering
\includegraphics[width=0.80\linewidth]{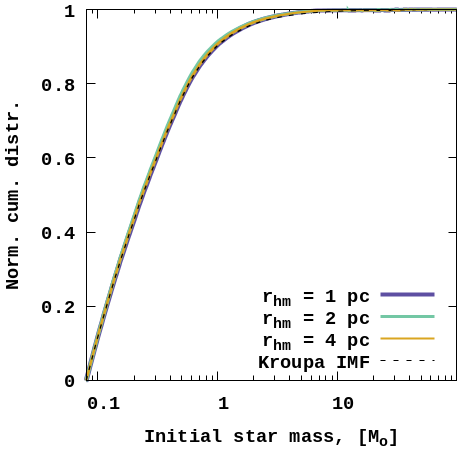}
\caption{Normalised cumulative distribution for initial star masses.}
\label{fig:cum-coll}
\end{figure}

In Fig. \ref{fig:cum-coll}, we present the cumulative distribution of initial star masses for different r$_{\rm hm}$ models with the selected \cite{Kroupa2001} initial mass function. As we can see, all the lines are fully consistent. To carry out the full $N$-body modelling for our three sets (in total 150 GC) of models, we used the same $N$-body code \PGPU, which was mentioned in Sect. \ref{sec:Mesh}, but with the up-to-date stellar evolution prescription \citep{Kamlah2022MNRAS}. 

%-------------------------------------------------------------------------%
%-------------------------------------------------------------------------%

Based on the initial mass function and total particle numbers described above, for each GC model, we expect to have 174 black holes (BHs), 723 neutron stars (NSs), and $\sim$8000 white dwarfs (WDs); however, the majority ($\sim$87$k$)  will still be made up of low-mass stars on the main sequence after 5 Gyr of evolution. As an initial model for the Galactic NSC, we included a special central proto-NSC object with a mass of $4\times10^6$ M$_\odot$ and a radius of 10 pc in our simulations. This object acts as an extra-gravitational term, which we include to our global Galactic potential {\tt 411324} TVP. During the $N$-body simulation, we assumed that a star from the GC has accreted to the NSC when the particular GC star is inside the 10 pc radius of the proto-NSC. When such a star was detected, we added the current mass of the star at the time of the pass to the NSC and considered the star to be accreted. In such a case, we will always have a growing NSC mass.  Such a simple star versus proto-NSC merging scenario (based only on distance) might overestimate the number of accretion cases with some high-velocity objects. To overcome this effect, we restricted our study to GC systems only with the orbital velocities less than 150 km/s, which lies in a range of the bounding velocity around proto-NSC with the masses and sizes considered above.

%%%%%%%%%%%%%%%%%%%%%%%%%%%%%%%%%%%%%%%%%%%%%%%%%%%%%%%%%%%%%%%%%%%%
\section{Analysis of the GCs stellar accretion processes on to proto-NSC}\label{sec:analysis}
%%%%%%%%%%%%%%%%%%%%%%%%%%%%%%%%%%%%%%%%%%%%%%%%%%%%%%%%%%%%%%%%%%%%

In this section, we discuss in general terms the stellar accretion rate onto the NSC during the full duration of simulations (5 Gyr) for our three sets of models. We illustrate the accretion rate in the context of the (i) orbital {\tt ecc}, (ii) initial phase space conditions, and (iii) contribution from the GC's high-mass stellar remnants.

%%%%%%%%%%%%%%%%%%%%%%%%%%%%%%%%%%%%%%%%%%%%%%%%%%%%%%%%%%%%%%%%%%%%
\subsection{Global accretion rate onto proto-NSC}\label{subsec:global-rate}
%%%%%%%%%%%%%%%%%%%%%%%%%%%%%%%%%%%%%%%%%%%%%%%%%%%%%%%%%%%%%%%%%%%%

\begin{figure*}[htb!]
\centering
\includegraphics[width=0.99\linewidth]{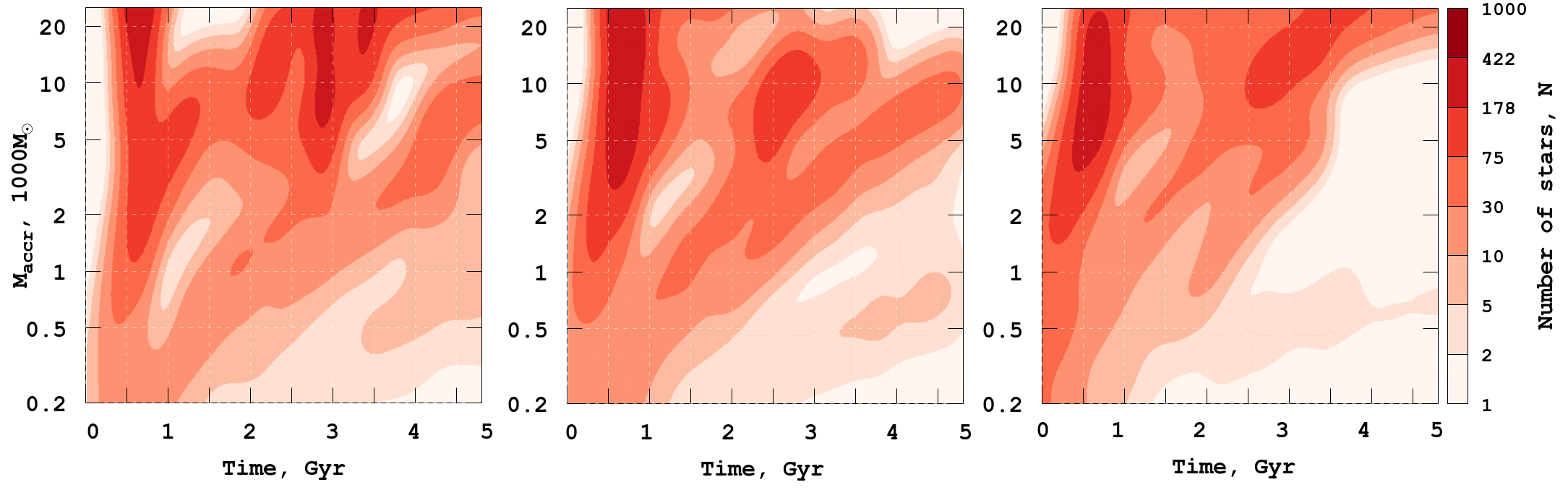}
\caption{Stellar accretion rates onto the NSC for 50 GCs models for three different values of  r$_{\rm hm}$ = 1, 2, and 4 pc, respectively.}
\label{fig:accr}
\end{figure*}

In Fig. \ref{fig:accr}, we show the dynamics of stellar accretion rates onto NSC for all 50 GCs models with three different values of r$_{\rm hm}$ = 1, 2, and 4 pc. The data are presented as a 2D histogram of simulation time versus cumulative accreted mass onto NSC. To enhance the analysis and reveal trends and regions with the highest contributions, the histogram bins were smoothed using a Gaussian kernel. The plot was constructed using a 2D histogram with 100 $\times$ 100 bins, interpolated over five points, and smoothed using a Gaussian kernel with $\sigma$ = 0.2. The used Gaussian kernel affects only the sharpness of the contours, not the overall distributions on the plots.

%-------------------------------------------------------------------------%
%-------------------------------------------------------------------------%

%-------------------------------------------------------------------------%
\begin{figure}[htb!]
\centering
\includegraphics[width=0.99\linewidth]{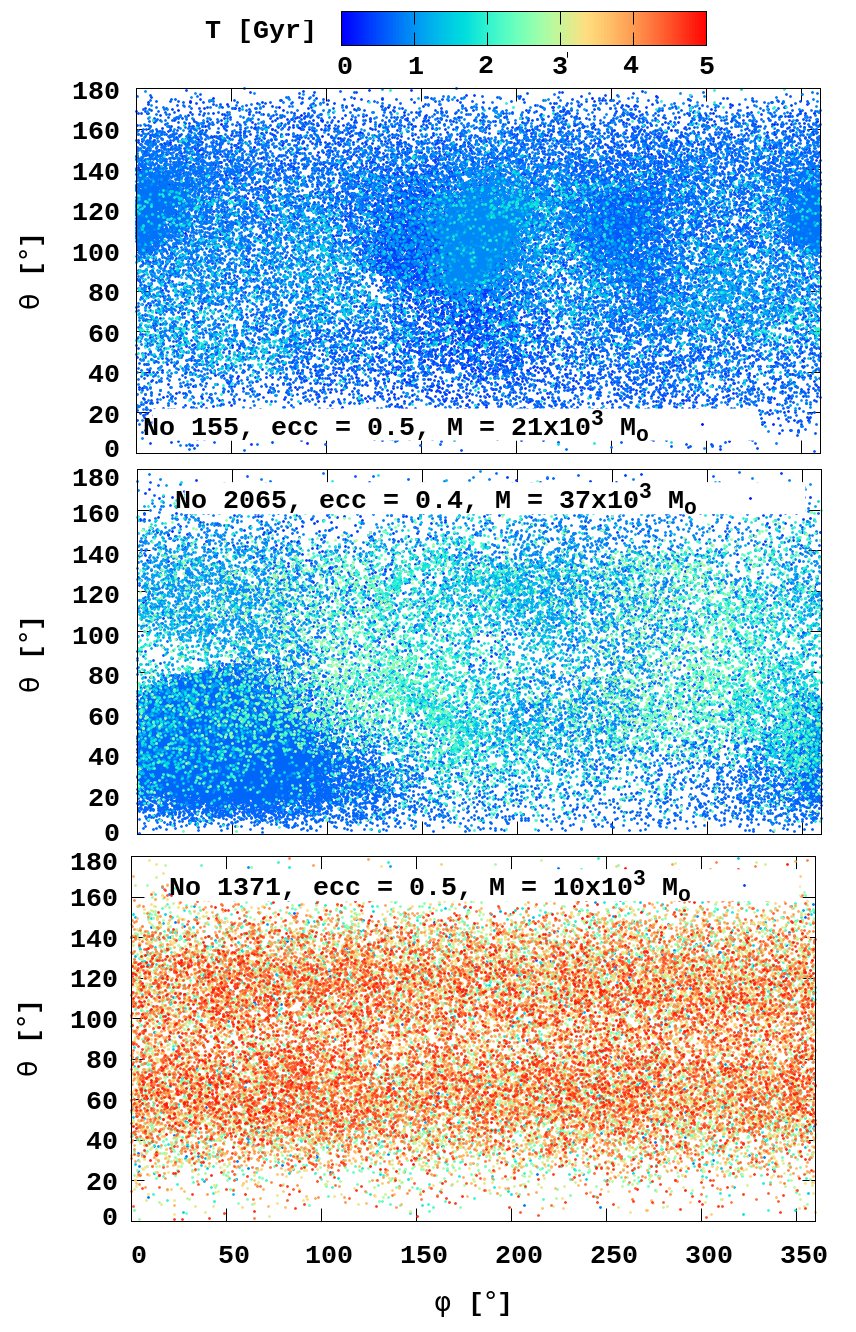}
\caption{Polar and azimuthal angles of particles at the moment of accretion in the Cartesian Galactocentric coordinate system. The accretion radius sphere of the proto-NSC was set to 10 pc. The points individual colour coding shows the moment of accretion. Each panel corresponds to the accretion from different GC.}
\label{fig:ugol-combo}
\end{figure}
%-------------------------------------------------------------------------%

Figure \ref{fig:accr} clearly illustrates the difference in the timing of major accretion episodes onto the proto-NSC as a function of the model cluster's density (parametrized by half-mass). As shown, for clusters with r$_{\rm hm}$ = 1 pc, accretion events occur throughout the integration time interval. Models with r$_{\rm hm}$ = 2 pc have a maximum accretion rate during the first 1 Gyr and later around 2.5--3 Gyr. The last set of models with r$_{\rm hm}$ = 4 pc also includes an early accretion phase that almost ended after 4 Gyr. For all three sets of models, the proto-NSC accretion process has a clear maximum set around 0.5 Gyr. 

To illustrate the accretion process onto the central NSC, in Fig. \ref{fig:ugol-combo}, we present the polar and azimuthal angles of particles at the moment of accretion in the Cartesian Galactocentric coordinate system, assuming a proto-NSC accretion radius of 10 pc. For the presentation, we chose three models with high accretion rates, but with different cluster lifetimes. Two models (No 2065 and 155 in Table \ref{tab:stat}) have a short dissolution time of around 0.6--0.9 Gyr. The third model (No 1371) is one of our long-lifetime clusters, but they  expel almost all the stars during the 5 Gyr evolution time. The short-lifetime clusters show a strong asymmetric and patchy distribution of the accretion angle. They both have a strong accretion events during the first few hundred megayears (Myr; large blue regions). In contrast, the long-lifetime cluster has a more regular and uniform accretion angle distribution. The accretion events (each dots in these figures are individual accretion moments) are distributed equally in time (from blue to red). Besides the patchy structures in some of these plots, we can generally conclude that the accretion processes on our central 10 pc sphere are quite random and stochastic.    

In Table \ref{tab:results}, we present the total number of accreted stars and their total mass for each of all 50 GC models for our three sets of GCs. As can be seen, the model with r$_{\rm hm}$ = 1 pc contributed more accreted masses compared to the other models. Based on the values presented in the table, we can conclude that our extended set of 150 numerical simulations provides an average lower limit of the mass contribution to the NSC from all investigated GCs of $\approx$ 6\%.

%-------------------------------------------------------------------------%
\begin{table}[h]
\caption{Total mass accretion the the NSC for initial sets of the r$_{\rm hm}$.}
\centering
\begin{tabular}{c|cc}
\hline
\hline 
r$_{\rm hm}$ & N$_\text{accr}$  & M$_\text{accr}$, M$_{\odot}$ \\
\hline
\hline
1 pc  & 398 358 & 152 758 \\
2 pc  & 309 987 & 117 981 \\
4 pc  & 312 964 & 124 344 \\
\hline
\end{tabular}
\label{tab:results}
\end{table} 
%-------------------------------------------------------------------------%

In Table \ref{tab:stat}, we present detailed statistical stellar accretion information for all 50 GCs for each of our three sets. As can be seen, at least 5 GCs are completely dissolutes during 5 Gyr of simulations. We define the dissolution of a cluster as occurring when the remaining tidal mass decreases to less than 5 percent of the cluster initial mass. With initial r$_{\rm hm}$ = 1 pc we have five dissolved GCs during this 5 Gyr integration period. With an initial r$_{\rm hm}$ = 2 pc, we have 9 GCs, and with r$_{\rm hm}$ = 4 pc -- 42, respectively. In most cases, the dissolution of the GCs occurs in the models with the orbital {\tt ecc} in a range of 0.3--0.6. 

%-------------------------------------------------------------------------%
\begin{figure}[htb!]
\centering
\includegraphics[width=0.99\linewidth]{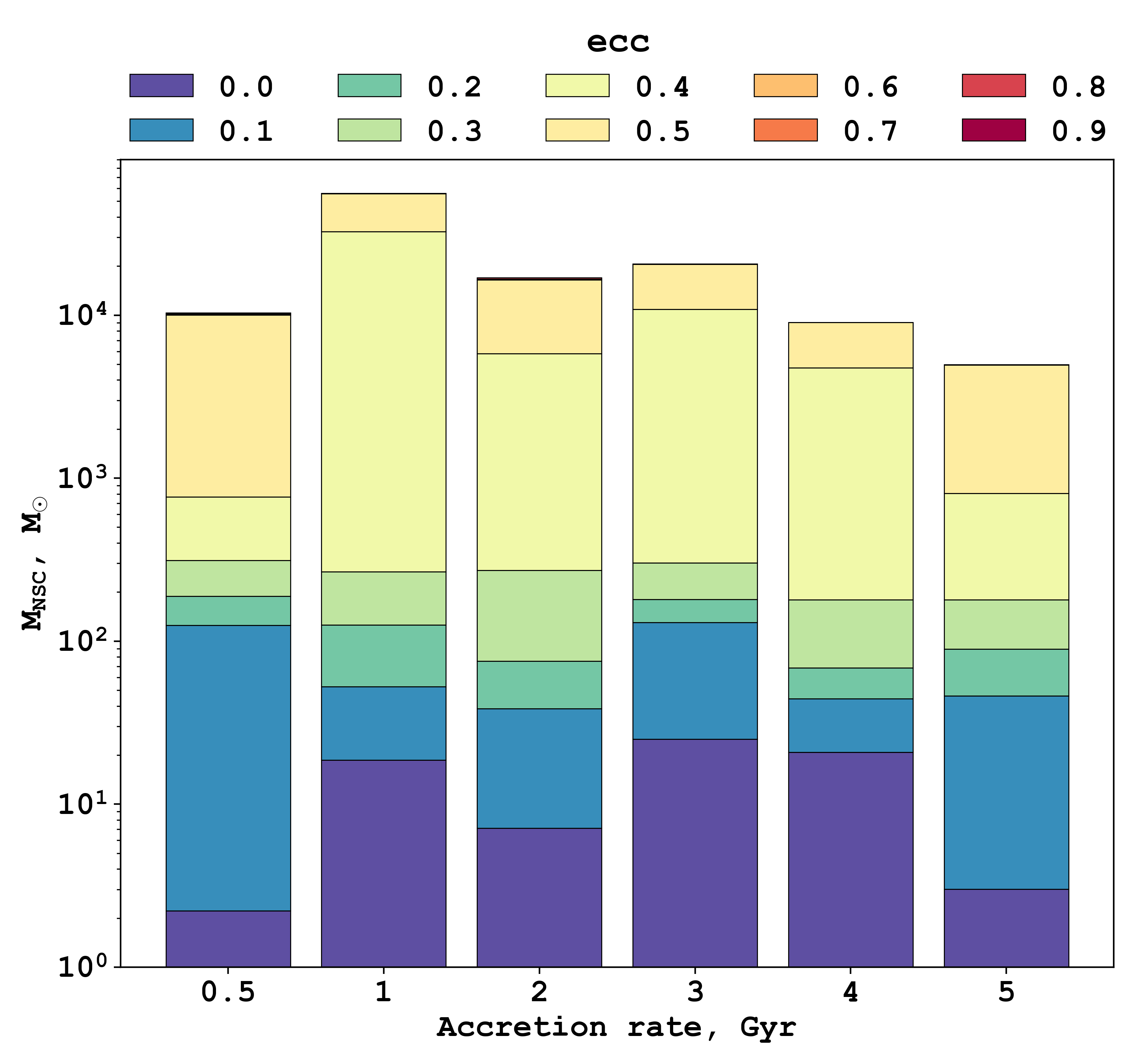}
\caption{Contribution of GCs stars to the mass of the NSC over 5 Gyr for different orbital eccentricities for r$_{\rm hm}$ = 2 pc case. The colours represent the orbital eccentricity, as shown in the legend. The mass is given in solar masses M$_\odot$ and is displayed on a logarithmic scale.}
\label{fig:accr_ecc_rhm2}
\end{figure}
%-------------------------------------------------------------------------%

%-------------------------------------------------------------------------%
\begin{figure}[htb!]
\centering
\includegraphics[width=0.99\linewidth]{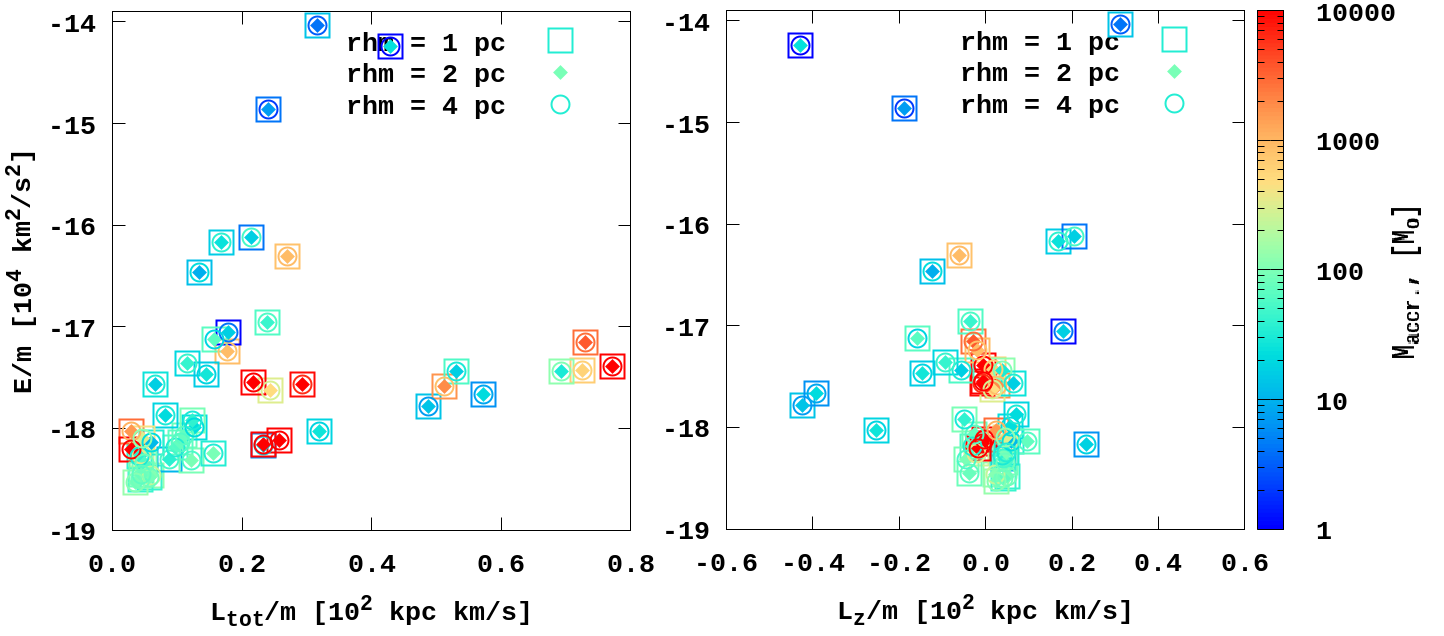}
\caption{Initial distribution for 50 theoretical GCs in E -- L$_{\rm tot}$, E -- L$_{\rm perp}$, E -- L$_{\rm z}$ at -10 look-back time. Different symbols show the r$_{\rm hm}$ values. Colour represent accreted mass from each GC to the NSC over 5 Gyr of integration time.}
\label{fig:e-tot-z}
\end{figure}
%-------------------------------------------------------------------------%

%-------------------------------------------------------------------------%
\begin{figure*}[htb!]
\centering
\includegraphics[width=0.99\linewidth]{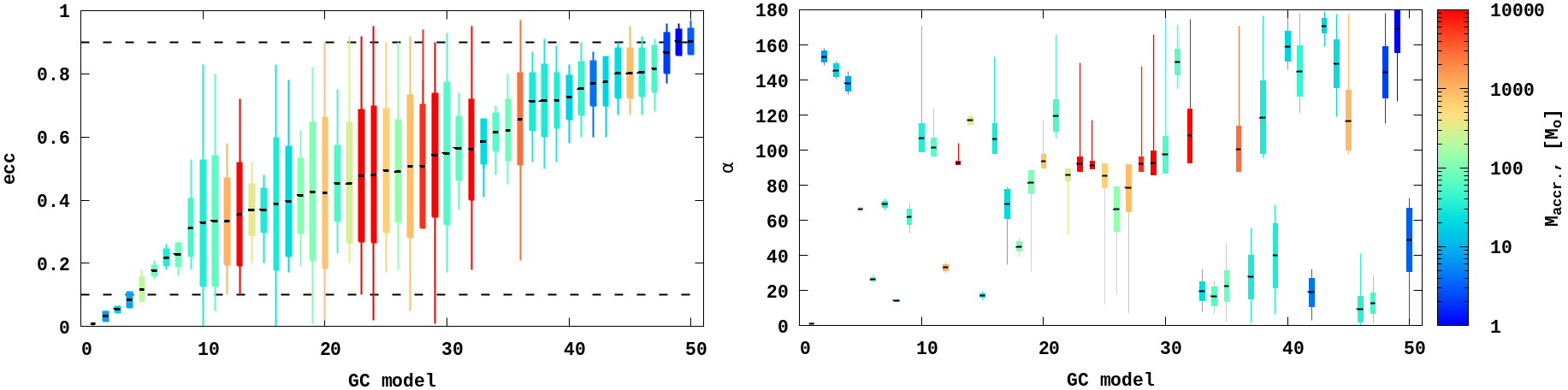}
\caption{Evolution of the mass accretion to the NSC for models with r$_{\rm hm}$ depends on the {\tt ecc} and inclination angle $\alpha$. Short black lines: Mean values, colour box -- $\pm$1$\sigma$ for {\tt ecc}, thin colour lines:\  minimum and maximum {\tt ecc,} and $\alpha$ values during 5 Gyr of integration. Colour:\ Accreted mass to the NSC.}
\label{fig:bar-a-e}
\end{figure*}
%-------------------------------------------------------------------------%

Figure \ref{fig:accr_ecc_rhm2} illustrates the contribution of GCs stars to the mass of the NSC over a period of 5 Gyr for the case of r$_{\rm hm}$ = 2 pc. The different colours represent the mean orbital eccentricity, as indicated in the legend. The mass is expressed in solar masses M$_\odot$ and is shown on a logarithmic scale. To better resolve the early accretion phase, we have intentionally divided the first Gyr into two intervals. This is because the highest accretion rates occur within this period for all models. This might indicate that NSC formation, at least in the cluster-inspiral scenario, occurs predominantly in the early stages of Galaxy assembling. The figure also shows that models with mean {\tt ecc} of 0.4 and 0.5 make the largest contribution in mass of the NSC. This may be attributed to the fact that GC models with these mean eccentricities are also characterized by pronounced orbital eccentricity oscillations over our 5 Gyr timescale, spanning a broad range of {\tt ecc} values (also see Fig. \ref{fig:bar-a-e}).

In Fig. \ref{fig:e-tot-z}, we present the mass accretion rate to the NSC as a function of specific energy and angular momentums: L$_{\rm tot}$, L$_{\rm perp}$, and L$_{\rm z}$ at the initial moment of time (i.e. 10 Gyr ago in {\tt 411321} external TVP potential). As we see from the plots, the highest mass accretion rate for GC models is a system, in which the initial phase space distribution has a nearly perpendicular orbital plane to the galactic disc, namely, of $\sim$0.0 in L$_{\rm z}$ (yellow-red range in palette colours).

%%%%%%%%%%%%%%%%%%%%%%%%%%%%%%%%%%%%%%%%%%%%%%%%%%%%%%%%%%%%%%%%%%%%
\subsection{Mass accretion to the NSC as function of GC orbital parameters distribution}\label{sec:int-inc}
%%%%%%%%%%%%%%%%%%%%%%%%%%%%%%%%%%%%%%%%%%%%%%%%%%%%%%%%%%%%%%%%%%%%

After running the full set of models, we analysed the orbital parameters of each individual run and checked the correlation between the model's orbital parameters and the amount of mass, accreted onto proto-NSC from each model. As main orbital parameters, we chose the orbits {\tt ecc} and inclination angle (defined as $\cos{(\alpha)}$ = L$_{\rm z}$/L$_{\rm tot}$). We chose these parameters because of the strong dependence of the accreted stellar mass on these values. Due to the time-variable nature of our external Galactic potential in each orbital revelation, we derived the {\tt ecc} and inclination angle, $\alpha$. We also prepared the statistics among each orbital revelation and derive the average and the standard deviation for these orbital parameters. As we can already see from Fig. \ref{fig:orb-evol}, the time evolution of the orbital shapes is quite complex and, as a consequence, we have quite significant standard deviations, especially with respect to {\tt ecc}.    

In Fig. \ref{fig:bar-a-e} we show the models' mass accretion onto NSC (colour coded) as a function of the model's initial {\tt ecc} and inclination angle, $\alpha$. Each model has been numbered (X-axis) according to its average {\tt ecc}. As we can see, the models with the highest mass accretion (model numbers from 10 to 37) also have the highest {\tt ecc} changes during the simulation. The most prominent GC mass donors have a {\tt ecc} change from 0.1 to 0.9, represented in red colour. 

One particularly notable feature of the accretion process in our set of models is the highest accretion for the runs with close to 90 degree of orbital inclination. These models (from 23 to 35) have almosy exactly  a 90 degree with small deviation (thick colour box). It is worth mentioning that these models have also a quite large of minimum--maximum inclination angles, during the time evolution (thin colour lines). It is also interesting that angle range have a quite strong asymmetry, namely, the indication to the significant counter rotation of these runs, from 90--180 degrees. 

For more deep illustration of the dynamical evolution for orbital parameters, such as {\tt ecc}, $\alpha$, and semi-major axis, we present several plots for the models with different accretion rates onto the proto-NSC (numbers in right panels) in Fig. \ref{fig:accr-10-100}. Here, in the middle panels, we see a certain compression of the semi-major axis of the orbit towards the GalC around 1.5--4 Gyr. The dynamics of the orbits is directly related to the time evolution of the Galactic potential, namely, to the halo mass and scale length for the same time period. For GC models that have evolution in the Galactic Pale, the Myamoto-Nagai scales $a_{\rm d}$ and $b_{\rm d}$ also have an influence, squeezing the scales for the disc (see Fig. \ref{fig:bar-a-e}). The left panels show the orbital {\tt ecc} evolution of the individual models. As we can see, the most mass-donor GCs have a wide range of {\tt ecc} changes, ranging from 0.1 to 0.9. 

Based on Figs. \ref{fig:bar-a-e} and \ref{fig:accr-10-100}, we can conclude that same most mass donor models have close to constant orbital inclination angle of almost 90 degrees. A combination of {\tt ecc} and inclination angle $\alpha$ gives us a best candidates for the GC models, which can be a source of the stellar mass accretion onto proto-NSC.

%%%%%%%%%%%%%%%%%%%%%%%%%%%%%%%%%%%%%%%%%%%%%%%%%%%%%%%%%%%%%%%%%%%%
\subsection{Accretion rate for the stellar remnants from GCs}\label{sec:int}
%%%%%%%%%%%%%%%%%%%%%%%%%%%%%%%%%%%%%%%%%%%%%%%%%%%%%%%%%%%%%%%%%%%%

We analysed the accretion of stellar remnants (WDs, NSs, and BHs) from 50 GCs for three different values of r$_{\rm hm}$ = 1, 2, and 4 pc. Table \ref{tab:results_rem} shows the total number and masses of stellar remnants accreted onto the NSC for these three sets of models.

%-------------------------------------------------------------------------%
\begin{figure}[htb!]
\centering
\includegraphics[width=0.99\linewidth]{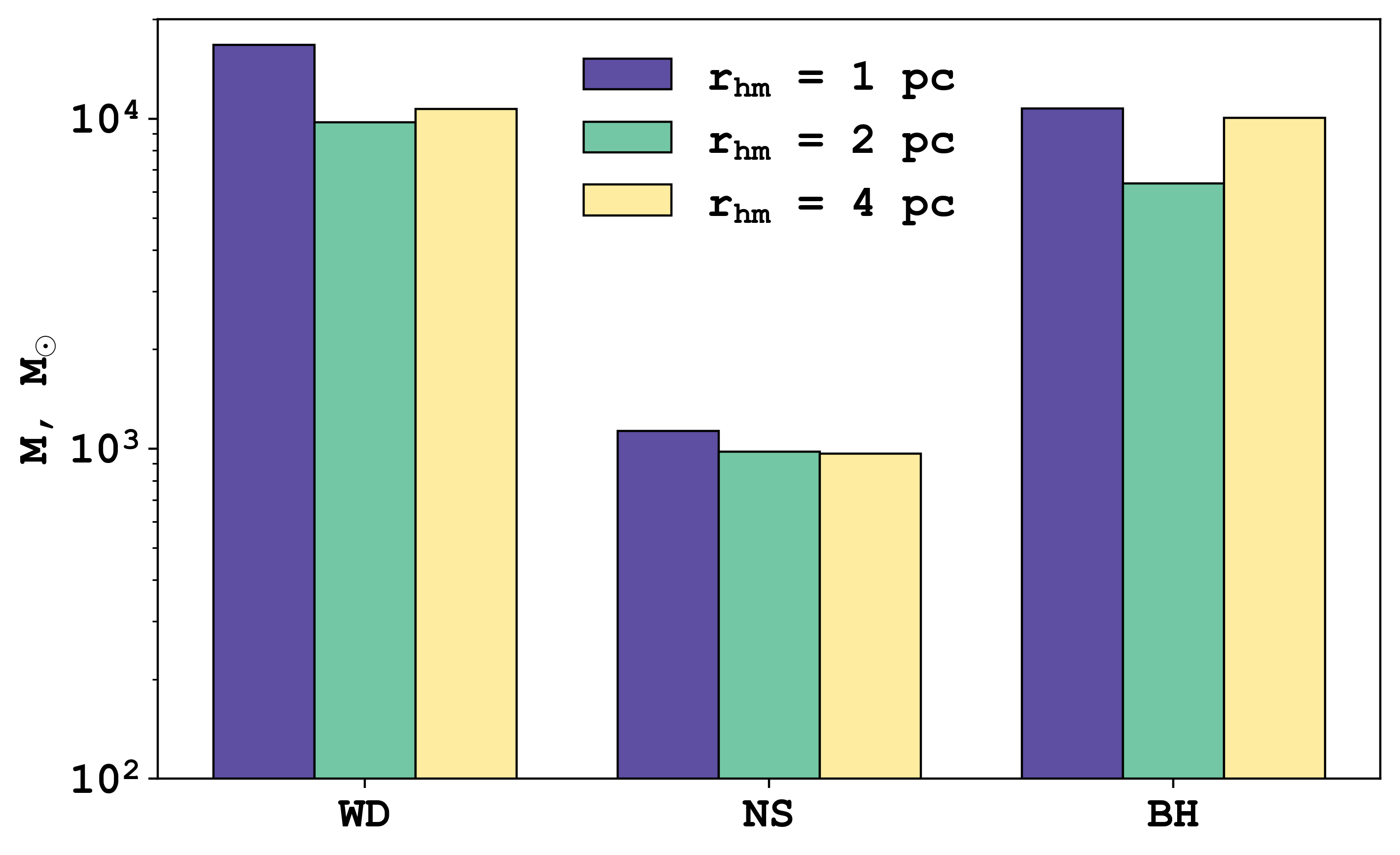}
\caption{Total accretion of the mass of stellar remnants (WD, NS, and BH) from 50 GCs for all the three cases: r$_{\rm hm}$ = 1, 2, and 4 pc. The mass in given in logarithmic scale.}
\label{fig:accr_remnants}
\end{figure}
%-------------------------------------------------------------------------%

Since the r$_{\rm hm}$ = 1 pc case has the highest amount of total accreted mass between three sets of models, it is not surprising that it also contributes the most mass to stellar remnants. Figure \ref{fig:accr_remnants} shows the total mass of accreted stellar remnants. For NSs and BHs, the mass contributions are comparable among all three models, reflecting their low absolute numbers in the stellar population. In contrast, the significantly higher mass of WDs accreted in the r$_{\rm hm}$ = 1 pc model likely results from their higher intrinsic abundance, as WDs are the dominant stellar remnants in clusters with the standard type of initial mass functions. The enhanced central density of the stellar remnants in such a compact clusters further amplifies their contribution.

%-------------------------------------------------------------------------%
\begin{table}[h]
\caption{Accretion of the stellar remnants onto the NSC for sets of GCs with different initial r$_{\rm hm}$, where the mass accretion is given in M$_{\odot}$.}
\centering
\begin{tabular}{c|cc|cc|cc}
\hline
\hline 
r$_{\rm hm}$ &  \multicolumn{2}{c|}{WD} & \multicolumn{2}{c|}{NS} & \multicolumn{2}{c}{BH}  \\
 & N$_\text{accr}$ & M$_\text{accr}$ & N$_\text{accr}$  & M$_\text{accr}$ & N$_\text{accr}$ & M$_\text{accr}$ \\
\hline
\hline
1 pc  & 19 765 & 16 719 & 909 & 1 131 & 546 & 10 752  \\
2 pc  & 10 957 & 9 732  & 785 & 979   & 336 & 6 359  \\
4 pc  & 12 275 & 10 688 & 779 & 965   & 508 & 10 072 \\
\hline
\end{tabular}
%}
\label{tab:results_rem}
\end{table} 
%-------------------------------------------------------------------------%

In Appendix \ref{sec:addi-accr-rem}, we present a more detailed analysis of the accretion of stellar remnants onto the NSC. As expected, for all three cases of r$_{\rm hm}$ = 1, 2, and 4 pc, the largest contribution to the NSC mass over 5 Gyr comes from stellar remnants accreted from GCs with mean orbital eccentricities around 0.4–0.5. In terms of the temporal evolution, the highest accretion rate for all types of remnants occurs in the interval between 0.5 and 1 Gyr. The analysis of individual accretion events shows that the bulk of the accreted objects are WDs and low-mass stars.

%%%%%%%%%%%%%%%%%%%%%%%%%%%%%%%%%%%%%%%%%%%%%%%%%%%%%%%%%%%%%%%%%%%%
\section{Discussions and conclusions}\label{sec:dis-con}
%%%%%%%%%%%%%%%%%%%%%%%%%%%%%%%%%%%%%%%%%%%%%%%%%%%%%%%%%%%%%%%%%%%%

In this section, we discuss the possible effects that can lead to the stellar accretion of GCs onto the proto-NSC during the long-term cosmological time scale dynamical evolution of the theoretical GCs system. In more detail, we discuss (i) the role of the GC mass loss itself and the possible accretion onto the GalC, (ii) the role of stellar remnants, and (iii) the physical effects of the time-variable potential on the accretion process.

%%%%%%%%%%%%%%%%%%%%%%%%%%%%%%%%%%%%%%%%%%%%%%%%%%%%%%%%%%%%%%%%%%%%
\subsection{Role of the GC's mass loss and accretion rate for low- and high-mass cases}\label{subsec:massloss-accrrate}
%%%%%%%%%%%%%%%%%%%%%%%%%%%%%%%%%%%%%%%%%%%%%%%%%%%%%%%%%%%%%%%%%%%%

\begin{figure*}[htb!]
\centering
\includegraphics[width=0.99\linewidth]{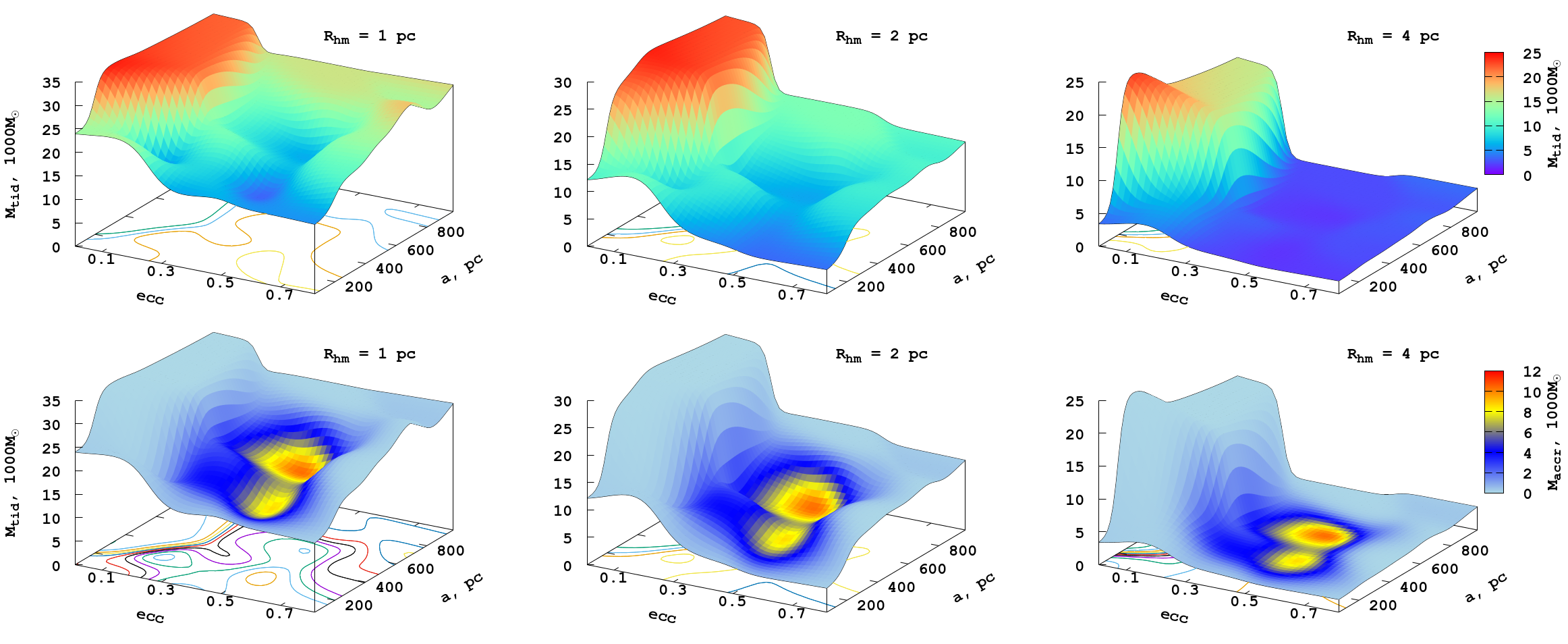}
\caption{Global tidal mass loss and accretion mass maps over all our theoretical GC models. The X and Y axes are the orbital parameters of our GC's. The $Z$ axis is a tidal mass of the clusters after 5 Gyr of dynamical integration. In the bottom row, the colour-coding is corresponds to the accreted mass onto proto-NSC from the individual clusters.}
\label{fig:accr-mtid}
\end{figure*}

As a basic GC model in our set of simulations, we choose systems with an initial mass of 60$\times10^3$ M$_{\odot}$. The general picture of the cluster's global mass loss for three different initial half-mass radii is presented in Fig. \ref{fig:accr-mtid}. The top row presents the global mass loss of clusters with different orbital parameters, while the bottom row (colour-coded) shows the proto-NSC-accreted mass from the GCs. A clear effect of the half-mass radius is visible: clusters with larger half-mass radii experience significantly greater total tidal mass loss across all orbital parameters (see the 'blue plateau' in the top-right panel). In general, the low-eccentricity models for all the different half-mass radii lose significantly less mass compared to the models with the higher {\tt ecc}.

%-------------------------------------------------------------------------%
%-------------------------------------------------------------------------%

The models with r$_{\rm hm}$ = 4 pc are almost completely dissolved; however, these dissolved clusters do not contribute significantly to the mass accreted onto the proto-NSC. In other words, the global tidal mass loss of some GC models does not necessarily translate into substantial stellar accretion onto the central region of the Galaxy. As mentioned in Sect. \ref{sec:integration}, we did not include any dynamical friction to the prescription of the accretion process. The possible dynamical friction effects can change the actual values of the accreted stellar masses (in the direction to the higher masses), but our general conclusions regarding the dependency of the NSC accreted mass from models half-mass radii, will be qualitatively valid even in this case.

%-------------------------------------------------------------------------%
\begin{figure}[htb!]
\centering
\includegraphics[width=0.99\linewidth]{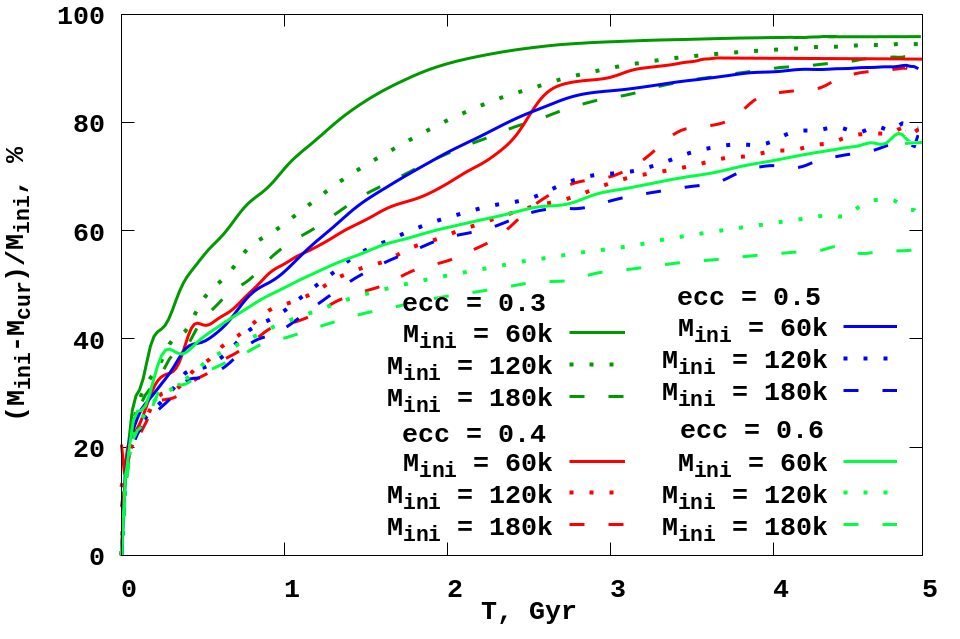}
\caption{Evolution of the mass loss in the GC models due to mass loss and orbital type after 5 Gyr of integration. Values M$_{\rm ini}$ are in M$_{\rm \odot}$.}
\label{fig:mass-loss}
\end{figure}
%-------------------------------------------------------------------------%

In addition to the main set of GC models, each with a a total initial mass of 60$\times10^3$ M$_{\odot}$, we additionally simulated the dynamical evolution of four models from the total set with r$_{\rm hm}$ = 2 pc and {\tt ecc} values from 0.3 to 0.6. For these models, higher initial masses of 120$\times10^3$ M$_{\odot}$ and 180$\times10^3$ M$_{\odot}$ were adopted (see Table \ref{tab:init-rhm-mass}). The general mass loss of these additional models is presented in Fig. \ref{fig:mass-loss}. As expected, the lower-mass models (among the set with the same initial half-mass systems) have a slightly larger general mass loss $\gtrsim$90\%, compared to the higher-mass models. The models with 120$\times10^3$ and 180$\times10^3$ M$_{\odot}$ exhibit very similar global mass loss behaviour, with values around $\sim$60--80\%. 

\begin{figure}[htb!]
\centering
\includegraphics[width=0.99\linewidth]{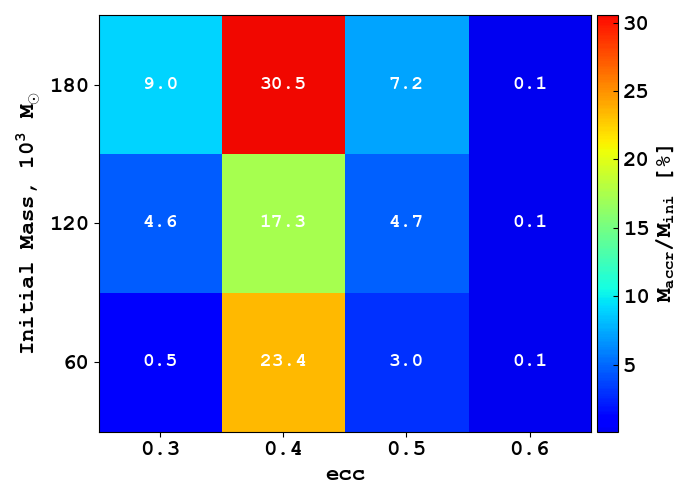}
\caption{Relative accretion from low- and high-mass GC models with different {\tt ecc} as percentages.}
\label{fig:accr_massive}
\end{figure}

The main results of the stellar accretion simulations for the more massive GC models are presented in Table \ref{tab:dif-mass}, where the relative fractions of the accreted mass and the number of particles are given as percentages of their initial values (also see Fig. \ref{fig:accr_massive}). As the initial mass of the GC models increases, their total mass contribution to the proto-NSC also grows. However, for the model with {\tt ecc} of 0.4, the relative fraction of accreted stars decreases when the mass increases from 60$\times10^3$ to 120$\times10^3$ M$_{\odot}$, although at 180$\times10^3$ M$_{\odot}$ the model no longer deviates from the established trend. It is worth noting that although the model with {\tt ecc} = 0.6 yields only a small number of stars, their average mass is approximately 4.4 M$_{\odot}$, compared to $\leq$3 M${\odot}$ for the other models.

As a side remark, we also ought to mention the generally low average mass of the accreted stars. For  models with {\tt ecc} 0.4 and 0.5, where the accretion statistics are more robust, the average masses of the accreted stars are $\approx$0.33 M$_{\odot}$. Compared with the average IMF mass of our clusters ($\approx$0.57 M$_{\odot}$), we see that during our 5 Gyr dynamical simulations, the accreted population is primarily dominated by low-mass stars. One possible reason for this mass selection is a significant mass segregation among the clusters,  resulting in a higher fraction of high-mass stars and remnants residing in the more tightly bound central region. As an example, we checked the average cumulative masses in our typical theoretical GCs ({\tt 2858} with r$_{\rm hm}$ = 1 pc) over the 5 Gyr of simulations and we find that after $\sim$1 Gyr the average cumulative mass within 0.1 pc is almost around $\sim$20 M$_\odot$ and it quickly drops below the level of $\sim$ 0.3 M$_\odot$ after 10 pc. 

%-------------------------------------------------------------------------%
\begin{table*}[h]
\caption{Accretion rates onto proto-NSC from low- and high-mass GCs.}
\centering
\resizebox{0.99\textwidth}{!}{
\begin{tabular}{cc|cccccc|cccccc|cccccc}
\hline
\hline 
 &  &  \multicolumn{6}{c|}{M$_{\rm ini}$ = 60$\times10^{3}$ M$_{\odot}$; N$_{\rm ini}$ = 105$k$} & \multicolumn{6}{c|}{M$_{\rm ini}$ = 120$\times10^{3}$ M$_{\odot}$; N$_{\rm ini}$ = 209$k$} & \multicolumn{6}{c}{M$_{\rm ini}$ = 180$\times10^{3}$ M$_{\odot}$; N$_{\rm ini}$ = 314$k$}  \\
\hline 
 No & {\tt ecc} & M$_\text{accr}$ & \% & N$_\text{accr}$ & \% & T & N$^{*}$ & M$_\text{accr}$ & \% & N$_\text{accr}$ & \% & T & N$^{*}$ & M$_\text{accr}$ & \% & N$_\text{accr}$ & \% & T & N$^{*}$ \\
  & & M$_{\odot}$ & & & & Gyr & 10$^{3}$ & M$_{\odot}$ & & & & Gyr & 10$^{3}$ & M$_{\odot}$ & & & & Gyr & 10$^{3}$ \\
\hline
\hline
270  & 0.3  & 299    & 0.5 & 669    & 0.6  & 5 & ds & 5 512  & 5   & 14 472 & 7   & 5 & ds  & 16 214 & 9   & 45 802  & 15  & 5 & ds \\
1370 & 0.4  & 14 020 & 23  & 41 714 & 40   & 5 & ds & 20 801 & 17  & 62 796 & 30  & 5 & 53  & 54 985 & 31  & 163 524 & 52  & 5 & 26   \\ 
1392 & 0.5  & 1 824  & 3   & 5 481  & 5    & 5 & 8  & 5 587  & 5   & 16 539 & 8   & 5 & 52  & 13 032 & 7   & 39 362  & 13  & 5 & 86   \\
175  & 0.6  & 70     & 0.1 & 13     & 0.1  & 5 & 30 & 116    & 0.1 & 29     & 0.1 & 5 & 113 & 201    & 0.1 & 47      & 0.1 & 5 & 210  \\
\hline
\end{tabular}
}
\tablefoot{N$^{*}$ is the number of particles remaining in the cluster. We consider a cluster to be dissoluted (ds) if only $\sim$5\% of its initial number of particles remains. For example, for 105$k$ this is approximately 5$k$ stars.}
\label{tab:dif-mass}
\end{table*} 
%-------------------------------------------------------------------------%

%-------------------------------------------------------------------------%
%-------------------------------------------------------------------------%

%%%%%%%%%%%%%%%%%%%%%%%%%%%%%%%%%%%%%%%%%%%%%%%%%%%%%%%%%%%%%%%%%%%%
\subsection{Role of the stellar remnants}\label{subsec:rem-role}
%%%%%%%%%%%%%%%%%%%%%%%%%%%%%%%%%%%%%%%%%%%%%%%%%%%%%%%%%%%%%%%%%%%%

Our analysis of stellar remnants accretion onto the NSC from GC models reveals a strong dependence of this process on the initial half-mass radius, r$_{\rm hm}$, and orbital eccentricities, {\tt ecc} (see Fig. \ref{fig:accr_remnants}). The time-resolved accretion history shows a pronounced early peak within the first Gyr (see Fig. \ref{fig:accr_remn_individ}). 

As seen from Fig. \ref{fig:accr_remnants}, the largest total contribution of remnants to the accreted mass comes from more compact GC models (r$_{\rm hm}$ = 1 pc, violet colour). It is worth noting that orbital eccentricity also plays a key role in the process of accretion efficiency. The highest rates remnant accretion are observed at intermediate values of {\tt ecc} = 0.4--0.6 (see Fig. \ref{fig:accr_remn_1_2_4}). The accretion models (Fig. \ref{fig:accr_remnants} and Fig. \ref{fig:accr_remn_individ}) also show a clear distinction among the remnant types in terms of both timing and mass contribution, with WDs being the dominant component across all scenarios. 

We also note the significantly higher accreted mass of BHs compared to that NS remnants in the global accreted stellar remnant mix (by almost a factor of 10; see Fig. \ref{fig:accr_remnants}). However, according to the adopted cluster IMF, the number of forming BHs is about four times smaller than the number of NS remnants. This apparent contradiction can be explained by our selective supernovae kick mechanisms \citep{Kamlah2022MNRAS}, where NS remnants always receive a significant velocity kick; meanwhile a substantial fraction of BHs, due to the fallback prescription, have a zero velocity kick at the moment of formation. These results highlight the relationship between NSC-GC interactions and the presence of stellar remnants in the NSC during its formation and growth.

%%%%%%%%%%%%%%%%%%%%%%%%%%%%%%%%%%%%%%%%%%%%%%%%%%%%%%%%%%%%%%%%%%%%
\subsection{Role of the external potential}\label{subsec:po-role}
%%%%%%%%%%%%%%%%%%%%%%%%%%%%%%%%%%%%%%%%%%%%%%%%%%%%%%%%%%%%%%%%%%%%

As we see from our study, the actual accretion onto the central galactic-scale proto-NSC is a complex dynamical process. In our picture of accretion onto the proto-NSC, we modelled the accretion (dynamical capture) of individual stars onto a forming the NSC. The detailed statistical picture of the dynamical accretion process is presented in Fig. \ref{fig:accr}. In Sect. \ref{subsec:global-rate}, we  describe the general accretion trends for the different sets of models. Here, we would like to discuss the direct influence of the modelled time-variable Galactic potential on the growth of the NSC. The mass and size evolution of our selected potential {\tt 411321} is shown in Fig. \ref{fig:subgalo}. 

The dark matter halo mass and characteristic scale exhibit a fairly monotonic growth over time. Since the halo characteristic scale is quite large (15-25 kpc in the selected integration interval from -10 to -5 Gyr), the direct influence of the extended halo potential is not significant for our set of the theoretical GC models (the maximum apocentre is 5 kpc). In contrast, the baryonic disc potential shows a markedly different behaviour. The ratio of $a_{\rm d}$ and $b_{\rm d}$ changes significantly over the entire integration interval, from $\sim$1 to 3. Such behaviour of the inner Galactic potential has a strong influence on our theoretical set of GCs. 

In Fig. \ref{fig:accr-10-100}, we see the direct effect of these changes on the evolution of the semi-major axis. Since these changes occur around 2--3 Gyr of integration time and we observe significant additional mass accretion in that same timeframe (see Fig. \ref{fig:accr}), we can directly connect these changes.  

%%%%%%%%%%%%%%%%%%%%%%%%%%%%%%%%%%%%%%%%%%%%%%%%%%%%%%%%%%%%%%%%%%%%
\subsection{General remarks}\label{subsec:general-rem}
%%%%%%%%%%%%%%%%%%%%%%%%%%%%%%%%%%%%%%%%%%%%%%%%%%%%%%%%%%%%%%%%%%%%

Based on the total accreted mass from all three sets of our theoretical 150 GCs during the first 5 Gyr of Galactic evolution, the forming NSC acquires $\sim$4$\times$10$^5$ M$_\odot$, see Table \ref{tab:results}. This mass corresponds to roughly 2\% of the current NSC mass of $\sim$2$\times$10$^7$ M$_\odot$ \citep{Bland-Hawthorn2016}. If we also include  the set of models with higher initial masses, 120 and 180$\times$10$^3$ M$_\odot$, in our global accretion estimate and assume the same frequency distribution of such clusters as for the lower-mass GC models (60$\times$10$^3$ M$_\odot$), then we can expect at least several times more stellar mass to be accreted onto the forming NSC. So, in the end, according to our set of three relatively low-mass GC models, we can gain around $\approx$ 6\% of the current NSC mass via pure stellar accretion from the theoretical GC models (very conservative, lower limit mass estimation).

At first glance, our findings seem to be in contradiction with the observationally motivated NSC accretion estimates (around 50\%) reported by \citet{Fahrion2022} for the MW NSC with a mass  $\sim$10$^7$ M$_\odot$. However, we need to take into account the fact that the initial GC masses adopted in our models of theoretical clusters are quite small. In this case, we can estimate indeed only a lower boundary of the NSC accretion mass. In the future, we will employ significantly higher initial GC mass models or a mixed population of low- and high-mass clusters and, in addition, we probably will be able to once again  achieve higher larger accretion rates by a factor of  few. In addition, in our next set of runs, we will include the simplified Chandrasekhar type of dynamical friction \citep{C1943a, C1943b} with some physically motivated theoretical parameters \citep{Just2011}.

In addition to the current set of relatively low-mass GC N-body models, which do not include possible dynamical friction, our estimate of the accreted mass represents a lower limit due to another simplification: we omitted the bulge component in the external time-variable potential model, which is currently based on the Illustris-TNG100 cosmological database.

Also, we need to take into account that probably none of the two accretion scenarios mentioned in the introduction (i.e., stellar or/and gas accretion followed by in situ star formation) are likely to operate in isolation. Therefore, in reality, we are dealing with a some combination of both processes \citep{Fahrion2022}. In our current scenario (low-mass GC destruction and star accretion), the remaining 80-90\% of the NSC mass is expected to form during a phase of significant gas accretion onto the NSC. This view is consistent with the high gas fraction in the GalC during the first few Gyrs of Galactic evolution.

%%%%%%%%%%%%%%%%%%%%%%%%%%%%%%%%%%%%%%%%%%%%%%%%%%%%%%%%%%%%%%%%%%%%%
\begin{acknowledgements}
%%%%%%%%%%%%%%%%%%%%%%%%%%%%%%%%%%%%%%%%%%%%%%%%%%%%%%%%%%%%%%%%%%%%%

The authors thank the anonymous referee for a very constructive report and suggestions that helped significantly improve the quality of the manuscript. This work was funded by the Science Committee of the Ministry of Education and Science of the Republic of Kazakhstan (Grant~No.~AP22787256). PB, MI, OV and OS thank the support from the special program of the Polish Academy of Sciences and the U.S. National Academy of Sciences under the Long-term program to support Ukrainian research teams grant No.~PAN.BFB.S.BWZ.329.022.2023.

\end{acknowledgements}

%%%%%%%%%%%%%%%%%%%%%%%%%%%%%%%%%%%%%%%%%%%%%%%%%%%%%%%%%%%%%%%%%%%%%
\bibliographystyle{mnras}  % style aa.bst
\bibliography{gc-lost}   % your references Yourfile.bib

@ARTICLE{VasAgama2019,
       author = {{Vasiliev}, Eugene},
        title = "{AGAMA: action-based galaxy modelling architecture}",
      journal = {\mnras},
         year = 2019,
        month = jan,
       volume = {482},
       number = {2},
        pages = {1525-1544},
          doi = {10.1093/mnras/sty2672},
archivePrefix = {arXiv},
       eprint = {1802.08239},
 primaryClass = {astro-ph.GA},
       adsurl = {https://ui.adsabs.harvard.edu/abs/2019MNRAS.482.1525V}
}

@BOOK{Binney2008,
       author = {{Binney}, James and {Tremaine}, Scott},
        title = "{Galactic Dynamics: Second Edition}",
         year = 2008,
    publisher = {Princeton, N.J.: Princeton University Press},
       adsurl = {https://ui.adsabs.harvard.edu/abs/2008gady.book.....B}
}

@ARTICLE{Fahrion2022,
       author = {{Fahrion}, Katja and {Leaman}, Ryan and {Lyubenova}, Mariya and {van de Ven}, Glenn},
        title = "{Disentangling the formation mechanisms of nuclear star clusters}",
      journal = {\aap},
         year = 2022,
        month = feb,
       volume = {658},
          eid = {A172},
        pages = {A172},
          doi = {10.1051/0004-6361/202039778},
archivePrefix = {arXiv},
       eprint = {2112.05610},
 primaryClass = {astro-ph.GA},
       adsurl = {https://ui.adsabs.harvard.edu/abs/2022A&A...658A.172F}
}

@ARTICLE{CN1942,
       author = {{Chandrasekhar}, S. and {von Neumann}, J.},
        title = "{The Statistics of the Gravitational Field Arising from a Random Distribution of Stars. I. The Speed of Fluctuations.}",
      journal = {\apj},
         year = 1942,
        month = may,
       volume = {95},
        pages = {489},
          doi = {10.1086/144420},
       adsurl = {https://ui.adsabs.harvard.edu/abs/1942ApJ....95..489C}
}

@ARTICLE{CN1943,
       author = {{Chandrasekhar}, S. and {von Neumann}, J.},
        title = "{The Statistics of the Gravitational Field Arising from a Random Distribution of Stars II}",
      journal = {\apj},
         year = 1943,
        month = jan,
       volume = {97},
        pages = {1},
          doi = {10.1086/144487},
       adsurl = {https://ui.adsabs.harvard.edu/abs/1943ApJ....97....1C}
}

@ARTICLE{C1943a,
       author = {{Chandrasekhar}, S.},
        title = "{Dynamical Friction. I. General Considerations: the Coefficient of Dynamical Friction.}",
      journal = {\apj},
         year = 1943,
        month = mar,
       volume = {97},
        pages = {255},
          doi = {10.1086/144517},
       adsurl = {https://ui.adsabs.harvard.edu/abs/1943ApJ....97..255C}
}

@ARTICLE{C1943b,
       author = {{Chandrasekhar}, S.},
        title = "{Dynamical Friction. II. The Rate of Escape of Stars from Clusters and the Evidence for the Operation of Dynamical Friction.}",
      journal = {\apj},
         year = 1943,
        month = mar,
       volume = {97},
        pages = {263},
          doi = {10.1086/144518},
       adsurl = {https://ui.adsabs.harvard.edu/abs/1943ApJ....97..263C}
}

@ARTICLE{NFW1997,
       author = {{Navarro}, Julio F. and {Frenk}, Carlos S. and {White}, Simon D.~M.},
        title = "{A Universal Density Profile from Hierarchical Clustering}",
      journal = {\apj},
         year = 1997,
        month = dec,
       volume = {490},
       number = {2},
        pages = {493-508},
          doi = {10.1086/304888},
archivePrefix = {arXiv},
       eprint = {astro-ph/9611107},
 primaryClass = {astro-ph},
       adsurl = {https://ui.adsabs.harvard.edu/abs/1997ApJ...490..493N}
}

@ARTICLE{Ishchenko2023a,
       author = {{Ishchenko}, Maryna and {Sobolenko}, Margaryta and {Berczik}, Peter and {Khoperskov}, Sergey and {Omarov}, Chingis and {Sobodar}, Olexander and {Makukov}, Maxim},
        title = "{Milky Way globular clusters on cosmological timescales. I. Evolution of the orbital parameters in time-varying potentials}",
      journal = {\aap},
         year = 2023,
        month = may,
       volume = {673},
          eid = {A152},
        pages = {A152 \hypertarget{I23a}{(Paper~II)}},
          doi = {10.1051/0004-6361/202245117},
archivePrefix = {arXiv},
       eprint = {2304.03547},
 primaryClass = {astro-ph.GA},
       adsurl = {https://ui.adsabs.harvard.edu/abs/2023A&A...673A.152I}
}

@ARTICLE{Kroupa2001,
       author = {{Kroupa}, Pavel},
        title = "{On the variation of the initial mass function}",
      journal = {\mnras},
         year = 2001,
        month = apr,
       volume = {322},
       number = {2},
        pages = {231-246},
          doi = {10.1046/j.1365-8711.2001.04022.x},
archivePrefix = {arXiv},
       eprint = {astro-ph/0009005},
 primaryClass = {astro-ph},
       adsurl = {https://ui.adsabs.harvard.edu/abs/2001MNRAS.322..231K}
}

@ARTICLE{Just2011,
       author = {{Just}, A. and {Khan}, F.~M. and {Berczik}, P. and {Ernst}, A. and {Spurzem}, R.},
        title = "{Dynamical friction of massive objects in galactic centres}",
      journal = {\mnras},
         year = 2011,
        month = feb,
       volume = {411},
       number = {1},
        pages = {653-674},
          doi = {10.1111/j.1365-2966.2010.17711.x},
archivePrefix = {arXiv},
       eprint = {1009.2455},
 primaryClass = {astro-ph.CO},
       adsurl = {https://ui.adsabs.harvard.edu/abs/2011MNRAS.411..653J}
}

@ARTICLE{Ishchenko2023b,
       author = {{Ishchenko}, Maryna and {Sobolenko}, Margaryta and {Kuvatova}, Dana and {Panamarev}, Taras and {Berczik}, Peter},
        title = "{Milky Way globular clusters on cosmological timescales. II. Interaction with the Galactic centre}",
      journal = {\aap},
         year = 2023,
        month = jun,
       volume = {674},
          eid = {A70},
        pages = {A70 \hypertarget{I23}{(Paper~I)}},
          doi = {10.1051/0004-6361/202245753},
archivePrefix = {arXiv},
       eprint = {2304.02311},
 primaryClass = {astro-ph.GA},
       adsurl = {https://ui.adsabs.harvard.edu/abs/2023A&A...674A..70I}
}

@ARTICLE{Ishchenko2024massloss,
       author = {{Ishchenko}, Maryna and {Berczik}, Peter and {Panamarev}, Taras and {Kuvatova}, Dana and {Kalambay}, Mukhagali and {Gluchshenko}, Anton and {Veles}, Oleksandr and {Sobolenko}, Margaryta and {Sobodar}, Olexander and {Omarov}, Chingis},
        title = "{Dynamical evolution of Milky Way globular clusters on the cosmological timescale: I. Mass loss and interaction with the nuclear star cluster}",
      journal = {\aap},
         year = 2024,
        month = sep,
       volume = {689},
          eid = {A178},
        pages = {A178},
          doi = {10.1051/0004-6361/202450399},
archivePrefix = {arXiv},
       eprint = {2406.18987},
 primaryClass = {astro-ph.GA},
       adsurl = {https://ui.adsabs.harvard.edu/abs/2024A&A...689A.178I}
}

@ARTICLE{Kamlah2022MNRAS,
       author = {{Kamlah}, A.~W.~H. and {Spurzem}, R. and {Berczik}, P. and {Arca Sedda}, M. and {Flammini Dotti}, F. and {Neumayer}, N. and {Pang}, X. and {Shu}, Q. and {Tanikawa}, A. and {Giersz}, M.},
        title = "{The impact of stellar evolution on rotating star clusters: the gravothermal-gravogyro catastrophe and the formation of a bar of black holes}",
      journal = {\mnras},
         year = 2022,
        month = nov,
       volume = {516},
       number = {3},
        pages = {3266-3283},
          doi = {10.1093/mnras/stac2281},
archivePrefix = {arXiv},
       eprint = {2205.04470},
 primaryClass = {astro-ph.GA},
       adsurl = {https://ui.adsabs.harvard.edu/abs/2022MNRAS.516.3266K}
}

@ARTICLE{Nelson2019,
       author = {{Nelson}, Dylan and {Springel}, Volker and {Pillepich}, Annalisa and {Rodriguez-Gomez}, Vicente and {Torrey}, Paul and {Genel}, Shy and {Vogelsberger}, Mark and {Pakmor}, Ruediger and {Marinacci}, Federico and {Weinberger}, Rainer and {Kelley}, Luke and {Lovell}, Mark and {Diemer}, Benedikt and {Hernquist}, Lars},
        title = "{The IllustrisTNG simulations: public data release}",
      journal = {Computational Astrophysics and Cosmology},
         year = 2019,
        month = may,
       volume = {6},
       number = {1},
          eid = {2},
        pages = {2},
          doi = {10.1186/s40668-019-0028-x},
archivePrefix = {arXiv},
       eprint = {1812.05609},
 primaryClass = {astro-ph.GA},
       adsurl = {https://ui.adsabs.harvard.edu/abs/2019ComAC...6....2N}
}

@INPROCEEDINGS{BSW2013,
   author = {{Berczik}, P. and {Spurzem}, R. and {Wang}, L.},
    title = "{Up to 700k GPU cores, Kepler, and the Exascale future for simulations of star clusters around black holes.}",
booktitle = {Third International Conference on High Performance Computing, HPC-UA 2013},
     year = 2013,
archivePrefix = "arXiv",
   eprint = {1312.1789},
 primaryClass = "astro-ph.IM",
    month = oct,
    pages = {52-59},
   adsurl = {http://adsabs.harvard.edu/abs/2013hpc..conf...52B}
}

@INPROCEEDINGS{Berczik2011,
       author = {{Berczik}, Peter and {Nitadori}, Keigo and {Zhong}, Shiyan and
         {Spurzem}, Rainer and {Hamada}, Tsuyoshi and {Wang}, Xiaowei and
         {Berentzen}, Ingo and {Veles}, Alexander and {Ge}, Wei},
        title = "{High performance massively parallel direct N-body simulations on large GPU clusters.}",
    booktitle = {International conference on High Performance Computing, HPC-UA 2011},
         year = "2011",
        month = "Oct",
        pages = {8-18},
       adsurl = {https://ui.adsabs.harvard.edu/abs/2011hpc..conf....8B}
}

@ARTICLE{Gravity2019,
       author = {{Gravity Collaboration} and {Abuter}, R. and {Amorim}, A. and {Baub{\"o}ck}, M. and {Berger}, J.~P. and {Bonnet}, H. and {Brandner}, W. and {Cl{\'e}net}, Y. and {Coud{\'e} Du Foresto}, V. and {de Zeeuw}, P.~T. and {Dexter}, J. and {Duvert}, G. and {Eckart}, A. and {Eisenhauer}, F. and {F{\"o}rster Schreiber}, N.~M. and {Garcia}, P. and {Gao}, F. and {Gendron}, E. and {Genzel}, R. and {Gerhard}, O. and {Gillessen}, S. and {Habibi}, M. and {Haubois}, X. and {Henning}, T. and {Hippler}, S. and {Horrobin}, M. and {Jim{\'e}nez-Rosales}, A. and {Jocou}, L. and {Kervella}, P. and {Lacour}, S. and {Lapeyr{\`e}re}, V. and {Le Bouquin}, J. -B. and {L{\'e}na}, P. and {Ott}, T. and {Paumard}, T. and {Perraut}, K. and {Perrin}, G. and {Pfuhl}, O. and {Rabien}, S. and {Rodriguez Coira}, G. and {Rousset}, G. and {Scheithauer}, S. and {Sternberg}, A. and {Straub}, O. and {Straubmeier}, C. and {Sturm}, E. and {Tacconi}, L.~J. and {Vincent}, F. and {von Fellenberg}, S. and {Waisberg}, I. and {Widmann}, F. and {Wieprecht}, E. and {Wiezorrek}, E. and {Woillez}, J. and {Yazici}, S.},
        title = "{A geometric distance measurement to the Galactic center black hole with 0.3\% uncertainty}",
      journal = {\aap},
         year = 2019,
        month = may,
       volume = {625},
          eid = {L10},
        pages = {L10},
          doi = {10.1051/0004-6361/201935656},
archivePrefix = {arXiv},
       eprint = {1904.05721},
 primaryClass = {astro-ph.GA},
       adsurl = {https://ui.adsabs.harvard.edu/abs/2019A&A...625L..10G}
}

@ARTICLE{Bland-Hawthorn2016,
       author = {{Bland-Hawthorn}, Joss and {Gerhard}, Ortwin},
        title = "{The Galaxy in Context: Structural, Kinematic, and Integrated Properties}",
      journal = {\araa},
         year = 2016,
        month = sep,
       volume = {54},
        pages = {529-596},
          doi = {10.1146/annurev-astro-081915-023441},
archivePrefix = {arXiv},
       eprint = {1602.07702},
 primaryClass = {astro-ph.GA},
       adsurl = {https://ui.adsabs.harvard.edu/abs/2016ARA&A..54..529B}
}

@ARTICLE{Kormendy2013,
       author = {{Kormendy}, John and {Ho}, Luis C.},
        title = "{Coevolution (Or Not) of Supermassive Black Holes and Host Galaxies: Supplemental Material}",
      journal = {arXiv e-prints},
         year = 2013,
        month = aug,
          eid = {arXiv:1308.6483},
        pages = {arXiv:1308.6483},
archivePrefix = {arXiv},
       eprint = {1308.6483},
 primaryClass = {astro-ph.CO},
       adsurl = {https://ui.adsabs.harvard.edu/abs/2013arXiv1308.6483K}
}

@ARTICLE{Baumgardt2021,
       author = {{Baumgardt}, H. and {Vasiliev}, E.},
        title = "{Accurate distances to Galactic globular clusters through a combination of Gaia EDR3, HST, and literature data}",
      journal = {\mnras},
         year = 2021,
        month = aug,
       volume = {505},
       number = {4},
        pages = {5957-5977},
          doi = {10.1093/mnras/stab1474},
archivePrefix = {arXiv},
       eprint = {2105.09526},
 primaryClass = {astro-ph.GA},
       adsurl = {https://ui.adsabs.harvard.edu/abs/2021MNRAS.505.5957B}
}

@ARTICLE{Miyamoto1975,
       author = {{Miyamoto}, M. and {Nagai}, R.},
        title = "{Three-dimensional models for the distribution of mass in galaxies.}",
      journal = {\pasj},
         year = 1975,
        month = jan,
       volume = {27},
        pages = {533-543},
       adsurl = {https://ui.adsabs.harvard.edu/abs/1975PASJ...27..533M}
}

@ARTICLE{Neumayer2020,
       author = {{Neumayer}, Nadine and {Seth}, Anil and {B{\"o}ker}, Torsten},
        title = "{Nuclear star clusters}",
      journal = {\aapr},
         year = 2020,
        month = jul,
       volume = {28},
       number = {1},
          eid = {4},
        pages = {4},
          doi = {10.1007/s00159-020-00125-0},
archivePrefix = {arXiv},
       eprint = {2001.03626},
 primaryClass = {astro-ph.GA},
       adsurl = {https://ui.adsabs.harvard.edu/abs/2020A&ARv..28....4N}
}

@ARTICLE{Schinnerer2006,
       author = {{Schinnerer}, Eva and {B{\"o}ker}, Torsten and {Emsellem}, Eric and {Lisenfeld}, Ute},
        title = "{Molecular Gas Dynamics in NGC 6946: A Bar-driven Nuclear Starburst ``Caught in the Act''}",
      journal = {\apj},
         year = 2006,
        month = sep,
       volume = {649},
       number = {1},
        pages = {181-200},
          doi = {10.1086/506265},
archivePrefix = {arXiv},
       eprint = {astro-ph/0605702},
 primaryClass = {astro-ph},
       adsurl = {https://ui.adsabs.harvard.edu/abs/2006ApJ...649..181S}
}

@ARTICLE{Tremaine1975,
       author = {{Tremaine}, S.~D. and {Ostriker}, J.~P. and {Spitzer}, L., Jr.},
        title = "{The formation of the nuclei of galaxies. I. M31.}",
      journal = {\apj},
         year = 1975,
        month = mar,
       volume = {196},
        pages = {407-411},
          doi = {10.1086/153422},
       adsurl = {https://ui.adsabs.harvard.edu/abs/1975ApJ...196..407T}
}

@ARTICLE{Antonini2013,
       author = {{Antonini}, Fabio},
        title = "{Origin and Growth of Nuclear Star Clusters around Massive Black Holes}",
      journal = {\apj},
         year = 2013,
        month = jan,
       volume = {763},
       number = {1},
          eid = {62},
        pages = {62},
          doi = {10.1088/0004-637X/763/1/62},
archivePrefix = {arXiv},
       eprint = {1207.6589},
 primaryClass = {astro-ph.GA},
       adsurl = {https://ui.adsabs.harvard.edu/abs/2013ApJ...763...62A}
}

@ARTICLE{Capuzzo-Dolcetta1993,
       author = {{Capuzzo-Dolcetta}, Roberto},
        title = "{The Evolution of the Globular Cluster System in a Triaxial Galaxy: Can a Galactic Nucleus Form by Globular Cluster Capture?}",
      journal = {\apj},
         year = 1993,
        month = oct,
       volume = {415},
        pages = {616},
          doi = {10.1086/173189},
archivePrefix = {arXiv},
       eprint = {astro-ph/9301006},
 primaryClass = {astro-ph},
       adsurl = {https://ui.adsabs.harvard.edu/abs/1993ApJ...415..616C}
}

@ARTICLE{Lotz2004,
       author = {{Lotz}, Jennifer M. and {Miller}, Bryan W. and {Ferguson}, Henry C.},
        title = "{The Colors of Dwarf Elliptical Galaxy Globular Cluster Systems, Nuclei, and Stellar Halos}",
      journal = {\apj},
         year = 2004,
        month = sep,
       volume = {613},
       number = {1},
        pages = {262-278},
          doi = {10.1086/422871},
archivePrefix = {arXiv},
       eprint = {astro-ph/0406002},
 primaryClass = {astro-ph},
       adsurl = {https://ui.adsabs.harvard.edu/abs/2004ApJ...613..262L}
}

@ARTICLE{Do2020,
       author = {{Do}, Tuan and {David Martinez}, Gregory and {Kerzendorf}, Wolfgang and {Feldmeier-Krause}, Anja and {Arca Sedda}, Manuel and {Neumayer}, Nadine and {Gualandris}, Alessia},
        title = "{Revealing the Formation of the Milky Way Nuclear Star Cluster via Chemo-dynamical Modeling}",
      journal = {\apjl},
         year = 2020,
        month = oct,
       volume = {901},
       number = {2},
          eid = {L28},
        pages = {L28},
          doi = {10.3847/2041-8213/abb246},
archivePrefix = {arXiv},
       eprint = {2009.02335},
 primaryClass = {astro-ph.GA},
       adsurl = {https://ui.adsabs.harvard.edu/abs/2020ApJ...901L..28D}
}

@ARTICLE{King1966,
       author = {{King}, Ivan R.},
        title = "{The structure of star clusters. III. Some simple dynamical models}",
      journal = {\aj},
         year = 1966,
        month = feb,
       volume = {71},
        pages = {64},
          doi = {10.1086/109857},
       adsurl = {https://ui.adsabs.harvard.edu/abs/1966AJ.....71...64K}
}

@ARTICLE{Blum2003,
       author = {{Blum}, R.~D. and {Ram{\'\i}rez}, Solange V. and {Sellgren}, K. and {Olsen}, K.},
        title = "{Really Cool Stars and the Star Formation History at the Galactic Center}",
      journal = {\apj},
         year = 2003,
        month = nov,
       volume = {597},
       number = {1},
        pages = {323-346},
          doi = {10.1086/378380},
archivePrefix = {arXiv},
       eprint = {astro-ph/0307291},
 primaryClass = {astro-ph},
       adsurl = {https://ui.adsabs.harvard.edu/abs/2003ApJ...597..323B}
}

@ARTICLE{Pfuhl2011,
       author = {{Pfuhl}, O. and {Fritz}, T.~K. and {Zilka}, M. and {Maness}, H. and {Eisenhauer}, F. and {Genzel}, R. and {Gillessen}, S. and {Ott}, T. and {Dodds-Eden}, K. and {Sternberg}, A.},
        title = "{The Star Formation History of the Milky Way's Nuclear Star Cluster}",
      journal = {\apj},
         year = 2011,
        month = nov,
       volume = {741},
       number = {2},
          eid = {108},
        pages = {108},
          doi = {10.1088/0004-637X/741/2/108},
archivePrefix = {arXiv},
       eprint = {1110.1633},
 primaryClass = {astro-ph.GA},
       adsurl = {https://ui.adsabs.harvard.edu/abs/2011ApJ...741..108P}
}

@ARTICLE{Lotz2001,
       author = {{Lotz}, Jennifer M. and {Telford}, Rosemary and {Ferguson}, Henry C. and {Miller}, Bryan W. and {Stiavelli}, Massimo and {Mack}, Jennifer},
        title = "{Dynamical Friction in DE Globular Cluster Systems}",
      journal = {\apj},
         year = 2001,
        month = may,
       volume = {552},
       number = {2},
        pages = {572-581},
          doi = {10.1086/320545},
archivePrefix = {arXiv},
       eprint = {astro-ph/0102079},
 primaryClass = {astro-ph},
       adsurl = {https://ui.adsabs.harvard.edu/abs/2001ApJ...552..572L}
}

@ARTICLE{Capuzzo-Dolcetta2009,
       author = {{Capuzzo-Dolcetta}, R. and {Mastrobuono-Battisti}, A.},
        title = "{Globular cluster system erosion in elliptical galaxies}",
      journal = {\aap},
         year = 2009,
        month = nov,
       volume = {507},
       number = {1},
        pages = {183-193},
          doi = {10.1051/0004-6361/200912255},
archivePrefix = {arXiv},
       eprint = {0904.0526},
 primaryClass = {astro-ph.CO},
       adsurl = {https://ui.adsabs.harvard.edu/abs/2009A&A...507..183C}
}

@ARTICLE{Oh2000,
       author = {{Oh}, K.~S. and {Lin}, D.~N.~C. and {Richer}, Harvey B.},
        title = "{Globular Clusters in the Fornax Dwarf Spheroidal Galaxy}",
      journal = {\apj},
         year = 2000,
        month = mar,
       volume = {531},
       number = {2},
        pages = {727-738},
          doi = {10.1086/308477},
       adsurl = {https://ui.adsabs.harvard.edu/abs/2000ApJ...531..727O}
}

@ARTICLE{Agarwal2011,
       author = {{Agarwal}, Meghann and {Milosavljevi{\'c}}, Milo{\v{s}}},
        title = "{Nuclear Star Clusters from Clustered Star Formation}",
      journal = {\apj},
         year = 2011,
        month = mar,
       volume = {729},
       number = {1},
          eid = {35},
        pages = {35},
          doi = {10.1088/0004-637X/729/1/35},
archivePrefix = {arXiv},
       eprint = {1008.2986},
 primaryClass = {astro-ph.CO},
       adsurl = {https://ui.adsabs.harvard.edu/abs/2011ApJ...729...35A}
}

@ARTICLE{Hartmann2011,
       author = {{Hartmann}, Markus and {Debattista}, Victor P. and {Seth}, Anil and {Cappellari}, Michele and {Quinn}, Thomas R.},
        title = "{Constraining the role of star cluster mergers in nuclear cluster formation: simulations confront integral-field data}",
      journal = {\mnras},
         year = 2011,
        month = dec,
       volume = {418},
       number = {4},
        pages = {2697-2714},
          doi = {10.1111/j.1365-2966.2011.19659.x},
archivePrefix = {arXiv},
       eprint = {1103.5464},
 primaryClass = {astro-ph.CO},
       adsurl = {https://ui.adsabs.harvard.edu/abs/2011MNRAS.418.2697H}
}

@ARTICLE{Boker2004,
       author = {{B{\"o}ker}, Torsten and {Sarzi}, Marc and {McLaughlin}, Dean E. and {van der Marel}, Roeland P. and {Rix}, Hans-Walter and {Ho}, Luis C. and {Shields}, Joseph C.},
        title = "{A Hubble Space Telescope Census of Nuclear Star Clusters in Late-Type Spiral Galaxies. II. Cluster Sizes and Structural Parameter Correlations}",
      journal = {\aj},
         year = 2004,
        month = jan,
       volume = {127},
       number = {1},
        pages = {105-118},
          doi = {10.1086/380231},
archivePrefix = {arXiv},
       eprint = {astro-ph/0309761},
 primaryClass = {astro-ph},
       adsurl = {https://ui.adsabs.harvard.edu/abs/2004AJ....127..105B}
}

@ARTICLE{Matthews1997,
       author = {{Matthews}, Lynn D. and {Gallagher}, John S., III},
        title = "{B and V CCD Photometry of Southern, Extreme Late-Type Spiral Galaxies}",
      journal = {\aj},
         year = 1997,
        month = nov,
       volume = {114},
        pages = {1899},
          doi = {10.1086/118613},
archivePrefix = {arXiv},
       eprint = {astro-ph/9709145},
 primaryClass = {astro-ph},
       adsurl = {https://ui.adsabs.harvard.edu/abs/1997AJ....114.1899M}
}

@ARTICLE{Boker2002,
       author = {{B{\"o}ker}, Torsten and {Laine}, Seppo and {van der Marel}, Roeland P. and {Sarzi}, Marc and {Rix}, Hans-Walter and {Ho}, Luis C. and {Shields}, Joseph C.},
        title = "{A Hubble Space Telescope Census of Nuclear Star Clusters in Late-Type Spiral Galaxies. I. Observations and Image Analysis}",
      journal = {\aj},
         year = 2002,
        month = mar,
       volume = {123},
       number = {3},
        pages = {1389-1410},
          doi = {10.1086/339025},
archivePrefix = {arXiv},
       eprint = {astro-ph/0112086},
 primaryClass = {astro-ph},
       adsurl = {https://ui.adsabs.harvard.edu/abs/2002AJ....123.1389B}
}

@ARTICLE{Ho1995,
       author = {{Ho}, L.~C. and {Filippenko}, A.~V. and {Sargent}, W.~L.},
        title = "{A Search for ``Dwarf'' Seyfert Nuclei. II. an Optical Spectral Atlas of the Nuclei of Nearby Galaxies}",
      journal = {\apjs},
         year = 1995,
        month = jun,
       volume = {98},
        pages = {477},
          doi = {10.1086/192170},
       adsurl = {https://ui.adsabs.harvard.edu/abs/1995ApJS...98..477H}
}

@ARTICLE{Genzel2003,
       author = {{Genzel}, R. and {Sch{\"o}del}, R. and {Ott}, T. and {Eisenhauer}, F. and {Hofmann}, R. and {Lehnert}, M. and {Eckart}, A. and {Alexander}, T. and {Sternberg}, A. and {Lenzen}, R. and {Cl{\'e}net}, Y. and {Lacombe}, F. and {Rouan}, D. and {Renzini}, A. and {Tacconi-Garman}, L.~E.},
        title = "{The Stellar Cusp around the Supermassive Black Hole in the Galactic Center}",
      journal = {\apj},
         year = 2003,
        month = sep,
       volume = {594},
       number = {2},
        pages = {812-832},
          doi = {10.1086/377127},
archivePrefix = {arXiv},
       eprint = {astro-ph/0305423},
 primaryClass = {astro-ph},
       adsurl = {https://ui.adsabs.harvard.edu/abs/2003ApJ...594..812G}
}

@ARTICLE{Cote2006,
       author = {{C{\^o}t{\'e}}, Patrick and {Piatek}, Slawomir and {Ferrarese}, Laura and {Jord{\'a}n}, Andr{\'e}s and {Merritt}, David and {Peng}, Eric W. and {Ha{\c{s}}egan}, Monica and {Blakeslee}, John P. and {Mei}, Simona and {West}, Michael J. and {Milosavljevi{\'c}}, Milo{\v{s}} and {Tonry}, John L.},
        title = "{The ACS Virgo Cluster Survey. VIII. The Nuclei of Early-Type Galaxies}",
      journal = {\apjs},
         year = 2006,
        month = jul,
       volume = {165},
       number = {1},
        pages = {57-94},
          doi = {10.1086/504042},
archivePrefix = {arXiv},
       eprint = {astro-ph/0603252},
 primaryClass = {astro-ph},
       adsurl = {https://ui.adsabs.harvard.edu/abs/2006ApJS..165...57C}
}

@ARTICLE{Wehner2006,
       author = {{Wehner}, Elizabeth H. and {Harris}, William E.},
        title = "{From Supermassive Black Holes to Dwarf Elliptical Nuclei: A Mass Continuum}",
      journal = {\apjl},
         year = 2006,
        month = jun,
       volume = {644},
       number = {1},
        pages = {L17-L20},
          doi = {10.1086/505387},
archivePrefix = {arXiv},
       eprint = {astro-ph/0603801},
 primaryClass = {astro-ph},
       adsurl = {https://ui.adsabs.harvard.edu/abs/2006ApJ...644L..17W}
}

@ARTICLE{Ferrarese2006,
       author = {{Ferrarese}, Laura and {C{\^o}t{\'e}}, Patrick and {Dalla Bont{\`a}}, Elena and {Peng}, Eric W. and {Merritt}, David and {Jord{\'a}n}, Andr{\'e}s and {Blakeslee}, John P. and {Ha{\c{s}}egan}, Monica and {Mei}, Simona and {Piatek}, Slawomir and {Tonry}, John L. and {West}, Michael J.},
        title = "{A Fundamental Relation between Compact Stellar Nuclei, Supermassive Black Holes, and Their Host Galaxies}",
      journal = {\apjl},
         year = 2006,
        month = jun,
       volume = {644},
       number = {1},
        pages = {L21-L24},
          doi = {10.1086/505388},
archivePrefix = {arXiv},
       eprint = {astro-ph/0603840},
 primaryClass = {astro-ph},
       adsurl = {https://ui.adsabs.harvard.edu/abs/2006ApJ...644L..21F}
}

@ARTICLE{Rossa2006,
       author = {{Rossa}, J{\"o}rn and {van der Marel}, Roeland P. and {B{\"o}ker}, Torsten and {Gerssen}, Joris and {Ho}, Luis C. and {Rix}, Hans-Walter and {Shields}, Joseph C. and {Walcher}, Carl-Jakob},
        title = "{Hubble Space Telescope STIS Spectra of Nuclear Star Clusters in Spiral Galaxies: Dependence of Age and Mass on Hubble Type}",
      journal = {\aj},
         year = 2006,
        month = sep,
       volume = {132},
       number = {3},
        pages = {1074-1099},
          doi = {10.1086/505968},
archivePrefix = {arXiv},
       eprint = {astro-ph/0604140},
 primaryClass = {astro-ph},
       adsurl = {https://ui.adsabs.harvard.edu/abs/2006AJ....132.1074R}
}

@ARTICLE{Seth2006,
       author = {{Seth}, Anil C. and {Dalcanton}, Julianne J. and {Hodge}, Paul W. and {Debattista}, Victor P.},
        title = "{Clues to Nuclear Star Cluster Formation from Edge-on Spirals}",
      journal = {\aj},
         year = 2006,
        month = dec,
       volume = {132},
       number = {6},
        pages = {2539-2555},
          doi = {10.1086/508994},
archivePrefix = {arXiv},
       eprint = {astro-ph/0609302},
 primaryClass = {astro-ph},
       adsurl = {https://ui.adsabs.harvard.edu/abs/2006AJ....132.2539S}
}

@ARTICLE{Capuzzo-Dolcetta2008ApJ,
       author = {{Capuzzo-Dolcetta}, R. and {Miocchi}, P.},
        title = "{Merging of Globular Clusters in Inner Galactic Regions. II. Nuclear Star Cluster Formation}",
      journal = {\apj},
         year = 2008,
        month = jul,
       volume = {681},
       number = {2},
        pages = {1136-1147},
          doi = {10.1086/588017},
archivePrefix = {arXiv},
       eprint = {0801.1072},
 primaryClass = {astro-ph},
       adsurl = {https://ui.adsabs.harvard.edu/abs/2008ApJ...681.1136C}
}

@ARTICLE{DeLorenzi2013,
       author = {{De Lorenzi}, F. and {Hartmann}, M. and {Debattista}, V.~P. and {Seth}, A.~C. and {Gerhard}, O.},
        title = "{Three-integral multicomponent dynamical models and simulations of the nuclear star cluster in NGC 4244}",
      journal = {\mnras},
         year = 2013,
        month = mar,
       volume = {429},
       number = {4},
        pages = {2974-2985},
          doi = {10.1093/mnras/sts545},
archivePrefix = {arXiv},
       eprint = {1208.2161},
 primaryClass = {astro-ph.GA},
       adsurl = {https://ui.adsabs.harvard.edu/abs/2013MNRAS.429.2974D}
}

@ARTICLE{Capuzzo-Dolcetta2008MNRAS,
       author = {{Capuzzo-Dolcetta}, R. and {Miocchi}, P.},
        title = "{Self-consistent simulations of nuclear cluster formation through globular cluster orbital decay and merging}",
      journal = {\mnras},
         year = 2008,
        month = jul,
       volume = {388},
       number = {1},
        pages = {L69-L73},
          doi = {10.1111/j.1745-3933.2008.00501.x},
archivePrefix = {arXiv},
       eprint = {0804.4421},
 primaryClass = {astro-ph},
       adsurl = {https://ui.adsabs.harvard.edu/abs/2008MNRAS.388L..69C}
}

@ARTICLE{Gnedin2014,
       author = {{Gnedin}, Oleg Y. and {Ostriker}, Jeremiah P. and {Tremaine}, Scott},
        title = "{Co-evolution of Galactic Nuclei and Globular Cluster Systems}",
      journal = {\apj},
         year = 2014,
        month = apr,
       volume = {785},
       number = {1},
          eid = {71},
        pages = {71},
          doi = {10.1088/0004-637X/785/1/71},
archivePrefix = {arXiv},
       eprint = {1308.0021},
 primaryClass = {astro-ph.CO},
       adsurl = {https://ui.adsabs.harvard.edu/abs/2014ApJ...785...71G}
}

@ARTICLE{Loose1982,
       author = {{Loose}, H.~H. and {Kruegel}, E. and {Tutukov}, A.},
        title = "{Bursts of star formation in the galactic centre}",
      journal = {\aap},
         year = 1982,
        month = jan,
       volume = {105},
       number = {2},
        pages = {342-350},
       adsurl = {https://ui.adsabs.harvard.edu/abs/1982A&A...105..342L}
}

@ARTICLE{Milosavljevic2004,
       author = {{Milosavljevi{\'c}}, Milo{\v{s}}},
        title = "{On the Origin of Nuclear Star Clusters in Late-Type Spiral Galaxies}",
      journal = {\apjl},
         year = 2004,
        month = apr,
       volume = {605},
       number = {1},
        pages = {L13-L16},
          doi = {10.1086/420696},
archivePrefix = {arXiv},
       eprint = {astro-ph/0310574},
 primaryClass = {astro-ph},
       adsurl = {https://ui.adsabs.harvard.edu/abs/2004ApJ...605L..13M}
}

@ARTICLE{Nayakshin2009,
       author = {{Nayakshin}, Sergei and {Wilkinson}, Mark I. and {King}, Andrew},
        title = "{Competitive feedback in galaxy formation}",
      journal = {\mnras},
         year = 2009,
        month = sep,
       volume = {398},
       number = {1},
        pages = {L54-L57},
          doi = {10.1111/j.1745-3933.2009.00709.x},
archivePrefix = {arXiv},
       eprint = {0907.1002},
 primaryClass = {astro-ph.CO},
       adsurl = {https://ui.adsabs.harvard.edu/abs/2009MNRAS.398L..54N}
}

@ARTICLE{Aharon2015,
       author = {{Aharon}, Danor and {Perets}, Hagai B.},
        title = "{Formation and Evolution of Nuclear Star Clusters with In Situ Star Formation: Nuclear Cores and Age Segregation}",
      journal = {\apj},
         year = 2015,
        month = feb,
       volume = {799},
       number = {2},
          eid = {185},
        pages = {185},
          doi = {10.1088/0004-637X/799/2/185},
archivePrefix = {arXiv},
       eprint = {1409.5121},
 primaryClass = {astro-ph.GA},
       adsurl = {https://ui.adsabs.harvard.edu/abs/2015ApJ...799..185A}
}

@ARTICLE{Seth2008,
       author = {{Seth}, Anil and {Ag{\"u}eros}, Marcel and {Lee}, Duane and {Basu-Zych}, Antara},
        title = "{The Coincidence of Nuclear Star Clusters and Active Galactic Nuclei}",
      journal = {\apj},
         year = 2008,
        month = may,
       volume = {678},
       number = {1},
        pages = {116-130},
          doi = {10.1086/528955},
archivePrefix = {arXiv},
       eprint = {0801.0439},
 primaryClass = {astro-ph},
       adsurl = {https://ui.adsabs.harvard.edu/abs/2008ApJ...678..116S}
}

@ARTICLE{Do2019,
       author = {{Do}, Tuan and {Hees}, Aurelien and {Ghez}, Andrea and {Martinez}, Gregory D. and {Chu}, Devin S. and {Jia}, Siyao and {Sakai}, Shoko and {Lu}, Jessica R. and {Gautam}, Abhimat K. and {O'Neil}, Kelly Kosmo and {Becklin}, Eric E. and {Morris}, Mark R. and {Matthews}, Keith and {Nishiyama}, Shogo and {Campbell}, Randy and {Chappell}, Samantha and {Chen}, Zhuo and {Ciurlo}, Anna and {Dehghanfar}, Arezu and {Gallego-Cano}, Eulalia and {Kerzendorf}, Wolfgang E. and {Lyke}, James E. and {Naoz}, Smadar and {Saida}, Hiromi and {Sch{\"o}del}, Rainer and {Takahashi}, Masaaki and {Takamori}, Yohsuke and {Witzel}, Gunther and {Wizinowich}, Peter},
        title = "{Relativistic redshift of the star S0-2 orbiting the Galactic Center supermassive black hole}",
      journal = {Science},
         year = 2019,
        month = aug,
       volume = {365},
       number = {6454},
        pages = {664-668},
          doi = {10.1126/science.aav8137},
archivePrefix = {arXiv},
       eprint = {1907.10731},
 primaryClass = {astro-ph.GA},
       adsurl = {https://ui.adsabs.harvard.edu/abs/2019Sci...365..664D}
}

@INPROCEEDINGS{Schodel2008,
       author = {{Sch{\"o}del}, Rainer and {Merritt}, David and {Eckart}, Andreas},
        title = "{The nuclear star cluster of the Milky Way}",
    booktitle = {Journal of Physics Conference Series},
         year = 2008,
       series = {Journal of Physics Conference Series},
       volume = {131},
        month = oct,
    publisher = {IOP},
          eid = {012044},
        pages = {012044},
          doi = {10.1088/1742-6596/131/1/012044},
archivePrefix = {arXiv},
       eprint = {0810.0204},
 primaryClass = {astro-ph},
       adsurl = {https://ui.adsabs.harvard.edu/abs/2008JPhCS.131a2044S}
}

@ARTICLE{Gillessen2009,
       author = {{Gillessen}, S. and {Eisenhauer}, F. and {Trippe}, S. and {Alexander}, T. and {Genzel}, R. and {Martins}, F. and {Ott}, T.},
        title = "{Monitoring Stellar Orbits Around the Massive Black Hole in the Galactic Center}",
      journal = {\apj},
         year = 2009,
        month = feb,
       volume = {692},
       number = {2},
        pages = {1075-1109},
          doi = {10.1088/0004-637X/692/2/1075},
archivePrefix = {arXiv},
       eprint = {0810.4674},
 primaryClass = {astro-ph},
       adsurl = {https://ui.adsabs.harvard.edu/abs/2009ApJ...692.1075G}
}

@ARTICLE{Paumard2006,
       author = {{Paumard}, T. and {Genzel}, R. and {Martins}, F. and {Nayakshin}, S. and {Beloborodov}, A.~M. and {Levin}, Y. and {Trippe}, S. and {Eisenhauer}, F. and {Ott}, T. and {Gillessen}, S. and {Abuter}, R. and {Cuadra}, J. and {Alexander}, T. and {Sternberg}, A.},
        title = "{The Two Young Star Disks in the Central Parsec of the Galaxy: Properties, Dynamics, and Formation}",
      journal = {\apj},
         year = 2006,
        month = jun,
       volume = {643},
       number = {2},
        pages = {1011-1035},
          doi = {10.1086/503273},
archivePrefix = {arXiv},
       eprint = {astro-ph/0601268},
 primaryClass = {astro-ph},
       adsurl = {https://ui.adsabs.harvard.edu/abs/2006ApJ...643.1011P}
}

@ARTICLE{Feldmeier-Krause2015,
       author = {{Feldmeier-Krause}, A. and {Neumayer}, N. and {Sch{\"o}del}, R. and {Seth}, A. and {Hilker}, M. and {de Zeeuw}, P.~T. and {Kuntschner}, H. and {Walcher}, C.~J. and {L{\"u}tzgendorf}, N. and {Kissler-Patig}, M.},
        title = "{KMOS view of the Galactic centre. I. Young stars are centrally concentrated}",
      journal = {\aap},
         year = 2015,
        month = dec,
       volume = {584},
          eid = {A2},
        pages = {A2},
          doi = {10.1051/0004-6361/201526336},
archivePrefix = {arXiv},
       eprint = {1509.04707},
 primaryClass = {astro-ph.GA},
       adsurl = {https://ui.adsabs.harvard.edu/abs/2015A&A...584A...2F}
}

@ARTICLE{Do2015,
       author = {{Do}, Tuan and {Kerzendorf}, Wolfgang and {Winsor}, Nathan and {St{\o}stad}, Morten and {Morris}, Mark R. and {Lu}, Jessica R. and {Ghez}, Andrea M.},
        title = "{Discovery of Low-metallicity Stars in the Central Parsec of the Milky Way}",
      journal = {\apj},
         year = 2015,
        month = aug,
       volume = {809},
       number = {2},
          eid = {143},
        pages = {143},
          doi = {10.1088/0004-637X/809/2/143},
archivePrefix = {arXiv},
       eprint = {1506.07891},
 primaryClass = {astro-ph.GA},
       adsurl = {https://ui.adsabs.harvard.edu/abs/2015ApJ...809..143D}
}

@ARTICLE{Trippe2008,
       author = {{Trippe}, S. and {Gillessen}, S. and {Gerhard}, O.~E. and {Bartko}, H. and {Fritz}, T.~K. and {Maness}, H.~L. and {Eisenhauer}, F. and {Martins}, F. and {Ott}, T. and {Dodds-Eden}, K. and {Genzel}, R.},
        title = "{Kinematics of the old stellar population at the Galactic centre}",
      journal = {\aap},
         year = 2008,
        month = dec,
       volume = {492},
       number = {2},
        pages = {419-439},
          doi = {10.1051/0004-6361:200810191},
archivePrefix = {arXiv},
       eprint = {0810.1040},
 primaryClass = {astro-ph},
       adsurl = {https://ui.adsabs.harvard.edu/abs/2008A&A...492..419T}
}

@ARTICLE{Schodel2009,
       author = {{Sch{\"o}del}, R. and {Merritt}, D. and {Eckart}, A.},
        title = "{The nuclear star cluster of the Milky Way: proper motions and mass}",
      journal = {\aap},
         year = 2009,
        month = jul,
       volume = {502},
       number = {1},
        pages = {91-111},
          doi = {10.1051/0004-6361/200810922},
archivePrefix = {arXiv},
       eprint = {0902.3892},
 primaryClass = {astro-ph.GA},
       adsurl = {https://ui.adsabs.harvard.edu/abs/2009A&A...502...91S}
}

@ARTICLE{Arca-Sedda2018,
       author = {{Arca-Sedda}, Manuel and {Kocsis}, Bence and {Brandt}, Timothy D.},
        title = "{Gamma-ray and X-ray emission from the Galactic centre: hints on the nuclear star cluster formation history}",
      journal = {\mnras},
         year = 2018,
        month = sep,
       volume = {479},
       number = {1},
        pages = {900-916},
          doi = {10.1093/mnras/sty1454},
archivePrefix = {arXiv},
       eprint = {1709.03119},
 primaryClass = {astro-ph.GA},
       adsurl = {https://ui.adsabs.harvard.edu/abs/2018MNRAS.479..900A}
}

@ARTICLE{Tsatsi2017,
       author = {{Tsatsi}, Athanasia and {Mastrobuono-Battisti}, Alessandra and {van de Ven}, Glenn and {Perets}, Hagai B. and {Bianchini}, Paolo and {Neumayer}, Nadine},
        title = "{On the rotation of nuclear star clusters formed by cluster inspirals}",
      journal = {\mnras},
         year = 2017,
        month = jan,
       volume = {464},
       number = {3},
        pages = {3720-3727},
          doi = {10.1093/mnras/stw2593},
archivePrefix = {arXiv},
       eprint = {1610.01162},
 primaryClass = {astro-ph.GA},
       adsurl = {https://ui.adsabs.harvard.edu/abs/2017MNRAS.464.3720T}
}

@ARTICLE{Leigh2015,
       author = {{Leigh}, Nathan W.~C. and {Georgiev}, Iskren Y. and {B{\"o}ker}, Torsten and {Knigge}, Christian and {den Brok}, Mark},
        title = "{Nuclear star cluster formation in energy-space}",
      journal = {\mnras},
         year = 2015,
        month = jul,
       volume = {451},
       number = {1},
        pages = {859-869},
          doi = {10.1093/mnras/stv1012},
archivePrefix = {arXiv},
       eprint = {1505.01158},
 primaryClass = {astro-ph.GA},
       adsurl = {https://ui.adsabs.harvard.edu/abs/2015MNRAS.451..859L}
}

@ARTICLE{Bekki2004,
       author = {{Bekki}, Kenji and {Couch}, Warrick J. and {Beasley}, Michael A. and {Forbes}, Duncan A. and {Chiba}, Masashi and {Da Costa}, Gary S.},
        title = "{Explaining the Mysterious Age Gap of Globular Clusters in the Large Magellanic Cloud}",
      journal = {\apjl},
         year = 2004,
        month = aug,
       volume = {610},
       number = {2},
        pages = {L93-L96},
          doi = {10.1086/423372},
archivePrefix = {arXiv},
       eprint = {astro-ph/0406443},
 primaryClass = {astro-ph},
       adsurl = {https://ui.adsabs.harvard.edu/abs/2004ApJ...610L..93B}
}

@INPROCEEDINGS{Bekki2010,
       author = {{Bekki}, Kenji},
        title = "{Star cluster dynamics in galaxies}",
    booktitle = {Star Clusters: Basic Galactic Building Blocks Throughout Time and Space},
         year = 2010,
       editor = {{de Grijs}, Richard and {L{\'e}pine}, Jacques R.~D.},
       series = {IAU Symposium},
       volume = {266},
        month = jan,
        pages = {219-230},
          doi = {10.1017/S1743921309991086},
       adsurl = {https://ui.adsabs.harvard.edu/abs/2010IAUS..266..219B}
}

@ARTICLE{Bekki2010MNRAS,
       author = {{Bekki}, Kenji},
        title = "{Dynamical friction of star clusters against disc field stars in galaxies: implications on stellar nucleus formation and globular cluster luminosity functions}",
      journal = {\mnras},
         year = 2010,
        month = feb,
       volume = {401},
       number = {4},
        pages = {2753-2762},
          doi = {10.1111/j.1365-2966.2009.15874.x},
       adsurl = {https://ui.adsabs.harvard.edu/abs/2010MNRAS.401.2753B}
}

@ARTICLE{Miocchi2006,
       author = {{Miocchi}, P. and {Capuzzo Dolcetta}, R. and {Di Matteo}, P. and {Vicari}, A.},
        title = "{Merging of Globular Clusters in Inner Galactic Regions. I. Do They Survive the Tidal Interaction?}",
      journal = {\apj},
         year = 2006,
        month = jun,
       volume = {644},
       number = {2},
        pages = {940-953},
          doi = {10.1086/503663},
archivePrefix = {arXiv},
       eprint = {astro-ph/0501618},
 primaryClass = {astro-ph},
       adsurl = {https://ui.adsabs.harvard.edu/abs/2006ApJ...644..940M}
}

@ARTICLE{Capuzzo-Dolcetta2005,
       author = {{Capuzzo-Dolcetta}, R. and {Vicari}, A.},
        title = "{Dynamical friction on globular clusters in core-triaxial galaxies: is it a cause of massive black hole accretion?}",
      journal = {\mnras},
         year = 2005,
        month = jan,
       volume = {356},
       number = {3},
        pages = {899-912},
          doi = {10.1111/j.1365-2966.2004.08433.x},
archivePrefix = {arXiv},
       eprint = {astro-ph/0309488},
 primaryClass = {astro-ph},
       adsurl = {https://ui.adsabs.harvard.edu/abs/2005MNRAS.356..899C}
}

@ARTICLE{Pesce1992,
       author = {{Pesce}, E. and {Capuzzo-Dolcetta}, R. and {Vietri}, M.},
        title = "{Dynamical friction in a non-rotating triaxial galaxy : the effect on box orbits.}",
      journal = {\mnras},
         year = 1992,
        month = feb,
       volume = {254},
        pages = {466-476},
          doi = {10.1093/mnras/254.3.466},
       adsurl = {https://ui.adsabs.harvard.edu/abs/1992MNRAS.254..466P}
}

@ARTICLE{Perets2014,
       author = {{Perets}, Hagai B. and {Mastrobuono-Battisti}, Alessandra},
        title = "{Age and Mass Segregation of Multiple Stellar Populations in Galactic Nuclei and their Observational Signatures}",
      journal = {\apjl},
         year = 2014,
        month = apr,
       volume = {784},
       number = {2},
          eid = {L44},
        pages = {L44},
          doi = {10.1088/2041-8205/784/2/L44},
archivePrefix = {arXiv},
       eprint = {1401.1824},
 primaryClass = {astro-ph.GA},
       adsurl = {https://ui.adsabs.harvard.edu/abs/2014ApJ...784L..44P}
}

@ARTICLE{Gao2024,
       author = {{Gao}, Yuan and {Li}, Hui and {Zhang}, Xiaojia and {Su}, Meng and {Ng}, Stephen Chi Yung},
        title = "{Globular clusters contribute to the nuclear star clusters and galaxy centre {\ensuremath{\gamma}}-ray excess, moderated by galaxy assembly history}",
      journal = {\mnras},
         year = 2024,
        month = jan,
       volume = {527},
       number = {3},
        pages = {7731-7742},
          doi = {10.1093/mnras/stad3585},
archivePrefix = {arXiv},
       eprint = {2311.17071},
 primaryClass = {astro-ph.GA},
       adsurl = {https://ui.adsabs.harvard.edu/abs/2024MNRAS.527.7731G}
}

@ARTICLE{Partmann2025,
       author = {{Partmann}, Christian and {Naab}, Thorsten and {Lah{\'e}n}, Natalia and {Rantala}, Antti and {Hirschmann}, Michaela and {Hislop}, Jessica M. and {Petersson}, Jonathan and {Johansson}, Peter H.},
        title = "{The importance of nuclear star clusters for massive black hole growth and nuclear star formation in simulated low-mass galaxies}",
      journal = {\mnras},
         year = 2025,
        month = feb,
       volume = {537},
       number = {2},
        pages = {956-977},
          doi = {10.1093/mnras/staf002},
archivePrefix = {arXiv},
       eprint = {2409.18096},
 primaryClass = {astro-ph.GA},
       adsurl = {https://ui.adsabs.harvard.edu/abs/2025MNRAS.537..956P}
}

@ARTICLE{Pagnini2023,
       author = {{Pagnini}, G. and {Di Matteo}, P. and {Khoperskov}, S. and {Mastrobuono-Battisti}, A. and {Haywood}, M. and {Renaud}, F. and {Combes}, F.},
        title = "{The distribution of globular clusters in kinematic spaces does not trace the accretion history of the host galaxy}",
      journal = {\aap},
         year = 2023,
        month = may,
       volume = {673},
          eid = {A86},
        pages = {A86},
          doi = {10.1051/0004-6361/202245128},
archivePrefix = {arXiv},
       eprint = {2210.04245},
 primaryClass = {astro-ph.GA},
       adsurl = {https://ui.adsabs.harvard.edu/abs/2023A&A...673A..86P}
}

@ARTICLE{Kravtsov2005,
       author = {{Kravtsov}, Andrey V. and {Gnedin}, Oleg Y.},
        title = "{Formation of Globular Clusters in Hierarchical Cosmology}",
      journal = {\apj},
         year = 2005,
        month = apr,
       volume = {623},
       number = {2},
        pages = {650-665},
          doi = {10.1086/428636},
archivePrefix = {arXiv},
       eprint = {astro-ph/0305199},
 primaryClass = {astro-ph},
       adsurl = {https://ui.adsabs.harvard.edu/abs/2005ApJ...623..650K}
}
%%%%%%%%%%%%%%%%%%%%%%%%%%%%%%%%%%%%%%%%%%%%%%%%%%%%%%%%%%%%%%%%%%%%%

\begin{appendix}
\onecolumn

%%%%%%%%%%%%%%%%%%%%%%%%%%%%%%%%%%%%%%%%%%%%%%%%%%%%%%%%%%%%%%%%%%%%%
\section{GCs orbital evolution for different {\tt ecc}} \label{app:orbits}
%%%%%%%%%%%%%%%%%%%%%%%%%%%%%%%%%%%%%%%%%%%%%%%%%%%%%%%%%%%%%%%%%%%%% 
%-------------------------------------------------------------------------%
 In these plots we show GCs orbital evolution during 5 Gyr of forward integration in {\tt 411321} external time-variable potential. Each GC was integrated as one physical particle. We present only 11 cases from 50 GCs to demonstrate the evolution of orbital {\tt ecc} from 0.0 to 0.9. We discuss grey colour models (with dynamical friction) in App. \ref{app:dyn-fric}.

\begin{figure*}[hbp!]
\centering
\includegraphics[width=0.32\linewidth]{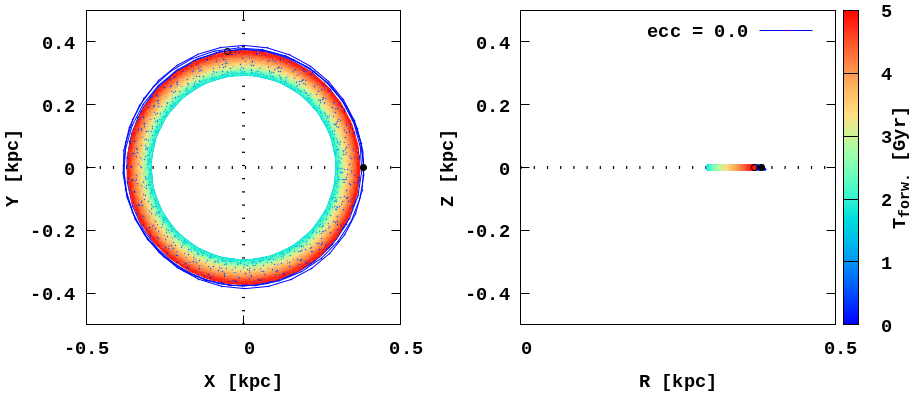}
\includegraphics[width=0.30\linewidth]{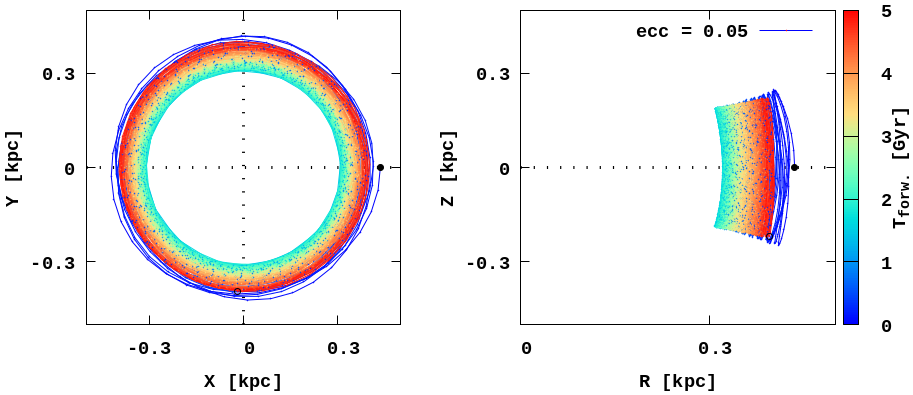}
\includegraphics[width=0.32\linewidth]{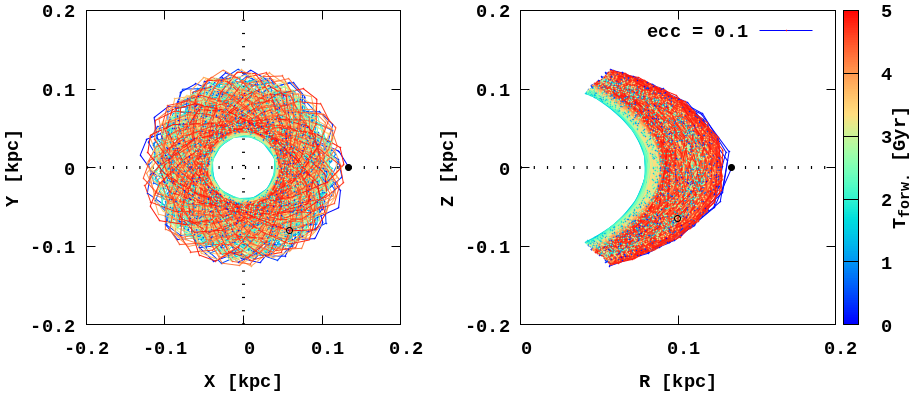}

\includegraphics[width=0.30\linewidth]{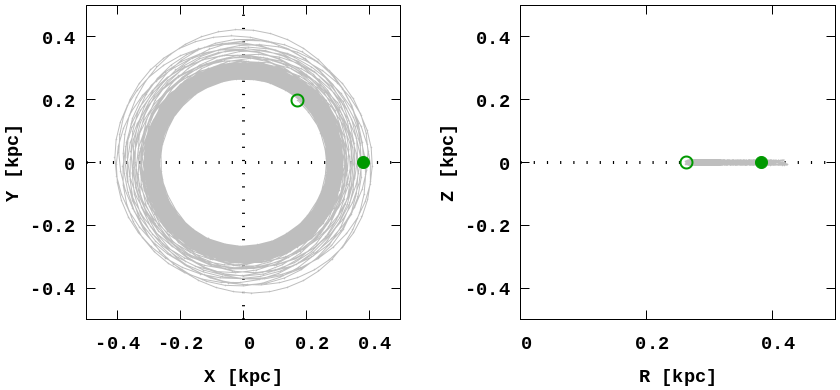}
\includegraphics[width=0.30\linewidth]{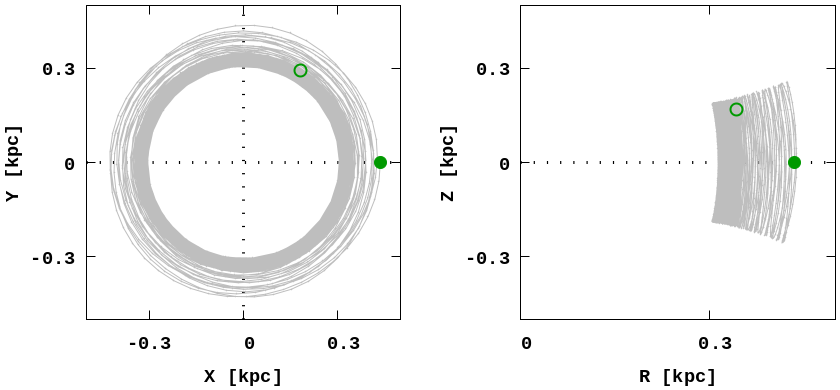}
\includegraphics[width=0.30\linewidth]{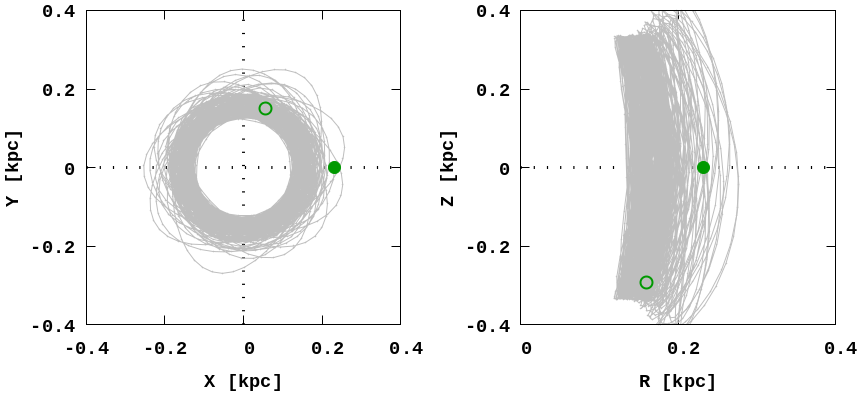}

\includegraphics[width=0.32\linewidth]{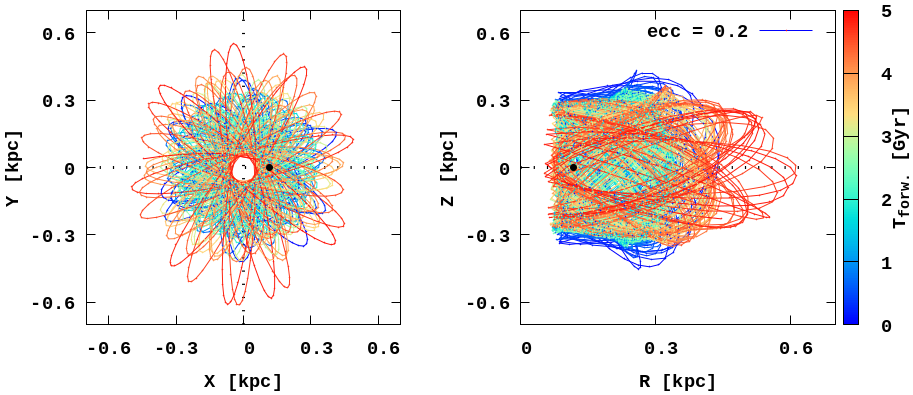}
\includegraphics[width=0.30\linewidth]{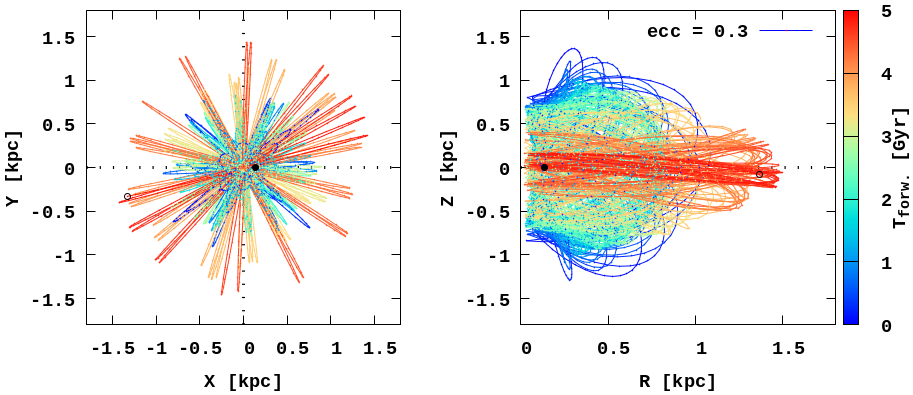}
\includegraphics[width=0.30\linewidth]{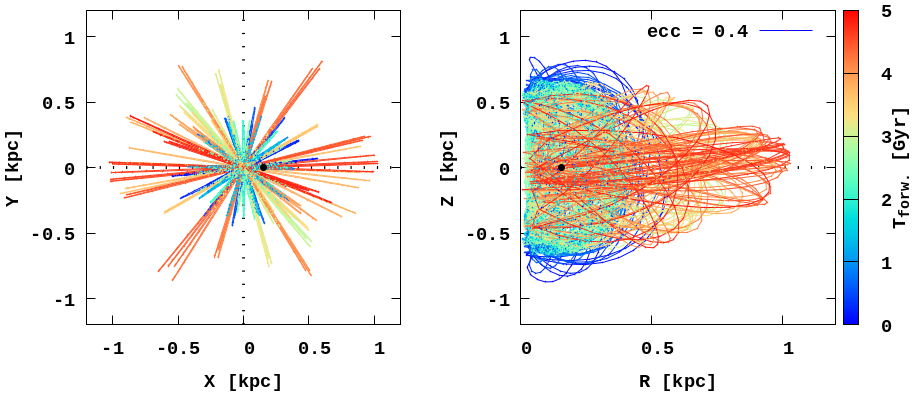}

\includegraphics[width=0.30\linewidth]{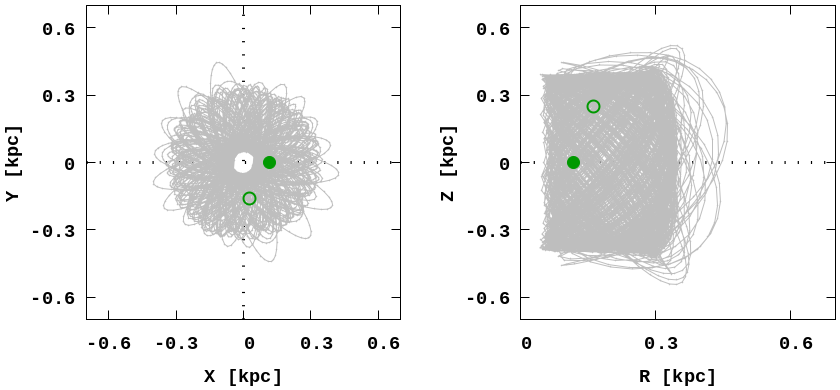}
\includegraphics[width=0.30\linewidth]{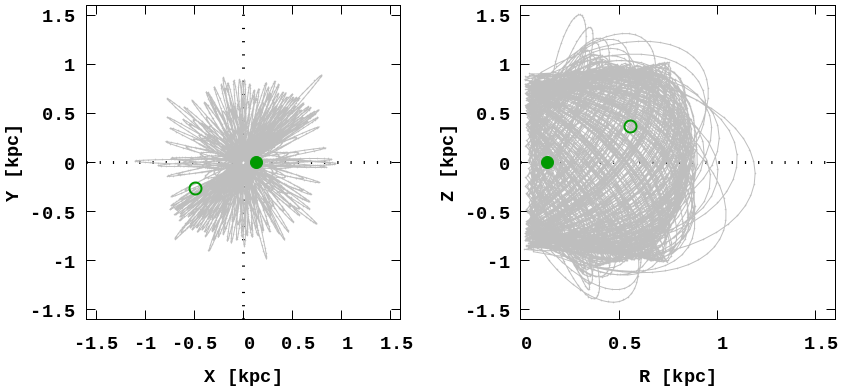}
\includegraphics[width=0.30\linewidth]{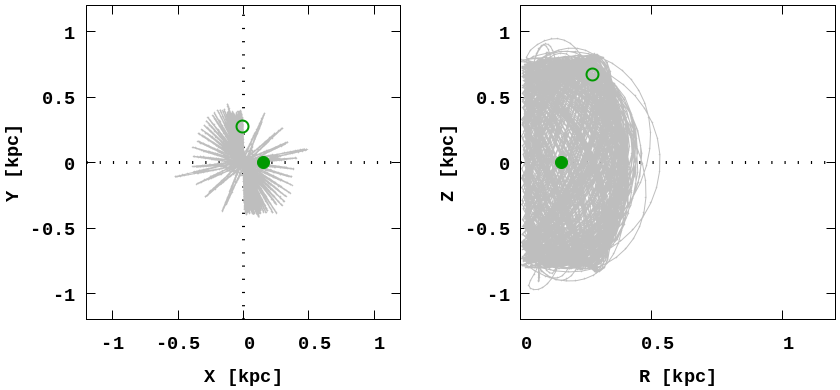}

\includegraphics[width=0.30\linewidth]{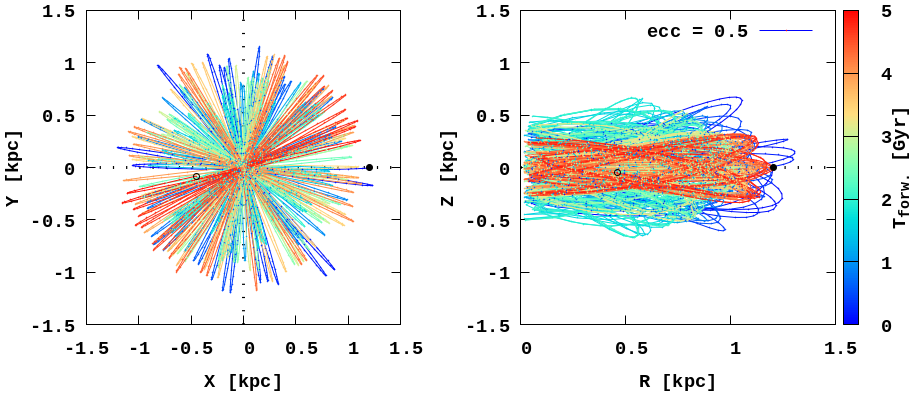}
\includegraphics[width=0.32\linewidth]{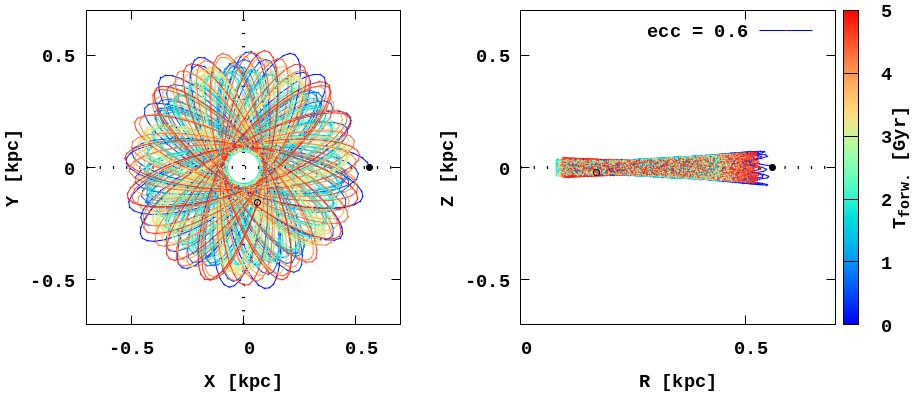}
\includegraphics[width=0.32\linewidth]{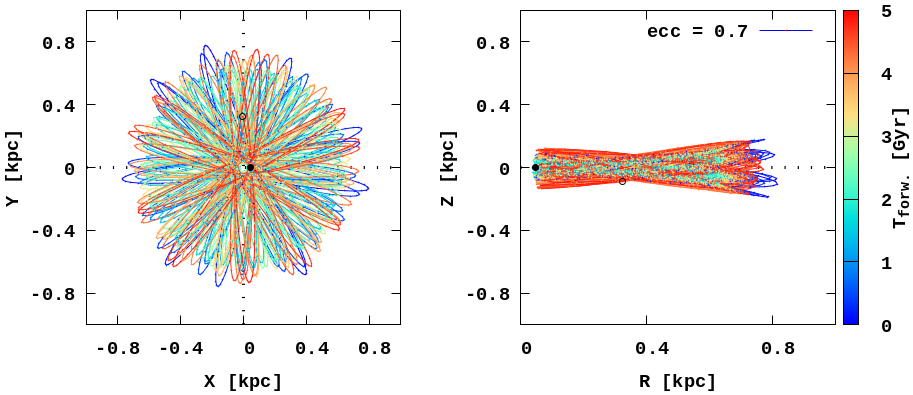}

\includegraphics[width=0.30\linewidth]{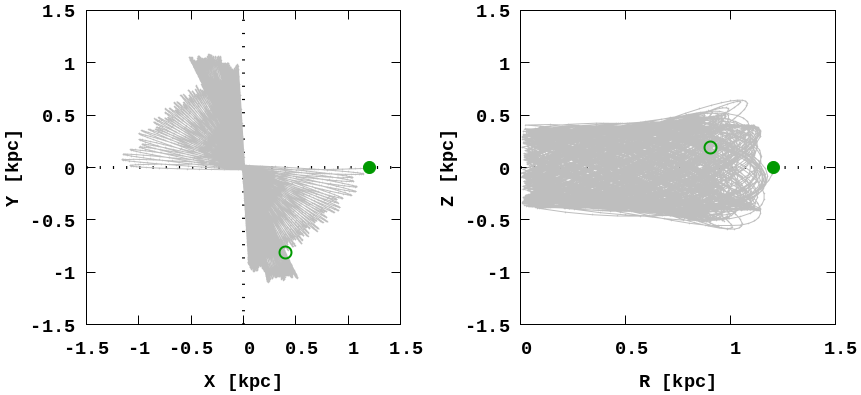}
\includegraphics[width=0.30\linewidth]{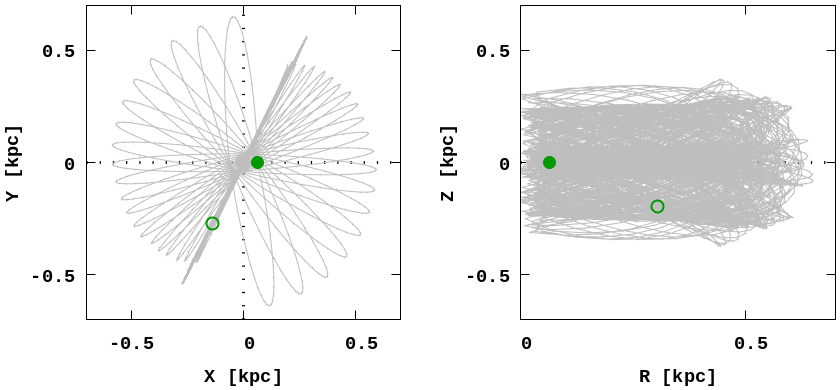}
\includegraphics[width=0.30\linewidth]{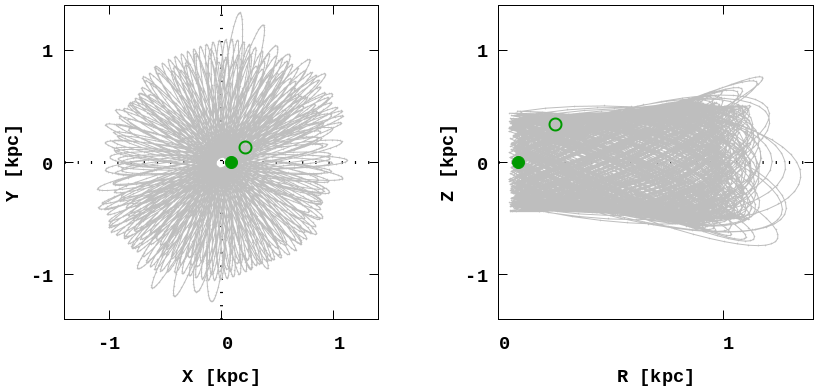}

\includegraphics[width=0.34\linewidth]{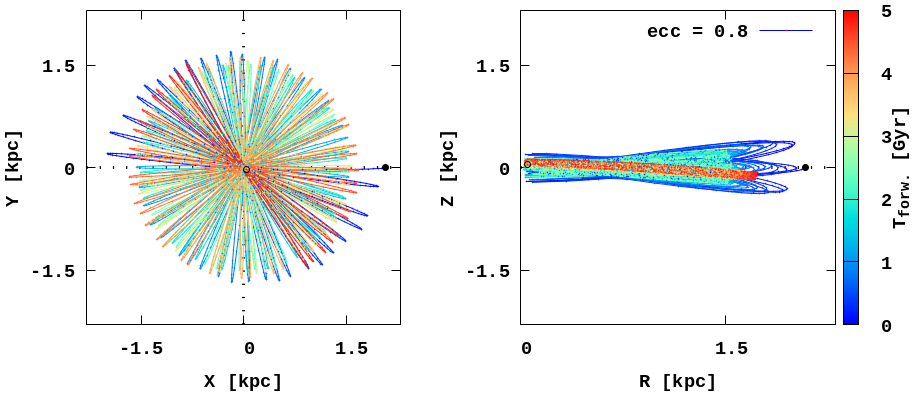}
\includegraphics[width=0.34\linewidth]{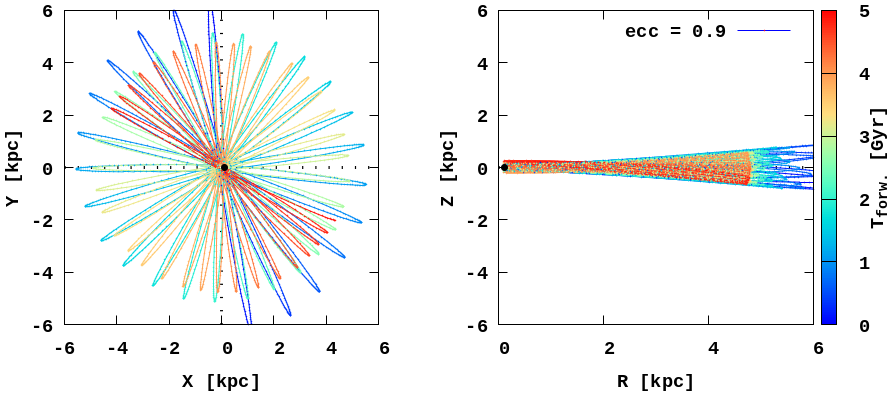}

\includegraphics[width=0.32\linewidth]{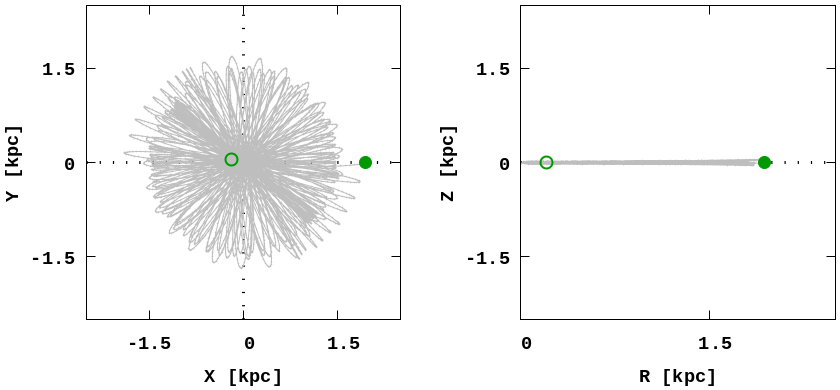}
\includegraphics[width=0.32\linewidth]{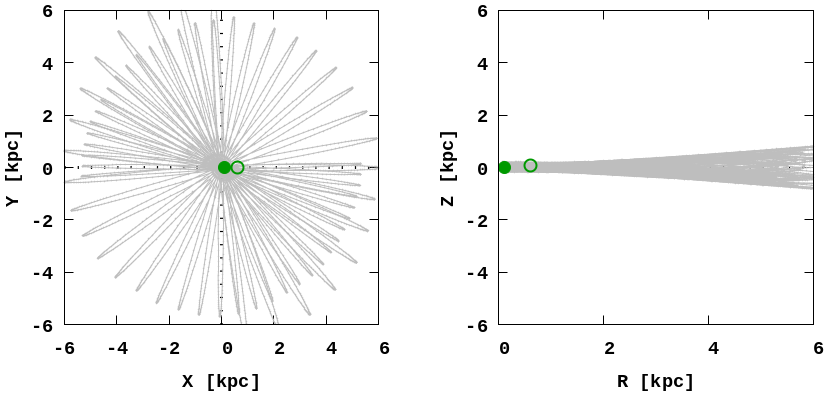}

\caption{Orbital evolution during 5 Gyr of integration, from -10 Gyr, represents orbits with different {\tt ecc}, from 0.0 to 0.9. Orbits with dynamical friction (grey colour) we discuss in App. \ref{app:dyn-fric}.}
\label{fig:orb-evol}
\end{figure*}
%-------------------------------------------------------------------------%

%%%%%%%%%%%%%%%%%%%%%%%%%%%%%%%%%%%%%%%%%%%%%%%%%%%%%%%%%%%%%%%%%%%%%
\section{Detailed statistic for GC models} \label{app:detail-table}
%%%%%%%%%%%%%%%%%%%%%%%%%%%%%%%%%%%%%%%%%%%%%%%%%%%%%%%%%%%%%%%%%%%%%

Table \ref{tab:stat} presents the mean orbital parameters -- semi-major axis (a), eccentricity ({\tt ecc}), and pericentre (r$_{\rm per}$) -- of all our 50 GC models with M$_{\rm ini}$ = 60$\times10^{3}$ M$_{\odot}$, obtained from their preliminary backward integration (to -10 Gyr) as point objects, as well as the results of subsequent integration (to -5 Gyr) of theoretical full $N$-body GC models with different half-mass radii r$_{\rm hm}$ = 1, 2, and 4 pc. As can be seen, the largest contribution to the proto-NSC comes from models with {\tt ecc} values of 0.4 and 0.5. It can also be noted that the more diffuse the cluster, the higher the probability of its complete dissolution (ds). For example, at r$_{\rm hm}$ = 4 pc, more than 80\% of the GC models are fully dissoluted (we consider a GC dissoluted if less than 5\% of its initial mass remains).

\begin{table*}[h]
\caption{Detailed statistics on 50 theoretical GCs (initial orbital parameters and\ accretion) for r$_{\rm hm}$ 1, 2, and 4 pc.}
\centering
%\sisetup{separate-uncertainty}
\resizebox{0.81\textwidth}{!}{
\begin{tabular}{c|ccc|cccc|cccc|cccc}

\hline
\hline
  &  &  &  & \multicolumn{4}{c|}{r$_{\rm hm}$ = 1 pc} & \multicolumn{4}{c|}{r$_{\rm hm}$ = 2 pc} & \multicolumn{4}{c}{r$_{\rm hm}$ = 4 pc} \\ \cline{5-16}
N$_{\rm mod}$  & a & {\tt ecc} & r$_{\rm per}$ & T & N* & N$_{\rm accr}$ & M$_{\rm accr}$& T & N* & N$_{\rm accr}$ & M$_{\rm accr}$ & T & N* & N$_{\rm accr}$ & M$_{\rm accr}$ \\

 & pc &  & pc & Gyr & 10$^3$ &  & M$_\odot$ & Gyr & 10$^3$ &  & M$_\odot$ & Gyr & 10$^3$ &  &  M$_\odot$ \\
 (1) & (2) & (3) & (4) & (5) & (6) & (7) & (8) & (9) & (10) & (11) & (12) & (13) & (14) & (15) & (16) \\
\hline
\hline
194   & 326$\pm$23   & 0.009$\pm$0.001  & 323$\pm$23  & 5 & 96   & 5      & 6      & 5 & 65   & 7      & 17     & 5 & 78   & 7      & 17      \\
684   & 494$\pm$36   & 0.033$\pm$0.015  & 478$\pm$36  & 5 & 90   & 6      & 13     & 5 & 83   & 6      & 13     & 5 & 52   & 6      & 7       \\
205   & 377$\pm$26   & 0.054$\pm$0.009  & 356$\pm$25  & 5 & 90   & 6      & 18     & 5 & 78   & 6      & 26     & 5 & 52   & 6      & 18      \\
936   & 519$\pm$37   & 0.085$\pm$0.024  & 475$\pm$36  & 5 & 85   & 4      & 6      & 5 & 70   & 5      & 21     & 5 & 81   & 1      & 8       \\
\hline
149   & 223$\pm$24   & 0.130$\pm$0.198 & 151$\pm$51  & 5 & 43   & 21     & 56     & 5 & 33   & 24     & 58     & 5 & ds    & 19     & 33     \\ 
19    & 112$\pm$6    & 0.117$\pm$0.038  & 99$\pm$8    & 5 & 55   & 31     & 125    & 5 & 19   & 25     & 104    & 5 & ds   & 47     & 223     \\
20    & 113$\pm$5    & 0.176$\pm$0.017  & 93$\pm$4    & 5 & 70   & 19     & 42     & 5 & 48   & 16     & 88     & 5 & 15   & 17     & 81      \\
381   & 332$\pm$41   & 0.178$\pm$0.209 & 208$\pm$84  & 5 & 66   & 16     & 94     & 5 & 30   & 9      & 38     & 5 & ds   & 7      & 30      \\
247   & 269$\pm$33   & 0.127$\pm$0.206 & 184$\pm$72  & 5 & 62   & 47     & 77     & 5 & 29   & 65     & 71     & 5 & ds   & 48     & 64     \\
\hline  
1730  & 502$\pm$63   & 0.208$\pm$0.218 & 297$\pm$136 & 5 & 45   & 422    & 121    & 4 & ds   & 44     & 30     & 5 & ds   & 228    & 100     \\
96    & 231$\pm$14   & 0.217$\pm$0.026  & 181$\pm$13  & 5 & 16   & 17     & 33     & 5 & ds   & 15     & 93     & 5 & 16   & 17     & 33      \\
415   & 311$\pm$36   & 0.22$\pm$0.173   & 191$\pm$66  & 5 & 73   & 7      & 22     & 5 & 41   & 8      & 22     & 5 & 6    & 9      & 21     \\
7     & 87$\pm$3     & 0.227$\pm$0.036  & 67$\pm$4    & 5 & 61   & 22     & 43     & 5 & 25   & 20     & 67     & 5 & ds   & 30     & 81     \\
6     & 90$\pm$8     & 0.296$\pm$0.118 & 54$\pm$14   & 5 & 37   & 28     & 132    & 5 & 7    & 26     & 89     & 3 & ds   & 27     & 110     \\
\hline
47    & 179$\pm$15   & 0.313$\pm$0.092  & 124$\pm$24  & 5 & 63   & 8      & 10     & 5 & 32   & 12     & 36     & 5 & ds   & 15     & 50      \\
3685  & 721$\pm$88   & 0.324$\pm$0.223 & 333$\pm$179 & 5 & 66   & 124    & 58     & 5 & 31   & 135    & 43     & 4 & ds   & 84     & 56      \\
270   & 291$\pm$31   & 0.332$\pm$0.160 & 149$\pm$54  & 5 & 9    & 1 211  & 411    & 5 & ds   & 669    & 299    & 1.5 & ds & 159    & 125    \\
34   & 210$\pm$20   & 0.333$\pm$0.119 & 116$\pm$30  & 5 & 47   & 22     & 94     & 5 & 11   & 21     & 94     & 4 & ds   & 17     & 53     \\
90    & 96$\pm$6     & 0.368$\pm$0.069  & 61$\pm$9    & 5 & 61   & 76     & 40     & 5 & ds   & 50     & 38     & 5 & ds   & 40     & 30     \\
78    & 108$\pm$7    & 0.369$\pm$0.081  & 68$\pm$12   & 5 & 38   & 710    &  242   & 5 & 11   & 690    & 275    & 5 & ds   & 779    & 343     \\
\hline
2457  & 578$\pm$76   & 0.423$\pm$0.238  & 346$\pm$172 & 5 & 52   & 2 846  & 870    & 5 & 27   & 2 631  & 876    & 3 & ds   & 2 019  & 779     \\
1059  & 447$\pm$58   & 0.454$\pm$0.189  & 249$\pm$104 & 5 & 61   & 826    & 310    & 5 & 25   & 1 072  & 476    & 5 & ds   & 367    & 280     \\
264   & 69$\pm$8     & 0.469$\pm$0.136 & 47$\pm$14   & 5 & 14   & 8 568  & 2 771  & 5 & 3    & 4 438  & 1 549  & 3 & ds   & 2 589  & 1 021  \\
1370  & 483$\pm$65   & 0.477$\pm$0.209  & 257$\pm$119 & 3 & ds   & 78 011 & 30 325 & 5 & 5    & 41 714 & 14 020 & 3 & ds   & 39 453 & 15 796  \\
2065  & 540$\pm$67   & 0.481$\pm$0.216  & 289$\pm$141 & 0.6 & ds & 90 270 & 36 903 & 0.6 & ds & 90 879 & 36 567 & 0.6 & ds & 91 663 & 36 189  \\
1723  & 529$\pm$69   & 0.493$\pm$0.195  & 269$\pm$113 & 5 & 53   & 2 145  & 647    & 5 & 20   & 1 819  & 582    & 5 & ds & 1 714  & 618     \\
\hline
1371  & 482$\pm$64   & 0.506$\pm$0.194  & 239$\pm$103 & 5 & 22   & 29 573 & 8 896  & 5 & 9    & 30 756 & 9 961  & 3.5 & ds & 17 174 & 5 940   \\
1392  & 463$\pm$52   & 0.507$\pm$0.224  & 233$\pm$120 & 5 & 27   & 5 351  & 1 637  & 5 & 8    & 5 481  & 1 824  & 3 & ds   & 2 150  & 844     \\
2858  & 615$\pm$59   & 0.511$\pm$0.145 & 213$\pm$93  & 5 & 35   & 8 739  & 2 673  & 5 & 10   & 10 486 & 3 511  & 3 & ds   & 7 378  & 2 669   \\
155   & 261$\pm$26   & 0.519$\pm$0.163 & 171$\pm$55  & 2.4 & ds & 56 975 & 23 665 & 0.9 & ds & 51 467 & 22 122 & 1 & ds   & 45 755 & 18 917  \\
156   & 288$\pm$34   & 0.541$\pm$0.195  & 136$\pm$67  & 3 & ds   & 77 522 & 29 848 & 0.9 & ds & 35 685 & 12 698 & 2 & ds   & 75 090 & 28 654  \\
82    & 251$\pm$26   & 0.560$\pm$0.159  & 112$\pm$45  & 5 & ds   & 31 946 & 11 261 & 2 & ds  & 28 980 & 10 945 & 1 & ds   & 23 752 & 9 759   \\
\hline
1689  & 519$\pm$45   & 0.603$\pm$0.113 & 150$\pm$64  & 5 & 63   & 35     & 55     & 5 & 14   & 24     & 16     & 5 & ds   & 16     & 30     \\
175   & 290$\pm$21   & 0.615$\pm$0.060  & 112$\pm$21  & 5 & 67   & 10     & 66     & 5 & 30   & 13     & 70     & 5 & ds   & 13     & 60     \\
46    & 202$\pm$16   & 0.621$\pm$0.096  & 77$\pm$24   & 5 & 25   & 16     & 90     & 5 & ds   & 10     & 57     & 2 & ds   & 17     & 78     \\       
600   & 376$\pm$29   & 0.628$\pm$0.091 & 110$\pm$38  & 5 & 41   & 33     & 18     & 5 & 11   & 14     & 21     & 4 & ds   & 11     & 30     \\
12    & 150$\pm$12   & 0.663$\pm$0.100 & 66$\pm$19   & 5 & 29   & 26     & 80     & 5 & ds   & 24     & 76     & 1.8 & ds & 12     & 66     \\
167   & 200$\pm$14   & 0.657$\pm$0.072 & 83$\pm$18   & 5 & 38   & 14     & 34     & 5 & 6    & 12     & 11     & 3 & ds   & 7      & 22     \\
\hline
1404  & 394$\pm$33   & 0.715$\pm$0.087  & 114$\pm$40  & 5 & 55   & 70     & 25     & 5 & 30   & 30     & 16     & 3.5 &  ds & 23     & 32     \\
1636  & 527$\pm$37   & 0.726$\pm$0.069  & 145$\pm$40  & 5 & 60   & 4      & 15     & 5 & 23   & 5      & 27     & 5 & ds   & 6      & 18     \\
2022  & 538$\pm$42   & 0.754$\pm$0.083  & 134$\pm$50  & 5 & 57   & 10     & 19     & 5 & 15   & 6      & 41     & 5 & ds   & 7      & 42     \\
3353  & 690$\pm$52   & 0.770$\pm$0.071  & 160$\pm$55  & 5 & 77   & 1      & 1      & 5 & 28   & 4      & 16     & 5 & ds   & 3      & 4      \\
2799  & 650$\pm$49   & 0.776$\pm$0.077  & 147$\pm$56  & 5 & 75   & 8      & 64     & 5 & 40   & 0      & 0      & 5 & ds   & 6      & 18     \\
\hline
5906  & 838$\pm$61   & 0.803$\pm$0.077  & 167$\pm$70  & 5 & 61   & 7      & 12     & 5 & 24   & 7      & 9      & 5 & ds   & 6      & 22     \\
6446  & 898$\pm$66   & 0.803$\pm$0.079  & 180$\pm$77  & 5 & 72   & 2 533  & 792    & 5 & 15   & 2 559  & 891    & 5 & ds   & 2 160  & 882    \\
7513  & 957$\pm$65   & 0.806$\pm$0.076  & 188$\pm$77  & 5 & 73   & 3      & 15     & 5 & 15   & 1      & 24     & 5 & ds   & 2      & 37     \\
7531  & 973$\pm$67   & 0.815$\pm$0.074  & 182$\pm$77  & 5 & 76   & 3      & 4      & 5 & 30   & 4      & 16     & 5 & ds   & 4      & 51     \\
13707 & 1629$\pm$119 & 0.866$\pm$0.064  & 220$\pm$109 & 5 & 76   & 8      & 4      & 5 & 42   & 9      & 7      & 5 & ds   & 2      & 2      \\
\hline
15946 & 2526$\pm$205 & 0.901$\pm$0.041  & 251$\pm$103 & 5 & 82   & 1      & 1      & 5 & 51   & 1      & 23     & 5 & ds   & 0      & 0      \\
16768 & 2698$\pm$229 & 0.902$\pm$0.040  & 264$\pm$111 & 5 & 82   & 2      & 13     & 5 & 51   & 3      & 4      & 5 & 7    & 2      & 3      \\
\hline
\end{tabular}
}
\label{tab:stat}
\end{table*} 
%-------------------------------------------------------------------------%

%%%%%%%%%%%%%%%%%%%%%%%%%%%%%%%%%%%%%%%%%%%%%%%%%%%%%%%%%%%%%%%%%%%%%
\section{Evolution of the ecc, a, and inclination angle} \label{app:orb-param}
%%%%%%%%%%%%%%%%%%%%%%%%%%%%%%%%%%%%%%%%%%%%%%%%%%%%%%%%%%%%%%%%%%%%% 

\begin{figure*}[h!]
\centering
\includegraphics[width=0.7\linewidth]{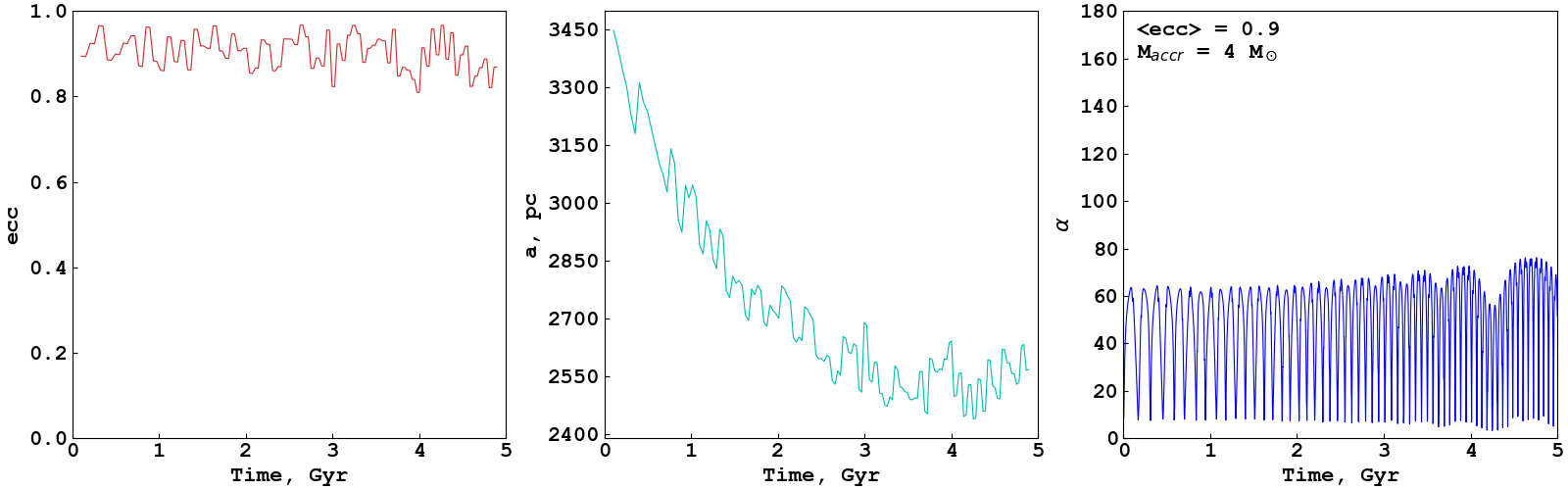}
\includegraphics[width=0.7\linewidth]{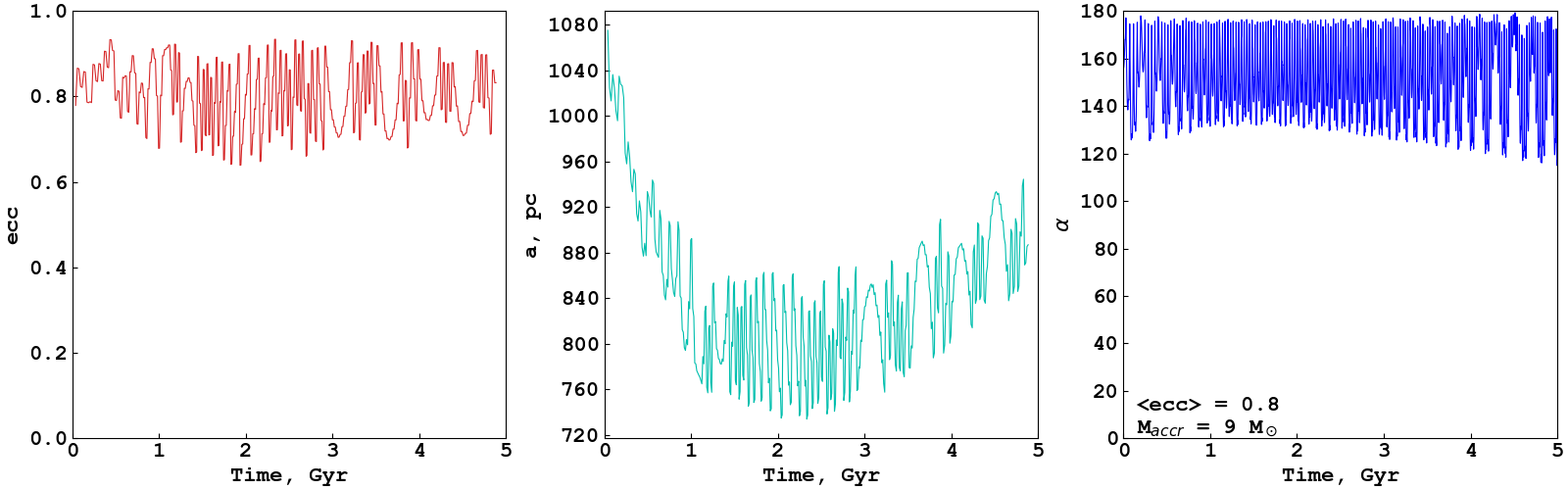}
\includegraphics[width=0.7\linewidth]{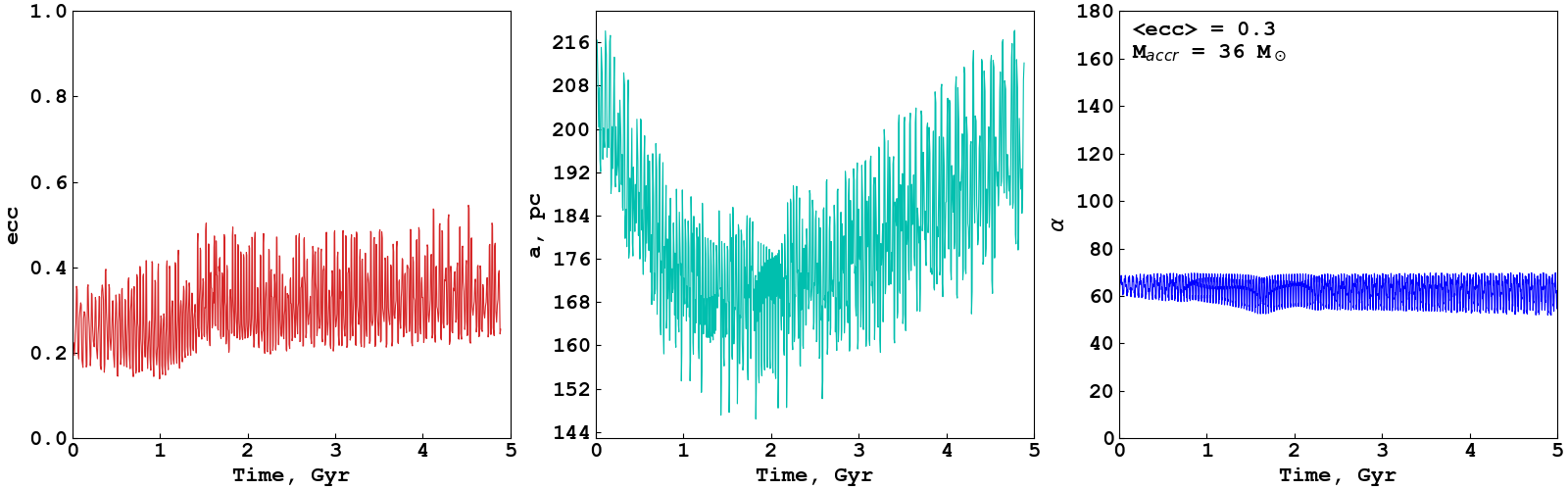}
\includegraphics[width=0.7\linewidth]{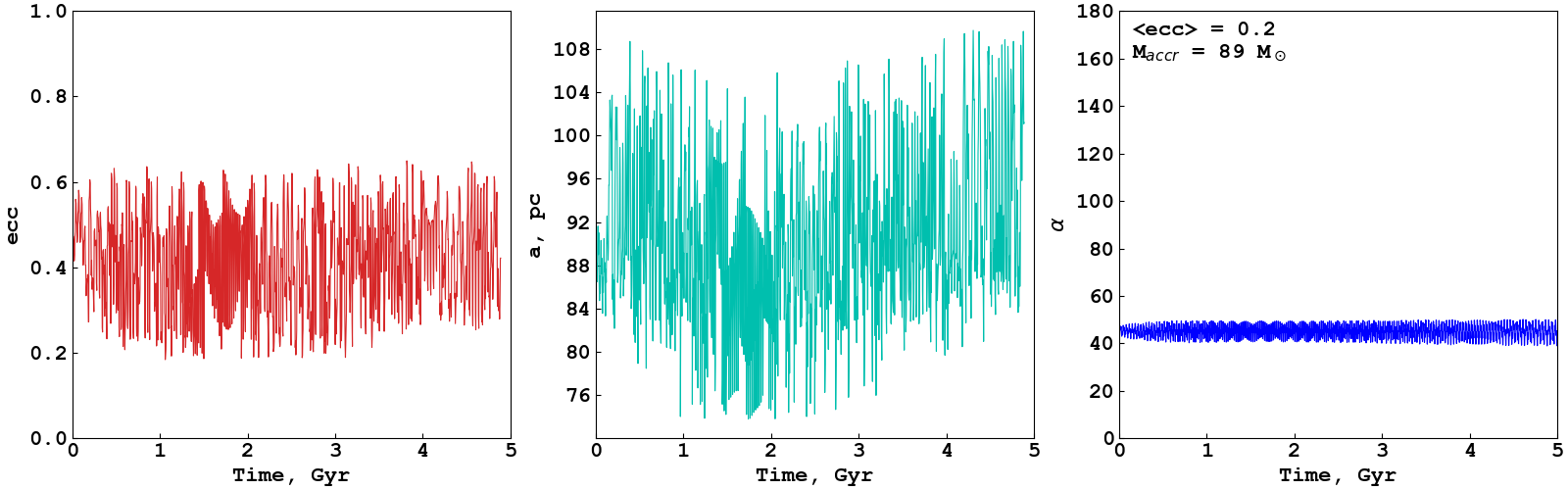}
\includegraphics[width=0.7\linewidth]{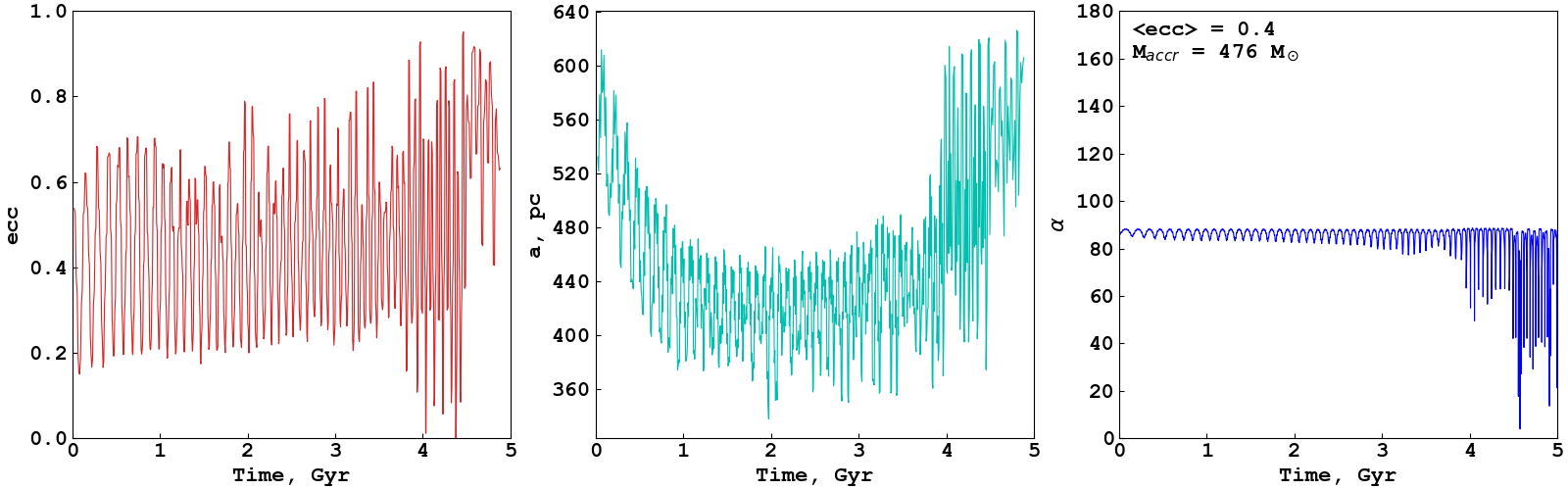}
\includegraphics[width=0.7\linewidth]{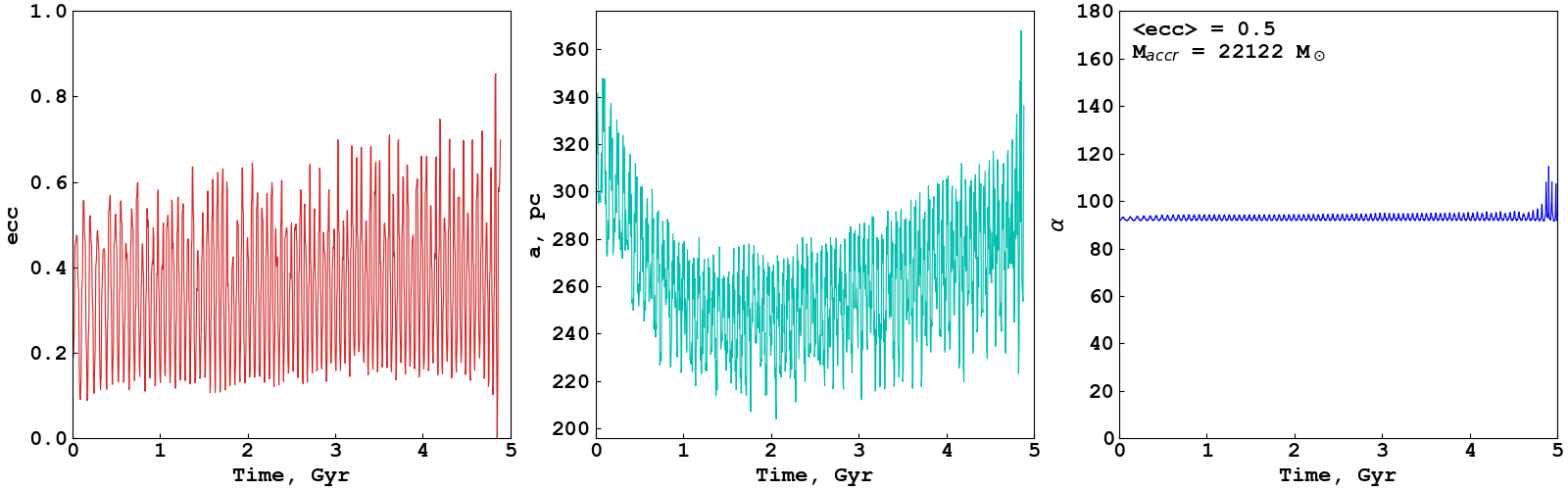}
\caption{Evolution of the orbital parameters {\tt ecc}, a, and $\alpha$ over 5 Gyr of GCs simulation. \\ First two plots represent models with accretion less then 10 M$_\odot$, two middle panels -- 10--100 M$_\odot$ and two bottom -- more than 100 M$_\odot$.}
\label{fig:accr-10-100}
\end{figure*}

Fig. \ref{fig:accr-10-100} shows the evolution of orbital parameters (eccentricity, semi-major axis, inclination angle) over 5 Gyr for representative GC models that contributed less than 10 M$_\odot$, between 10 and 100 M$_\odot$, and more than 100 M$_\odot$. As shown, models whose eccentricities and semi-major axes oscillated over a broad range of values made a larger contribution to the accreted mass. However, this trend does not extend to the inclination angle.

%-------------------------------------------------------------------------%

%-------------------------------------------------------------------------%

\clearpage
%%%%%%%%%%%%%%%%%%%%%%%%%%%%%%%%%%%%%%%%%%%%%%%%%%%%%%%%%%%%%%%%%%%%
\section{Additional analysis of the accretion rate for the stellar remnants from GCs}\label{sec:addi-accr-rem}
%%%%%%%%%%%%%%%%%%%%%%%%%%%%%%%%%%%%%%%%%%%%%%%%%%%%%%%%%%%%%%%%%%%%

\begin{figure*}[htb!]
\centering
\includegraphics[width=0.32\linewidth]{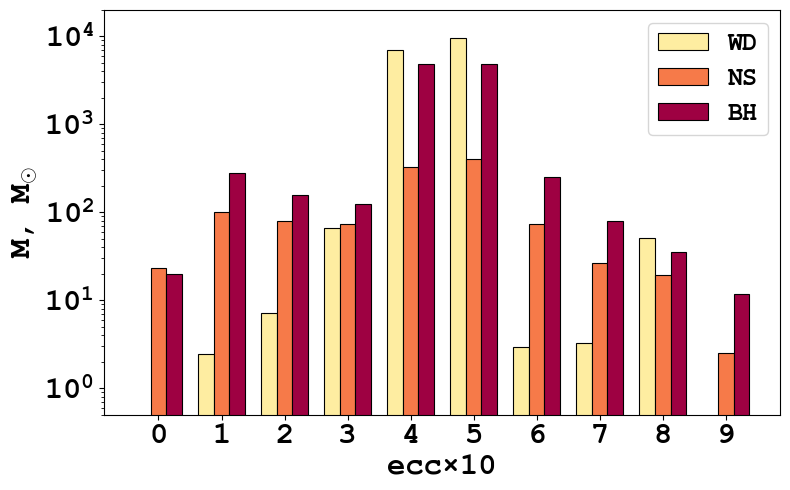}
\includegraphics[width=0.32\linewidth]{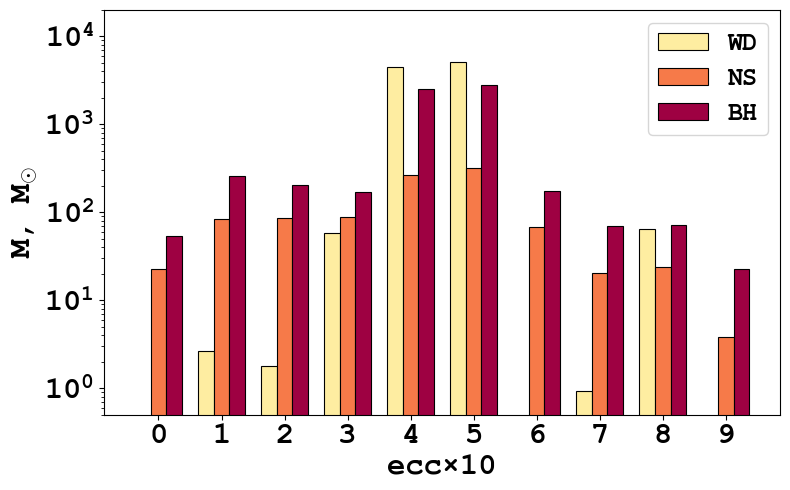}
\includegraphics[width=0.32\linewidth]{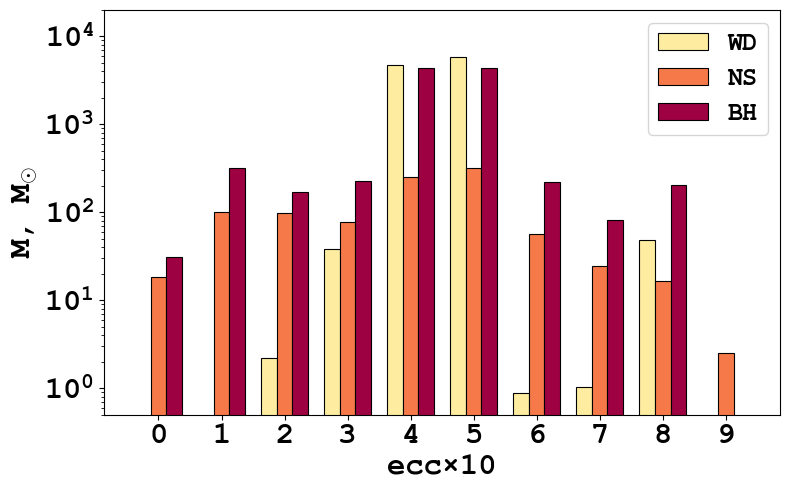}
\caption{Masses of accreted stellar remnants (WDs, NSs, and BHs, indicated by colours) from GCs models as a function of the orbital eccentricity of their host clusters, shown for r$_{\rm hm} $ = 1, 2, and 4 pc (from left to right).}
\label{fig:accr_remn_1_2_4}
\end{figure*}

In Fig. \ref{fig:accr_remn_1_2_4}, we present the distribution of the NSC accreted masses as a function of the GCs initial orbital {\tt ecc}. At all evolutionary stages, the largest contribution to accretion comes from GC models with mean orbital eccentricities of 0.4 and 0.5. These trends are clear for all three cases of r$_{\rm hm}$ = 1, 2, and 4 pc. Interestingly, for white dwarfs in all three cases, the accreted mass drops sharply beyond the peak at eccentricities 0.6 and 0.7, then it increases sharply at eccentricity 0.8. In contrast, on the lower-eccentricity side, the mass decreases more gradually, without a pronounced dip.

%-------------------------------------------------------------------------%
%-------------------------------------------------------------------------%

\begin{figure*}[h!]
\centering
\includegraphics[width=0.3\linewidth]{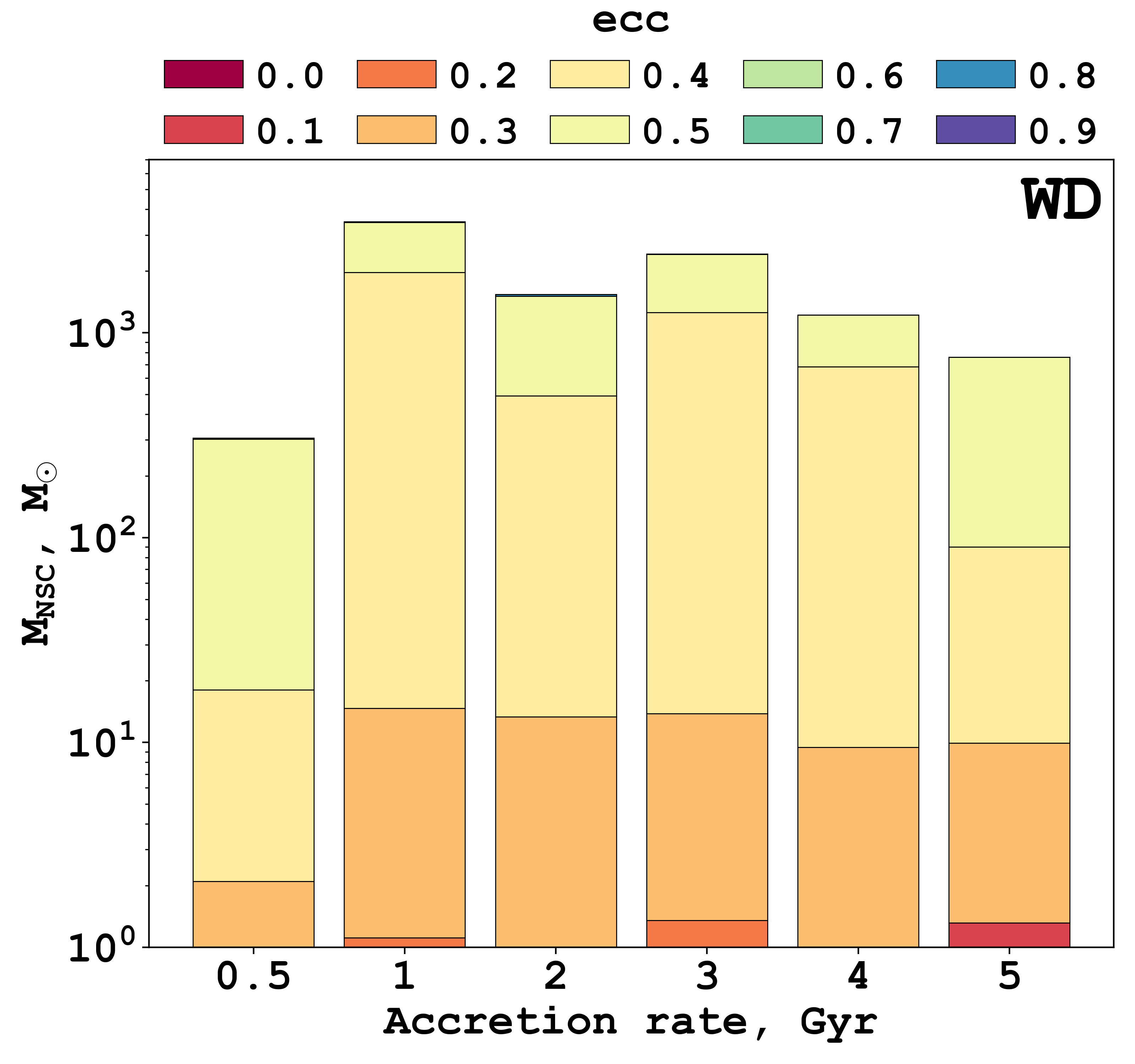}
\includegraphics[width=0.3\linewidth]{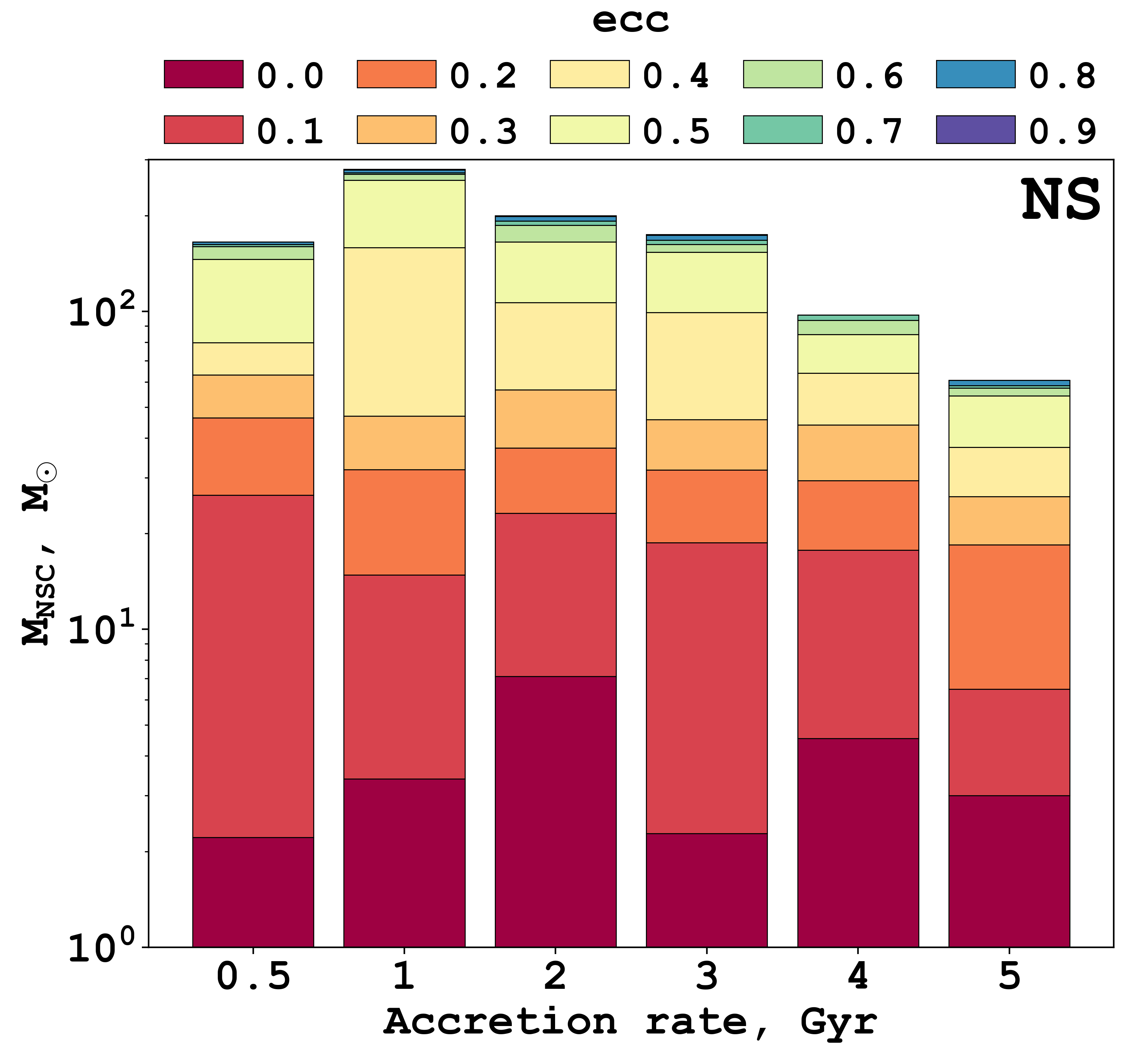}
\includegraphics[width=0.3\linewidth]{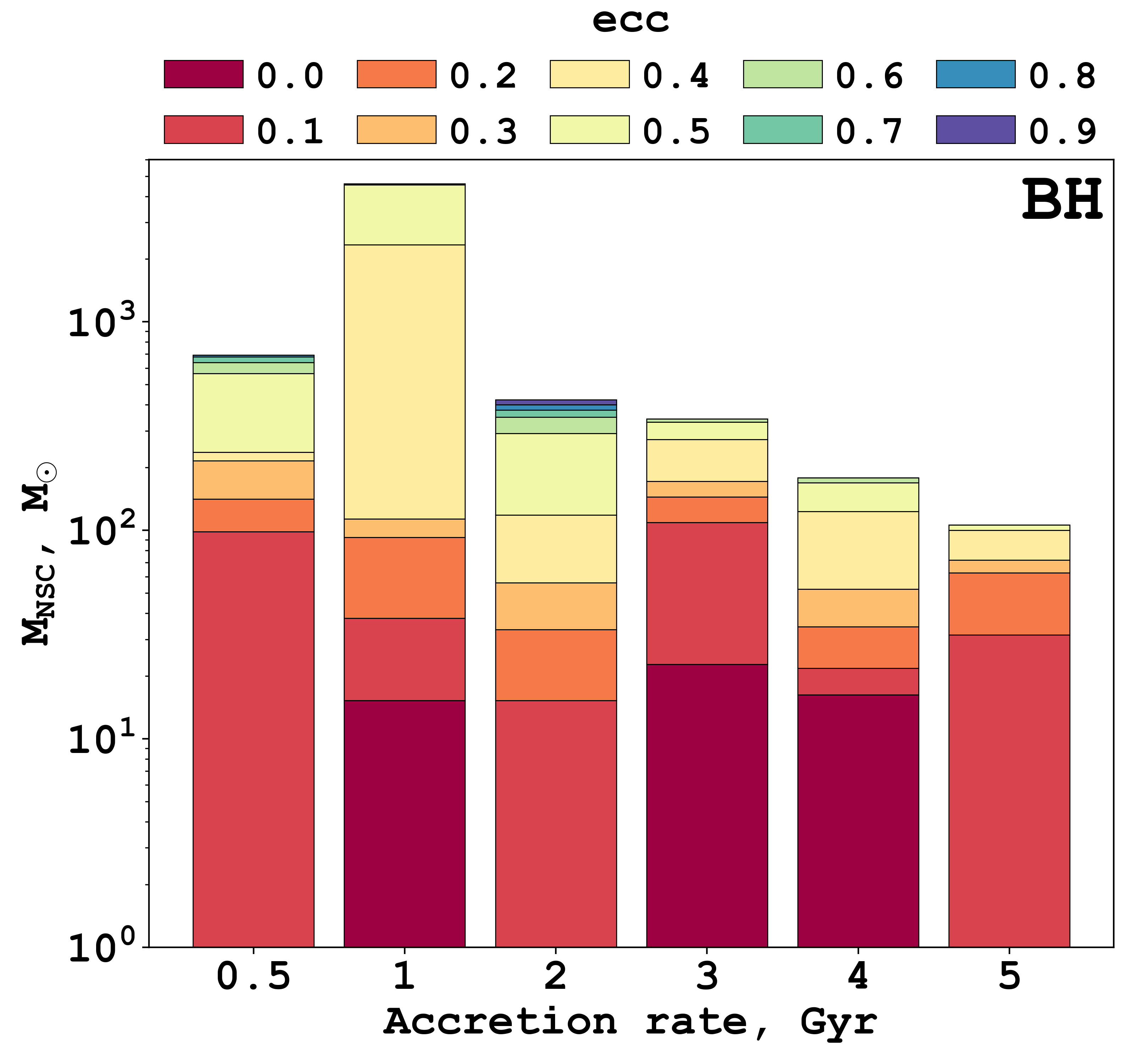}
\caption{Contribution of individual types of stellar remnants to the NSC mass (from left to right: WDs, NSs, and BHs) from all 50 GC models over 5 Gyr for different orbital eccentricities in the case of r$_{\rm hm}$ = 2 pc.}
\label{fig:accr_remn_individ}
\end{figure*}

In Fig. \ref{fig:accr_remn_individ} we illustrate the contribution of individual types of stellar remnant to the NSC mass over 5 Gyr (in different time bins), based on the 50 GC models with initial r$_{\rm hm}$ = 2 pc. The results are shown for different mean orbital eccentricities. The highest accretion rate for all stellar remnants occurs in the time interval between 0.5 and 1 Gyr. Then it gradually declines, with a smoother decrease in the case of WDs and a more pronounced drop for NSs and BHs. 

%-------------------------------------------------------------------------%
%-------------------------------------------------------------------------%

\begin{figure*}[h!]
\centering
\includegraphics[width=0.45\linewidth]{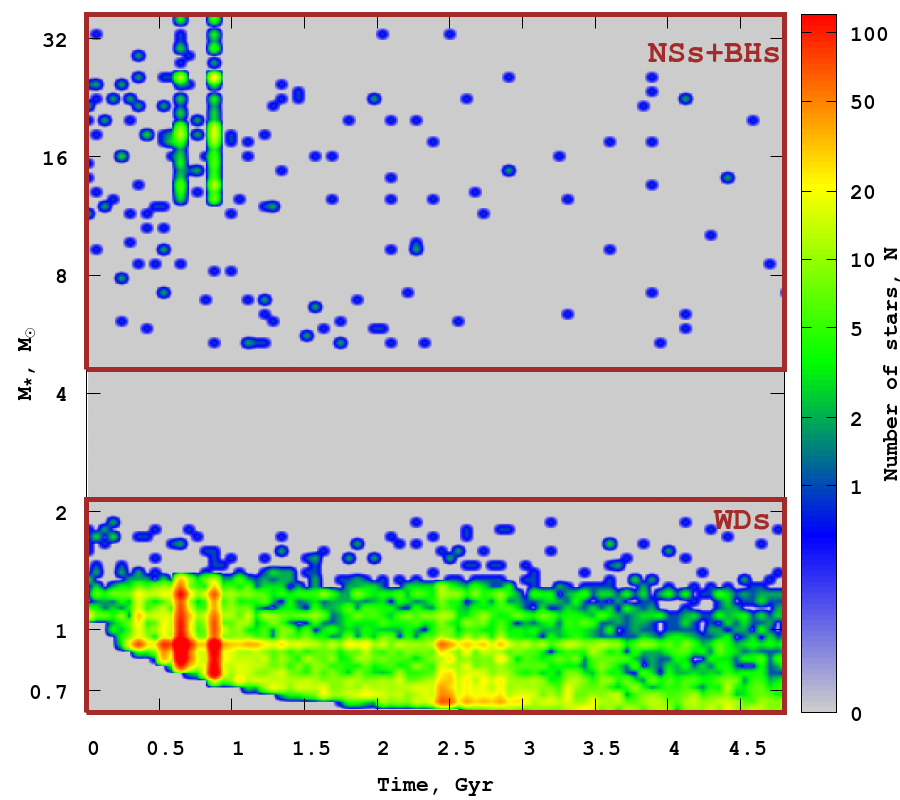}
\includegraphics[width=0.45\linewidth]{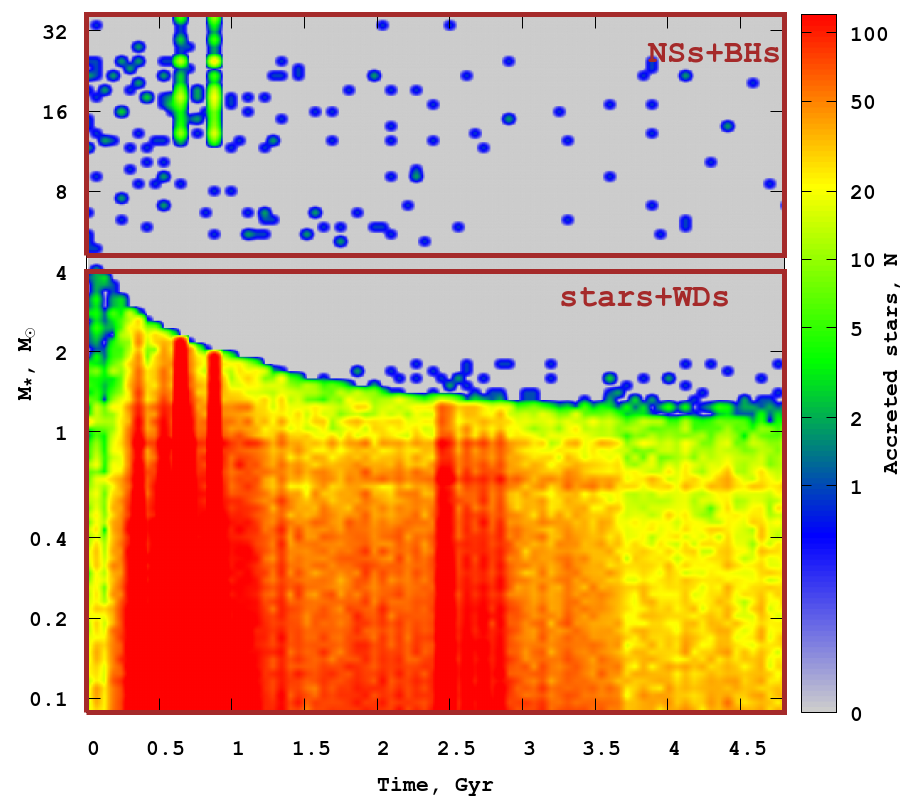}
\caption{Stellar remnants accretion rate onto the NSC for 50 GCs models for r$_{\rm hm}$ = 2 pc. Left: the mass–time distribution for stellar remnants (WDs, NSs, and BHs); right: for all stars. From the left graph, the boundary between white dwarfs at the bottom and neutron stars and black holes at the top is clearly visible (compact object mass gap). It is also clearly seen in the right figure that the main accretion number contribution comes from low-mass stars, approximately up to $\sim$1.4 solar masses (according to the Chandrasekhar limit).}
\label{fig:accr-2}
\end{figure*}

Fig. \ref{fig:accr-2} presents the dynamics of individual stellar accretion events onto the NSC for 50 GCs in the case with r$_{\rm hm}$ = 2 pc. For models with initial r$_{\rm hm}$ = 1 and 4 pc, we have similar accretion plots. The data are shown as a 2D histogram of time versus the mass of the accreted star, with a bin size of 100 for both axes. The left panel highlights the moments of accretion only of the stellar remnants, while the right panel includes all stars. Based on the stellar mass, we can clearly distinguish the relative frequencies and accretion timings of black holes, neutron stars, and white dwarfs. The left panel in Fig. \ref{fig:accr-2} clearly illustrates the boundary separating white dwarfs at the lower end from neutron stars and black holes at higher masses. The right panel demonstrates that the dominant contribution arises from low-mass stars with masses up to approximately 1.4 solar masses, corresponding to the Chandrasekhar limit.

%-------------------------------------------------------------------------%

%%%%%%%%%%%%%%%%%%%%%%%%%%%%%%%%%%%%%%%%%%%%%%%%%%%%%%%%%%%%%%%%%%%%%
\section{Influence of the dynamical friction on the accretion rate to the proto-NSC} \label{app:dyn-fric}
%%%%%%%%%%%%%%%%%%%%%%%%%%%%%%%%%%%%%%%%%%%%%%%%%%%%%%%%%%%%%%%%%%%%%

In our current work, we apply additional calculations to determine the influence of dynamical friction on our orbital parameter analysis for the GCs. Dynamical friction in astrophysical context indicates the collective deceleration exerted on a moving massive body by the fluctuating force of field stars. The existence of such an effect was first demonstrated by Chandrasekhar and von Neumann \citep{CN1942, CN1943} in their pioneering works. Later, Chandrasekhar in the studies of \citep{C1943a, C1943b} developed a more quantitative theory of dynamical friction. The resulting drag acceleration acting on the GC can be expressed as \citep{Binney2008}:

\begin{equation}
\frac{{\rm d} {\bf V}_{\rm GC} }{\rm dt} = -\frac{4\pi G^2\rho M_{\rm GC}}{V_{\rm GC}^3} \; \chi \cdot 
        \ln\Lambda \cdot {\bf V}_{\rm GC} 
    \quad \mathrm{with}\quad 
    \chi=\frac{\rho(<V_{\rm GC})}{\rho}.
    \label{dynfric}
\end{equation}

In general, the functions, $\chi$, and the Coulomb logarithm, $\Lambda,$ depend on the velocity of the massive object and on the properties of the background system. For more details on this dynamical friction description, we refer the reader to \cite{Just2011}. Based on the results of our previous study \cite{Just2011}, in our current calculations we use a fixed Coulomb logarithm value $\ln\Lambda$ = 5 and a fixed value for $\chi$ = 0.5.

In Fig. \ref{fig:orb-evol} we present the comparison of the evolution of the GC orbits over 5 Gyr of integration time for models including dynamical friction (grey colour) and without it (coloured lines). As can be seen from the plot, dynamical friction only slightly affects the orbits. In our opinion, applying dynamical friction in its present form does not significantly change the GC orbits, and so the position of lost stars from these GCs on the phase-space diagrams.

Also we would like to mention that one of the key points in our investigation is a quite strong dependence of mass accretion from clusters onto the proto-NSC for orbits with the high inclination angles, see Sect. \ref{sec:int-inc} and Fig. \ref{fig:accr-10-100}. For such high-$Z$ orbits, the effect of dynamical friction is also much weaker compared to models where the GC orbits lie inside the Galactic disk.

\end{appendix}

%%%%%%%%%%%%%%%%%%%%%%%%%%%%%%%%%%%%%%%%%%%%%%%%%%%%%%%%%%%%%%%%%%%%%
\end{document}